\documentclass[twocolumn]{aastex63} 
\usepackage{graphics,graphicx}
\hypersetup{urlcolor=blue}
\usepackage{natbib}
\usepackage[outdir=./]{epstopdf}
\usepackage{textcomp,gensymb}
\usepackage{xfrac}
\usepackage{xcolor}
\usepackage{multirow}

\usepackage{savesym}
\savesymbol{tablenum}
\usepackage{siunitx}
\restoresymbol{SIX}{tablenum}

\renewcommand\deg{\ensuremath{^\circ}}

\newcommand{\water}{H$_2$O}
\newcommand{\kepler}{{\it Kepler}}

\newcommand{\um}{$\mu$m}

\newcommand{\teff}{\ensuremath{T_{\text{eff}}}}

\newcommand{\gaia}{{\textit Gaia}}

\usepackage[outdir=./]{epstopdf}
\usepackage{ulem}
\usepackage{mathrsfs}
\usepackage{fontawesome}
\usepackage{comment}

\newcommand{\tess}{\textit{TESS}}
\newcommand{\ktwo}{{\textit K2}}

\newcommand{\jwst}{\textit{JWST}}
\newcommand{\hst}{\textit{HST}}

\newcommand{\cotwo}{CO$_2$}
\newcommand{\sotwo}{SO$_2$}

\newcommand{\tspot}{$T_{\rm{spot}}$}

\newcommand{\plname}{HIP 67522 b}
\newcommand{\starname}{HIP 67522}

\newcommand{\newedit}[1]{\textcolor{black}{#1}}

\pdfoutput=1

\shorttitle{That's no Hot Jupiter!} 
\shortauthors{Thao et al. 2024}

\begin{document}

\title{\textbf{The Featherweight Giant: Unraveling the Atmosphere of a 17 Myr Planet with \jwst\,}}

\correspondingauthor{Pa Chia Thao}
\email{pachia@live.unc.edu} 

\author[0000-0001-5729-6576]{Pa Chia Thao}
\altaffiliation{NSF GRFP}
\altaffiliation{NC Space Grant GRFP}
\affiliation{Department of Physics and Astronomy, The University of North Carolina at Chapel Hill, Chapel Hill, NC 27599, USA} 

\author[0000-0003-3654-1602]{Andrew W. Mann}
\affiliation{Department of Physics and Astronomy, The University of North Carolina at Chapel Hill, Chapel Hill, NC 27599, USA}

\author[0000-0002-9464-8101]{Adina D. Feinstein}
\altaffiliation{NHFP Sagan Fellow}
\affiliation{Laboratory for Atmospheric and Space Physics, University of Colorado Boulder, UCB 600, Boulder, CO 80309}
\affiliation{Department of Physics and Astronomy, Michigan State University, East Lansing, MI 48824, USA}

\author[0000-0002-8518-9601]{Peter Gao}
\affiliation{Earth and Planets Laboratory, Carnegie Institution for Science, 5241 Broad Branch Road, NW, Washington, DC 20015, USA}

\author[0000-0002-5113-8558]{Daniel Thorngren}
\affiliation{Department of Physics \& Astronomy, Johns Hopkins University, Baltimore, MD, USA}

\author[0000-0003-4459-9054]{Yoav Rotman}
\affiliation{School of Earth and Space Exploration, Arizona State University, Tempe, AZ 85281, USA}

\author[0000-0003-0156-4564]{Luis Welbanks}
\altaffiliation{NHFP Sagan Fellow}
\affiliation{School of Earth and Space Exploration, Arizona State University, Tempe, AZ 85281, USA}

\author[0000-0003-2631-3905]{Alexander Brown}
\affiliation{Center for Astrophysics and Space Astronomy, University of Colorado, 389 UCB, Boulder, CO 80309, USA}

\author[0000-0002-7119-2543]{Girish M. Duvvuri}
\affiliation{Department of Physics and Astronomy, Vanderbilt University, Nashville, TN 37235, USA}

\author[0000-0002-1002-3674]{Kevin France}
\affiliation{Laboratory for Atmospheric and Space Physics, University of Colorado Boulder, UCB 600, Boulder, CO 80309}
\affiliation{Department of Astrophysical and Planetary Sciences, University of Colorado, UCB 389, Boulder, CO 80309, USA}
\affiliation{Center for Astrophysics and Space Astronomy, University of Colorado, 389 UCB, Boulder, CO 80309, USA} 

\author[0009-0006-5319-8649]{Isabella Longo}
\affiliation{Laboratory for Atmospheric and Space Physics, University of Colorado Boulder, UCB 600, Boulder, CO 80309}

\author[0000-0003-1133-1027]{Angeli Sandoval}
\affiliation{Astrophysics Program, The CUNY Graduate Center, City University of New York, New York, NY 10016, USA}
\affiliation{Exoplanets and Stellar Astrophysics Laboratory, NASA Goddard Space Flight Center, Greenbelt, MD 20771, USA}

\author[0000-0002-5094-2245]{P. Christian Schneider}
\affiliation{Hamburger Sternwarte, Gojenbergsweg 112, D-21029, Hamburg, Germany}
\affiliation{Scientific Support Office, Directorate of Science, European Space Research and Technology Center (ESA/ESTEC)}

\author[0000-0001-9667-9449]{David J. Wilson}
\affiliation{Laboratory for Atmospheric and Space Physics, University of Colorado Boulder, UCB 600, Boulder, CO 80309}

\author[0000-0002-1176-3391]{Allison Youngblood}
\affiliation{Exoplanets and Stellar Astrophysics Laboratory, NASA Goddard Space Flight Center, Greenbelt, MD 20771, USA}

\author[0000-0001-7246-5438]{Andrew Vanderburg}
\affiliation{Department of Physics and Kavli Institute for Astrophysics and Space Research, Massachusetts Institute of Technology, Cambridge, MA 02139, USA}

\author[0000-0002-8399-472X]{Madyson G. Barber}
\affiliation{Department of Physics and Astronomy, The University of North Carolina at Chapel Hill, Chapel Hill, NC 27599, USA} 
\altaffiliation{NSF GRFP}

\author[0000-0001-7336-7725]{Mackenna L. Wood}%
\affiliation{Department of Physics and Astronomy, The University of North Carolina at Chapel Hill, Chapel Hill, NC 27599, USA} 
\affiliation{MIT Kavli Institute for Astrophysics and Space Research Massachusetts Institute of Technology, Cambridge, MA 02139, USA}

\author[0000-0003-1240-6844]{Natasha E. Batalha}
\affiliation{NASA Ames Research Center, Moffett Field, CA 94035, USA}

\author[0000-0001-9811-568X]{Adam L. Kraus}
\affiliation{Department of Astronomy, The University of Texas at Austin, Austin, TX 78712, USA}

\author[0000-0001-8504-5862]{Catriona Anne Murray}
\affiliation{Department of Astrophysical and Planetary Sciences, University of Colorado, Boulder, CO, USA}

\author[0000-0003-4150-841X]{Elisabeth R. Newton}
\affiliation{Department of Physics and Astronomy, Dartmouth College, Hanover, NH 03755, USA}

\author[0000-0001-9982-1332]{Aaron Rizzuto}
\affiliation{Mondo, Level 26/2 Southbank Blvd, Southbank VIC 3006, Australia}

\author[0000-0003-2053-07492]{Benjamin M. Tofflemire}
\affiliation{Department of Astronomy, The University of Texas at Austin, Austin, TX 78712, USA}
\altaffiliation{51 Pegasi b Fellow}

\author[0000-0002-8163-4608]{Shang-Min Tsai}
\affiliation{Department of Earth and Planetary Sciences, University of California, Riverside, CA, USA}

\author[0000-0003-4733-6532]{Jacob L.\ Bean}
\affiliation{Department of Astronomy \& Astrophysics, University of Chicago, Chicago, IL 60637, USA}

\author[0000-0002-3321-4924]{Zachory K. Berta-Thompson}
\affiliation{Department of Astrophysical and Planetary Sciences, University of Colorado, Boulder, CO, USA}

\author[0000-0001-5442-1300]{Thomas M. Evans-Soma}
\affiliation{School of Information and Physical Sciences, University of Newcastle, Callaghan, NSW, Australia}
\affiliation{Max Planck Institute for Astronomy, K\"{o}nigstuhl 17, D-69117 Heidelberg, Germany}

\author[0000-0001-8499-2892]{Cynthia S.\ Froning}
\affiliation{Southwest Research Institute, Div.\ 05, 6220 Culebra Rd., San Antonio, TX 78238, USA }

\author[0000-0002-1337-9051]{Eliza M.-R. Kempton}
\affiliation{Department of Astronomy, University of Maryland, 4296 Stadium Drive, College Park, MD 20742, USA}

\author[0000-0002-0747-8862]{Yamila Miguel}
\affiliation{Leiden Observatory, Leiden University, P.O. Box 9513, 2300 RA Leiden, NL}
\affiliation{SRON Netherlands Institute for Space Research, Niels Bohrweg 4, 2333 CA Leiden, NL}

\author[0000-0002-4489-0135]{J. Sebastian Pineda}
\affiliation{Laboratory for Atmospheric and Space Physics, University of Colorado Boulder, UCB 600, Boulder, CO 80309}


\begin{abstract} 
The characterization of young planets ($<$ 300 Myr) is pivotal for understanding planet formation and evolution. We present the 3-5µm transmission spectrum of the 17 Myr, Jupiter-size ($R$ $\sim$10$R_{\oplus}$) planet, HIP 67522b, observed with \jwst\ NIRSpec/G395H. To check for spot contamination, we obtain a simultaneous $g$-band transit with SOAR. The spectrum exhibits absorption features 30-50\% deeper than the overall depth, far larger than expected from an equivalent mature planet, and suggests that HIP~67522\,b's mass is $<$20 $M_{\oplus}$ irrespective of cloud cover and stellar contamination. \newedit{A Bayesian retrieval analysis returns a mass constraint of $13.8\pm1.0M_{\oplus}$.} This challenges the previous classification of HIP 67522b as a hot Jupiter and instead, positions it as a precursor to the more common sub-Neptunes. With a density of $<$0.10g/cm$^{3}$, HIP 67522 b is one of the lowest density planets known. \newedit{We find strong absorption from H$_{2}$O and CO$_{2}$ ($\ge7\sigma$), a modest detection of CO (3.5$\sigma$), and weak detections of H$_2$S and SO$_2$ ($\simeq2\sigma$).} Comparisons with radiative-convective equilibrium models suggest supersolar atmospheric metallicities and solar-to-subsolar C/O ratios, with photochemistry further constraining the inferred atmospheric metallicity to 3$\times$10 Solar due to the amplitude of the \sotwo\ feature. These results point to the formation of HIP 67522b beyond the water snowline, where its envelope was polluted by icy pebbles and planetesimals. The planet is likely experiencing substantial mass loss (0.01-0.03 M$_{\oplus}$ Myr$^{-1}$), sufficient for envelope destruction within a Gyr. This highlights the dramatic evolution occurring within the first 100 Myr of its existence.

\end{abstract}


\section{Introduction}\label{sec:intro}

A long-standing problem in exoplanet research is understanding how planets evolve. The simplest observational path to solving this problem is to compare the properties of planets as a function of age. While the majority of discovered planetary systems are either old ($>$1\,Gyr) or have unconstrained ages, the recent surveys conducted by the \ktwo\ and Transiting Exoplanet Survey Satellite (\tess) missions of nearby young clusters and associations have led to the discovery of a handful of \textit{young}, transiting exoplanets \citep[e.g.][]{bouma2020cluster, 2020MNRAS.498.5972N, Bouma2022, wood23}. Intriguingly, the radii of these worlds are larger than their older counterparts at similar orbital separations \citep{Vach2024}, suggesting that atmospheric loss and secular cooling shape planet populations over time \citep{owen2017evaporation}. 

Mass measurements are essential for providing a more quantitative test of planetary evolution models. While a young planet may seem similar in size to an older one, its mass may be significantly lower due to radius inflation from a high internal entropy \citep{Lopez2014}. However, such comparisons between young ($< 300$~Myr) and mature ($> 1$~Gyr) planets are complicated by the challenges of measuring young planet masses in the presence of strong stellar activity \citep{blunt2023overfitting, tran2021epoch}.

\citet{dewit13} proposed a method to deduce a planet's mass from its transmission spectrum by leveraging the correlation between the planet's surface gravity and atmospheric scale height ($H$). This relationship significantly influences the strength of the spectral features observed during transit: planets characterized by a low surface gravity and a low mean molecular weight will exhibit a larger scale height -- making them excellent targets for mass determination through their transmission spectrum.

In addition to the mass, characterizing young planetary atmospheres adds a crucial piece of information to the puzzle of planet formation, as their atmospheres may be one of the most pristine tracers of early formation and evolutionary processes. Using \hst\,/WFC3, four young exoplanets with well constrained ages have had their atmospheres characterized through transmission spectroscopy. The intermediate aged planets Kepler-51b and Kepler-51d \citep[500-700~Myr;][]{Libby-Roberts2022} both exhibit featureless transmission spectra, while the 10~Myr, K2-33b possesses a significant slope from the optical to the NIR wavelengths \citep{Thao2023}. The latter may be attributed to photochemical hazes \citep{Gao_hazes, Wang2019, Ohno2021}, or a circumplanetary dust ring \citep{ohno2022circumplanetary}. \newedit{The 23~Myr, V1298 Tau b exhibits a clear atmosphere with a significant strong \water\, absorption feature \citep{Barat2024_V1298Taub}.}

With \jwst\,'s ability to observe across the optical to mid-IR wavelengths, we are at a critical juncture for characterizing the atmospheres of very young planets ($<$ 100\, Myr) to understand their formation environment, evaluate their migration history, and refine atmospheric models. To this end, we explore the transmission spectrum of \plname, a young gas giant first discovered through photometry from \tess\ \citep{Rizzuto2020}. With a radius of $\simeq$10 $R_{\oplus}$ and a period of 6.96 days, the planet lands well within the nominal definition of a hot Jupiter; however, the lack of a precise mass measurement -- only an upper limit of $<5M_{J}$ from \cite{Rizzuto2020} -- introduces ambiguity into this classification. The young age ($\tau$= 17$\pm$2 Myr) is derived from the host star's kinematic membership in the Upper-Centaurus-Lupus (UCL) region of the Scorpius-Centarus OB association, establishing \plname\ as one the youngest gas giants known to transit.

In this paper, we analyze 14 transits of \plname, which were obtained between 2019 and 2023 using \tess\,, Southern Astrophysical Research Telescope (SOAR), and \jwst. The combined dataset spans from the visible to the infrared (0.5-5$\mu$m), allowing us to probe the planet's atmosphere, the extent of contamination from stellar activity, and put further constraints on its mass. The paper is presented as follows: Sections ~\ref{sec:tess} and \ref{sec:soar} describe our \tess\, and SOAR observations, along with details on transit fitting. Section~\ref{sec:sed} describes the process of constructing the spectral energy distribution (SED) of the host star. Section ~\ref{sec:jwst} presents details on our \jwst\, data reduction and creation of the transmission spectrum. In Section ~\ref{sec:interior} and \ref{sec:atmosphere}, we interpret the \jwst\, transmission spectrum using interior and atmospheric models to constrain the planet mass and atmospheric composition. \newedit{We perform a Bayesian retrieval analysis in Section ~\ref{sec:retreivals}.} We explore the impact of stellar surface inhomogeneities from the host star on the transmission spectrum in Section ~\ref{sec:spots}. We discuss the implications of our findings in Section ~\ref{sec:discuss} and state our conclusions in Section ~\ref{sec:summary}. 

\begin{figure}[ht]
    \centering
    \includegraphics[width=0.5\textwidth]{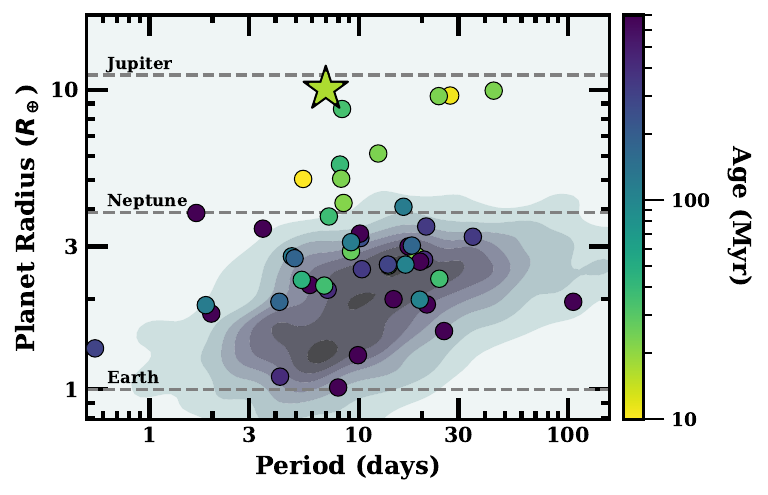}
    \caption{ Observed population of all planets from \kepler\,  as a function of the planet radius ($R_{\oplus}$) and orbital period (days) \citep{Dattilo2023}. Transiting planets with an age under 700 Myr are depicted as circles, with colors corresponding to their age derived from their host cluster or stellar association. \plname\, is outlined as a star and lands well within the mature hot Jupiter part of parameter space. Planet properties from the \citet{NASAexplanetarchive}. 
    \label{fig:young_pl}}
\end{figure}

\section{\tess\,} \label{sec:tess}

\subsection{Observations \& Set-up}
\plname\ (TIC 166527623; HD120411) was first discovered with the Transiting Exoplanet Survey Satellite \citep[\tess\,;][]{Ricker2014} in Sector 11, which was observed from 2019 April 22 to May 21. The target was pre-selected for short cadence (G011280; PI Rizzuto). Subsequently, \starname\,  was re-observed in Sector 38 (2021 April 28 to May 26) and Sector 64 (2023 April 06 to May 4th). For the Sector 38 observations in Cycle 3, the target was pre-selected for 120 second cadence observations for two programs: G03141 (PI: E. Newton) and G03130 (PI: A. Mann) due to the youth of the planet. For the Sector 64 observations in Cycle 5, the target was pre-selected for 20 second cadence for two programs: G05015 (PI: B. Hord) and G05106 (PI: E. Gillen) to search for companions around the planet and to observe stellar flare activity. In total, there were 12 transits observed by \tess\, in Sectors 11, 38, and 64. 


The data was processed by the Science Processing and Operations Center (SPOC) pipeline \citep{jenkins2016tess}. To enhance the data quality, we employed a custom extraction pipeline as outlined in \citet{vanderburg2019tess}. The process starts with the Simple Aperture Photometry curves \citep[SAP;][]{Twicken2010}, which we fit with a linear model that consists of a 0.3-day basis spline, with the mean and standard deviation derived from the spacecraft quaternion time series, seven co-trending vectors from the SPOC pipeline data conditioning, and a high-pass-filtered time series extracted from the SPOC background aperture. Errors for each sector of data were calculated using the standard deviation of the detrended and normalized out-of-transit light curve. This yielded errors of 6.7$\times10^{-4}$ and 1.4$\times10^{-3}$ for the 120-second cadence (Sectors 11 and 38) and 20-second cadence (Sector 64) data, respectively.

We conducted a visual inspection and removed $\sim$ 340 data points corresponding to a flare in Sector 64. This particular event falls within the time range of 2,460,059.57 and 2,460,059.65 BJD$_{TDB}$ and was subsequently excluded from the analysis.

\begin{figure*}[ht]
    \centering
    \includegraphics[width=0.970\textwidth]{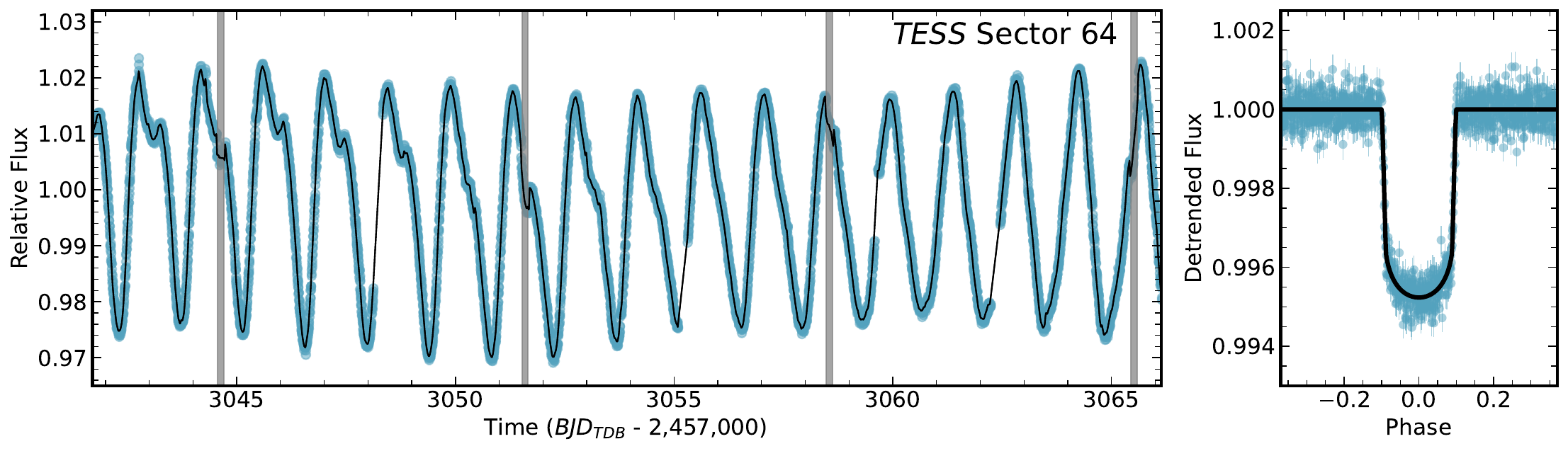}
    \caption{\textit{Left: } 
    \tess\, Sector 64 light curve (blue points) binned to 1.5 minute intervals and compared to a Gaussian process for the stellar variability (black). The transit times for \plname\ are shaded in gray. \textit{Right: } Phase folded light curve of \plname\, after the best fit stellar variability model has been removed. Data is binned to 1 minute intervals with the best-fit transit model shown in black. \label{fig:tesstransit}}
\end{figure*}

\subsection{Transit Fitting}

\newedit{Unlike the SOAR data, the \tess\ data was not simultaneous with the \jwst\ data. The spot pattern may change between transits, so the \tess\ data could not be fit simultaneously with the \jwst\ and SOAR datasets. Instead, our goal for analyzing the \tess\ data was to provide constraints on wavelength-independent parameters (e.g., orbital period and transit duration) to improve the \jwst\ fit. To this end, 
we} utilized \texttt{MISTTBORN} (MCMC Interface for Synthesis of Transits, Tomography, Binaries, and Others of a Relevant Nature)\footnote{\url{https://github.com/captain-exoplanet/misttborn}} to fit the transit photometry taken by \tess. \texttt{MISTTBORN} was first detailed in \citet{Mann2016a} with a significant expansion detailed in \citet{MISTTBORN}. It uses \texttt{BATMAN} \citep{Kreidberg2015} for generating the model transits, \texttt{emcee} \citep{Foreman-Mackey2013} to explore the transit parameter space using an affine-invariant Markov chain Monte Carlo (MCMC) algorithm, and \texttt{celerite} \citep{celerite} to model the stellar variability with a Gaussian process (GP). 

We fit for 12 parameters in total. The first four parameters were dedicated to regular transit features, which includes the time of inferior conjunction ($T_0$), the orbital period of the planet ($P$), the planet-to-star radius ratio ($R_p/R_\star$), and the impact parameter ($b$). We also fitted for the stellar density ($\rho_\star$) and included two parameters for limb darkening (u$_{1}$, $u_{2}$). Moreover, our model incorporates five additional parameters as part of the Gaussian Process (GP) Model. 

Our GP kernel is constructed from two stochastically driven damped simple harmonic oscillators (SHOs), characterized by two Fourier modes: one at the rotation period and one at half the rotation period. The GP parameters included the variability of the amplitude of the fundamental oscillation (log $A_{1}$), the quality factor of the half-period oscillator (log $Q_{2}$), the difference in quality factors between the two oscillators ($\Delta$=$Q_{1}-Q_{2}$), the primary signal period (log $P_{GP}$), and a mixture term describing the amplitude ratio of the two oscillators ($m$, where $A_{1}$/ $A_{2}$ = 1+e$^{-m}$). The inclusion of the mixture term $m$ facilitates sampling without the need for bounds, while ensuring that the oscillator has the largest amplitude at the full rotation period. 

We applied Gaussian priors on the limb-darkening coefficients based on the values from the \texttt{LDTK} toolkit \citep{2015MNRAS.453.3821P}, with errors accounting for uncertainties in stellar parameters and the difference between models used. A summary of the priors on the limb darkening coefficients is presented in Table ~\ref{tab:limb_darkening_priors}. We also applied a Gaussian prior on the stellar density, based on the stellar parameters given by the discovery paper \citep[$\rho_{*}$/$\rho_{\odot}$ = 0.46 $\pm$ 0.06;][]{Rizzuto2020}. All other parameters were sampled uniformly with physically motivated boundaries (e.g., $|b|<1+R_P/R_*$, $0<R_P/R_*<1$, and $\rho_*>0$). 

We ran the MCMC using 100 walkers for 200,000 steps including a burn-in of 20,000 steps. This was more than 50 times the autocorrelation time for all parameters, indicating it was more than sufficient for convergence. All output parameters from the \texttt{MISTTBORN} analysis are listed in Table~\ref{tab:tess_transitfit} with a subset of the resulting transit light curve fit shown in Figure ~\ref{fig:tesstransit}.

\begin{deluxetable} {lcr}
\tabletypesize{\scriptsize}
\tablecaption{Priors on Limb-darkening Coefficients \label{tab:limb_darkening_priors}}
\tablecolumns{3}
\tablewidth{0pt}
\tablehead{
\colhead{Filter} & 
\colhead{$g_{1}$} &  
\colhead{$g_{2}$}}
\startdata
\tess\, & $0.409\pm0.08$ & $0.169\pm0.04$\\ 
$g_{p}$ & $0.693\pm0.08$ & $0.101\pm0.04$\\ 
NIRSpec/G395H & $0.120\pm0.08$ & $0.098\pm0.04$\\ 
\enddata
\tablecomments{Limb-darkening priors are provided as the traditional linear and quadratic terms, but were fit using triangular sampling terms.}
\end{deluxetable}

\begin{deluxetable*}{lcr} 
\tabletypesize{\footnotesize} 
\tablecaption{\tess\, Transit Fitting Parameters \label{tab:tess_transitfit}}
\tablewidth{0pt}
\tablehead{
\colhead{Description} &
\colhead{Parameters} & 
\colhead{Value}}
\startdata
\multicolumn{3}{c}{System Parameters}\\
\hline
First mid-transit midpoint & $T_{0} (BJD)$ & $2458604.02388^{+0.00033}_{-0.00032}$ \\ 
Orbital Period & $P$ (days) & $6.9594718^{+2.3\times10^{-6}}_{-2.4\times10^{-6}}$ \\ 
Planet-to-star radius ratio & $R_P/R_{\star}$ & $0.0649^{+0.0017}_{-0.0018}$ \\ 
Impact parameter & $b$ & $0.16^{+0.12}_{-0.11}$ \\ 
Stellar density & $\rho_{\star}$ ($\rho_{\odot}$) & $0.436^{+0.015}_{-0.033}$ \\ 
Limb darkening & $q_{1}$ & $0.122^{+0.058}_{-0.055}$ \\ 
Limb darkening & $q_{2}$ & $0.354^{+0.04}_{-0.039}$ \\ 
\hline
\multicolumn{3}{c}{GP Parameters}\\
\hline
Amplitude of main peak & log$(Amp)$ & $-8.9^{+0.95}_{-0.76}$  \\ 
Decay timescale of the main peak & log$(Q_{1})$ & $7.26^{+1.1}_{-0.81}$ \\ 
Height of second peak relative to first peak & log$(m)$ & $-0.7^{+1.6}_{-1.2}$ \\ 
Decay timescale of the second peak & log$(Q_{2})$ & $0.0033^{+0.0054}_{-0.0025}$ \\ 
Rotation Period & log$(P_{GP})$ & $0.34902^{+0.0017}_{-0.00083}$ \\ 
\hline 
\multicolumn{3}{c}{Derived Parameters} \\ 
\hline
Ratio of semi-major axis to stellar radius & $a/R_{\star}$ & $11.63^{+0.13}_{-0.3}$ \\ 
Inclination & $i$ ($^{\circ}$) & $89.2^{+0.54}_{-0.61}$ \\ 
Transit depth & $\delta$ (\%) & $0.421 \pm 0.022$ \\ 
Planet radius\tablenotemark{a} & $R_P$ ($R_\oplus$) & $9.763 ^{+0.493}_{-0.504}$ \\ 
Semi-major axis & $a$ (AU) & $0.0747^{+0.0034}_{-0.0038}$ \\ 
Equilibrium Temperature \tablenotemark{b} & $T_{\mathrm{eq}}$ (K) & $1176.0^{+22.0}_{-17.0}$ \\ 
\enddata
\tablenotetext{a}{Assumes a stellar radius of $1.38\pm0.06R_\odot$.}
\tablenotetext{b}{Assumes an albedo of 0 and full heat redistribution}
\end{deluxetable*}

\section{SOAR} \label{sec:soar}

\subsection{Observations  \& Set-up}
We observed one transit of \plname\ using the Goodman High Throughput Spectrograph \citep{Clemens_2004} attached to the 4.1m Southern Astrophysical Research Telescope (SOAR) atop Cerro Pachon, Chile. This observation was taken simultaneously with our {\it JWST} observations on UT 2023 Feb 26. The primary goal of these observations were twofold: first, to check for signatures of significant unocculted star spots, which often manifest as a significantly deeper transit at bluer wavelengths \citep[e.g.,][]{Rackham2018,Thao2020}, and second, to check for flares, which tend to be more energetic at shorter wavelengths \citep{davenport2019evolution}. 

We configured our observations as follows. We used the red camera with a readout speed set to 750 Hz, ATTN 0, and the Sloan $g'$ filter in the spectrograph's imaging mode. In this mode, Goodman has a default \SI{7.2}{\arcminute} circular field of view with a pixel scale of \SI{0.15}{\arcsecond} pixel$^{-1}$. To decrease the readout time, the Region of Interest (ROI) setting was reduced to a 1500-pixel window in the read direction. Since the target is bright, we defocused the telescope slightly; significant defocusing was not possible as it impacts guiding. We opted for short exposures (2.5\,s), which also serves to help resolve out short-duration flares. In total, we took 1282 images of \starname. 

At approximately 9AM UTC, the rotator unexpectedly ceased its motion, resulting in the loss of guiding and causing our target of interest to move out of the field of view. While we were able to guide by placing the star at the center of the field, the broken rotator meant that all comparison stars rotated around the center. Some of these images were recoverable, albeit with lower photometric precision. Eventually, the rotation pushed the best comparison stars off the chip; we did not attempt to use any data past this, leaving us with the first 1028 images. We applied bias corrections on all images in our analysis. We built a median (dome) flat, but found that it did not improve the precision for data before the rotator failure and offered only marginal benefit post-failure.  

We performed aperture photometry on the target and eight nearby comparison stars. We used an optimal aperture (50 pixels) and estimated the sky background using an annulus around each star with an inner and outer radius of 120 and 220 pixels, respectively. We built a master comparison star light curve using the robust weighted mean. We estimate the uncertainties in each comparison star light curve by building a combined light curve from the other seven comparison stars to remove the overall trend, then estimate the residual scatter using the median absolute deviation. The assumption is that the residual scatter is dominated by noise (random or systematic) in a given comparison star. Since star-by-star uncertainties are needed to build each combined curve as well as the final master comparison star curve, this process was done iteratively until the final uncertainty converged (change of less than 1\%).

\subsection{Transit Fitting}

We fit the $g'$ data using the \texttt{BATMAN} model \citep{Kreidberg2015} and \texttt{emcee} MCMC optimization \citep{Foreman-Mackey2013}. We fixed the wavelength-independent parameters, planet inclination ($i_p$), the ratio of semi-major axis-to-stellar radius ($a_R*$), and the time of inferior conjunction ($T_0$) to the values from the \jwst\ broadband curve, the orbital period to the value derived from the \tess\ data, the eccentricity to zero, and the limb-darkening parameters to values from the \texttt{LDTK} code (see Section~\ref{sec:jwst}). The remaining transit parameter, $R_P/R_*$, was fit assuming a uniform prior. We tested fitting $i_p$, $a_R*$, $T_0$, and the limb-darkening parameters assuming Gaussian priors and found an almost negligible change in both the resulting $R_P/R_*$ and uncertainties. The lack of change is in large part because variations in these parameters can be absorbed by the polynomial coefficients and their uncertainties.

To handle stellar variability, we fit for a second-order polynomial in time. We tested adding airmass- or position-dependent terms, as well as a third-order polynomial, and found they offered no improvement over a second-order polynomial in time \citep[based on the Bayesian Information Criterion; ][]{schwarz1978_bic} \newedit{and shifted the $R_P/R_*$ by $<1\sigma$. Further, a polynomial in time was used for the \jwst\ analysis (Section~\ref{sec:jwst}), keeping these two datasets more consistent.}

\begin{figure}[ht]
    \centering
    \includegraphics[width=0.50\textwidth]{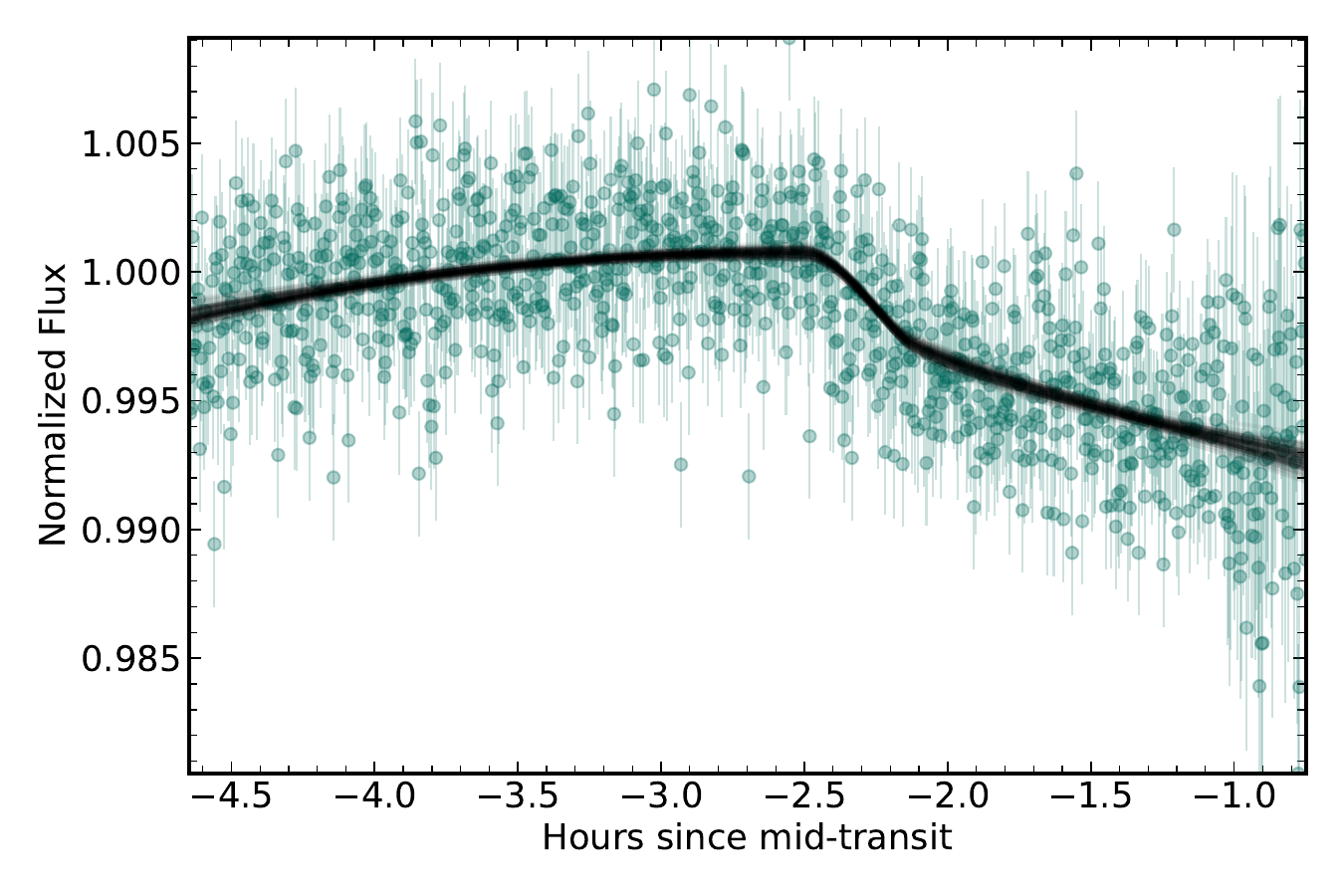}
    \caption{Transit photometry of \plname\, taken with SOAR using the Sloan $g'$ filter compared to 100 random draws from the fit posteriors (black). Photometric uncertainties are larger for the last $\simeq$30\,minutes due to the rotator failure. 
    \label{fig:gtransit}}
\end{figure}

We ran the fit for 10,000 steps with 32 walkers, and checked for convergence using the autocorrelation time. We repeated the fit excluding all data after the rotator failed, and found the change was insignificant for any parameter. The transit and best-fit model are shown in Figure~\ref{fig:gtransit}. The final planet-to-star radius ratio was \newedit{$0.0629\pm0.0040$}. We use this value in our exploration of the effect of unocculted spots (see Section~\ref{sec:spots}).

\section{Host Star SED} \label{sec:sed} 

Models of planetary atmospheres (Section~\ref{sec:modeldes}) require an input stellar spectral energy distribution (SED). In particular, atmospheric photochemistry is sensitive to the ultraviolet (UV) and visible flux ($<8000$\,\AA). We construct the required spectrum by combining direct observations of \starname\ from \hst\ and the \textit{Chandra} X-ray observatory with stellar models where needed. 

\starname\ shows strong high energy emission because of its youth and rapid rotation, which permits detailed measurement of its UV and X-ray flux. It has a spectral type of G0 IV and is still slightly more luminous than the zero-age main sequence. For the red end of the SED ($>$5690\,\AA), we used an updated spectrum from \citet{Rizzuto2020}. Briefly, their spectrum was built by fitting available photometry with a grid of spectral templates and BT-SETTL atmospheric models to fill in gaps (e.g., regions of high telluric contamination) and treating reddening as a free parameter. Our only change from \citet{Rizzuto2020} was to add in \gaia\ DR3 photometry to the fit, which resulted in a negligible overall change to the spectrum.

For 1162--5689\,\AA, we used spectra taken with \hst/STIS as part of a MUSCLES Extension program of \jwst\ Cycle 1 targets (HST-GO-16701; PI: A. Youngblood). \starname\ was observed with \hst/STIS on 2022 September 05 with the G140L, G230L, and G430L gratings with total exposure times of 14192\,s, 2103\,s, and 10\,s, respectively. We used standard STIS data reduction tools. The pipeline wavelength calibration was in error by $\approx$100 km s$^{-1}$ in each spectrum. We re-extracted each spectrum after supplying the correct SHIFTA1 keyword and disabling the WAVECORR keyword. Light from \starname\ was detected in essentially all wavelength bins except near detector edges, where sensitivity is generally low and/or noise is high. We co-added the six G140L exposures, weighted by the flux uncertainty, then spliced the three gratings together using the contribution from the bluer spectrum where the gratings overlapped.  

Corrections for interstellar absorption and extinction were applied in two ways. The effects of dust extinction (A$_V$ = 0.12$\pm$0.06 mag; \citealt{Rizzuto2020}) were removed assuming a standard extinction law and R$_V$ = 3.1. For typical interstellar dust this would correspond to a neutral hydrogen column density of 2.1 $\pm$ 1.1 $\times$ 10$^{20}$ cm$^{-2}$ \citep{Predehl1995}. The bright Lyman-$\alpha$ line at 1215.67 \AA\ is strongly affected by interstellar \ion{H}{1} absorption. We found that a reconstruction following the methods of \cite{Youngblood2016} was not possible, so we spliced into the spectrum an estimated Lyman-$\alpha$ profile using the scaling relations between Lyman-$\alpha$ surface flux and stellar rotation period from \cite{Wood2005}. This estimate agreed closely with an estimate based on a scaling relationship between Lyman-$\alpha$ and stellar age from \cite{Ribas2005}.

The X-ray spectrum of \starname\ was observed by the Chandra X-ray Observatory on 2021 Feb 08 using the ACIS-S4 detector (OBSID 24675, PI: G. Garmire). No obvious variability was detected in the 2.08 ks (35 minute) observation. The observed count rate is 0.093 $\pm$ 0.007 ct s$^{-1}$ and the observed 0.3-10.0 keV flux is 1.20 $\pm$ 0.09 $\times$ 10$^{-12}$ erg cm$^{-2}$ s$^{-1}$. This corresponds to a X-ray luminosity of 2.2 $\pm$ 0.2 $\times$ 10$^{30}$ erg s$^{-1}$ (log L$_x$ = 30.35 erg s$^{-1}$). The log ratio of X-ray luminosity in the ROSAT band (0.1-2.4 keV) to the bolometric luminosity is -3.53. This ratio is at the level corresponding to ``saturated" coronal activity \citep{VilhuWalter1987, Pizzolato2003,Wright2018} and coronal activity is likely to continue at this level for several hundred million years \citep{Johnstone2021A&A...649A..96J}. An XSPEC (V. 12.13.0) single temperature spectral fit to the data estimated a coronal temperature of 1.16 $\pm$ 0.08 keV ($\sim$13 MK). This Chandra X-ray spectrum was used as direct input to the DEM (differential emission measure) modeling.

The extreme ultraviolet (EUV) part of the SED is not directly observable, and we make use of the fact that emission lines observed in the \hst/STIS and \textit{Chandra} data form in the transition region and corona respectively. These regions of the stellar atmosphere contain plasma with temperatures from $10^4$ -- $10^8$ K and are also responsible for forming the unobserved stellar EUV emission from 123--1162 \AA. We use the differential emission measure technique as implemented in \citet{Duvvuri:2021, Duvvuri:2023} to model the temperature and density structure of the upper stellar atmosphere given the observed FUV and X-ray data as inputs. We combine this model with atomic data for transitions and continuum emission between 123 -- 1162 \AA\, from \texttt{CHIANTI v10.1} \citep{Dere1999, DelZanna:2021} to fill in this unobserved segment of the stellar SED. 

We merged the optical-IR template spectrum with the MUSCLES-constructed spectrum using an overlapping region from 4500-5700\AA. The offset between the two was small ($\simeq$1\%), and consistent with random uncertainties in the absolute calibration. Applying a correction had no impact on the final results. 

\begin{figure*}[ht]
    \centering
    \includegraphics[width=\textwidth]{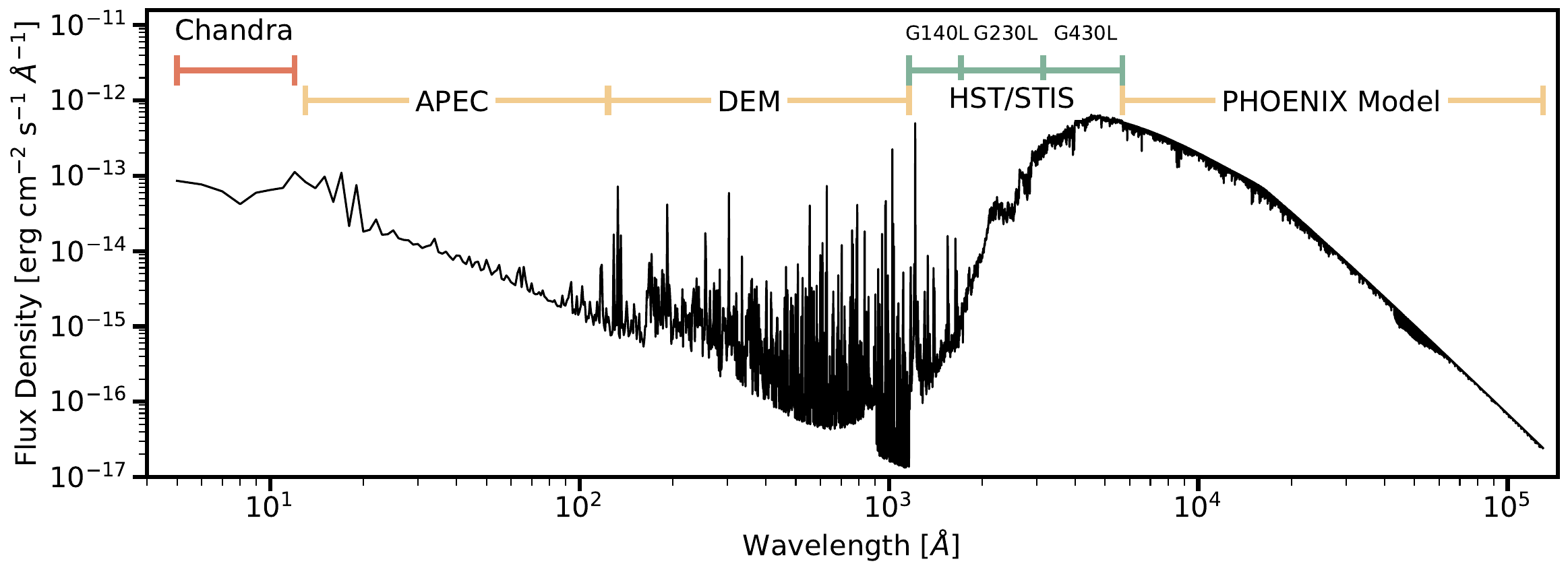}
    \caption{The spectral energy distribution (SED) for \starname\, (black). We label each component of the SED based on how it was derived as described in Section~\ref{sec:sed}. For our photochemical model, we used a mix of empirical templates and PHOENIX models longward 5690\AA, but show the pure model version here (and note the differences were small).
    \label{fig:sed}}
\end{figure*}

The spectrum will be available on the MUSCLES HLSP website (\url{https://archive.stsci.edu/prepds/muscles/}). We present the SED in Figure~\ref{fig:sed}, showing how each component of the SED was derived.

\section{\jwst} \label{sec:jwst}

\subsection{Observations \& Set-up}

We obtained a single transit observation of \plname\ on 2023 Feb 26 UTC using the Near Infrared Spectrograph (NIRSpec) in the Bright Object Time Series (BOTS) mode on \jwst. This observation was part of our General Observer Proposal in Cycle 1 (GO: 2498; PI: A. Mann). The observations were taken with the G395H grating, F290LP filter, and recorded with a \SI{1.6}{\arcsecond}$\times$\SI{1.6}{\arcsecond} fixed slit aperture with the spectra dispersed across both the NRS1 and NRS2 detectors. The SUB2048 subarray and NRSRAPID readout pattern were utilized. The observation lasted approximately 8.9 hours. We acquired a total of 2703 integrations, with 9 groups per integration, resulting in an effective integration time of $\sim$ 21943~s. 

Since the target's brightness exceeded the limit for the Wide Aperture Target Acquisition (WATA) mode, we selected a nearby acquisition source: Gaia DR3 36113920619134017152. For this acquisition, the target was taken with a clear filter, and recorded in the SUB32 subarray, employing the NRSRAPID readout pattern. We acquired 3 groups per integration, resulting in an effective exposure time of 0.045~s.   

The \jwst\ data encompassed a pre-transit baseline, but missed the egress. This is due to a previously undetected transit-timing variation (TTV). As a result of the TTV, \newedit{a fit to \tess\ and prior ground-based data yielded an erroneously low period. Propagated to the \jwst\ epoch, the predicted transit time was $\simeq$30\,min early, and} the \jwst\ transit came 10\,min late even compared to a `correct' linear ephemeris. 

The TTV exhibits a long period, causing it to be missed in the original \tess\ data (the sinusoid is linear over that time frame). However, this has been independently confirmed from other ground- and space-based data and will be characterized further in a forthcoming upcoming paper \citep{Thao_inprep}. 

We extracted the time-series spectra and measured transit depths using three different procedures (with overlap), which we deem Pipeline 1 (Section ~\ref{subsec:thao}), Pipeline 2 (Section ~\ref{subsec:feinstein}), and Pipeline 3 (Section~\ref{subsec:mann}). Table ~\ref{tab:transitfit}  provides a summary comparing the methods employed by the three pipelines for reducing the data and fitting the transit light curves.

\subsection{Pipeline 1: JWST Science Calibration Pipeline 
+ ExoTiC-JEDI }\label{subsec:thao}
\subsubsection{Reduction}

We processed the data following the methods detailed in \citet{grant2023detection}, employing a combination of the JWST Science Calibration Pipeline and custom open-source routines from \texttt{ExoTiC-JEDI}\footnote{\url{https://github.com/Exo-TiC/ExoTiC-JEDI}} \citep[v.0.1-beta;][]{Alderson2023}. Following the JWST standard pipeline, the data reduction workflow is divided into two distinct stages: In Stage 1, the uncalibrated raw data (\texttt{uncal.fits}) is converted into the count rate data (\texttt{rate.fits}); in Stage 2, the pipeline processes these outputs and extracts the calibrated spectra. Following this reduction process, we employed custom codes to fit the light curves. 

For Stage 1, we leverage the JWST pipeline's default processing steps, commencing with the \texttt{uncal} files and executing all of the STScI Steps designated for NIRSpec Stage 1 data. These steps include: \texttt{group\_scale}, \texttt{dq\_init}, \texttt{saturation}, \texttt{refpix}, \texttt{linearity}, \texttt{dark\_current}, \texttt{jump}, \texttt{ramp\_fit}, and \texttt{gain\_scale}. Instead of relying on the superbias provided by the JWST pipeline, which uses a fixed detector bias reference image, we opted for the \texttt{Exotic-JEDI} pipeline's custom bias approach. This custom bias subtraction method involves utilizing a bias image generated from the median of the first groups. Subsequently, we conducted tests with both the custom bias and the STScI superbias; however, we elected to use the custom bias as it resulted in a reduced median absolute deviation. After the jump detection step (\texttt{jump}), we proceed to destripe the data at the group level to mitigate the 1/f noise \citep{rustamkulov2023early}. This process involves subtracting the column-wise median background value from the pixels within each column. To compute the median background values, we exclude pixels flagged due to poor data quality, which includes \texttt{dq\_bits= 0}, \texttt{1}, \texttt{2}, \texttt{10}, \texttt{11}, \texttt{13}, or \texttt{19}.

For stage 2, we initially applied the JWST pipeline's default processing steps, which include \texttt{stsci\_assign\_wcs}, \texttt{stsci\_extract\_2}, \texttt{stsci\_srctype}, and \texttt{stsci\_wavecorr}. Following this, we used \texttt{ExoTiC-JEDI}'s code to perform outlier cleaning and destripe the rate images. Subsequently, we repeated the outlier cleaning step once more for further refinement. The reduced stellar spectra were then extracted using a box extraction method, employing a width equal to six times the measured standard deviation of the point spread function. We also used the optimal extraction method; however, we found that the box extraction method yielded slightly better results. 

\subsubsection{White Light Curve}
We perform additional outlier removal steps for the light curves from NRS1 and NRS2. Initially, we removed bad columns (spectra), identified as having a RMS scatter $>5\sigma$ from the average of the four nearest columns. This process results in the removal of 7 columns from NRS1 and 20 columns from NRS2. Additionally, any data points with deviation $>$3.5$\sigma$ in the individual light curves are excluded. This step removed 1 point from NRS1 and 3 points from NRS2. We then merge the NRS1 and NRS2 light curves into a single dataset, sorting the combined data by time.

We fit the broadband curve with 20 free parameters in total. The first four parameters were associated with the transit model: $T_0$, the planet-to-star radius ratio ($R_{p}/R_{*}$), $a/R_{*}$, $i_p$ and two limb-darkening parameters ($u_{1}$ and $u_{2}$). Five parameters ($a_{1}$, $b_{1}$, $c_{1}$, $d_{1}$, and $e_{1}$) served as coefficients in a fourth-order polynomial used to model the out-of-transit stellar variability, which is expressed as: 

\begin{equation}\label{eqn:var}
    LC = LC \times a_{1} + b_{1}\times(t) + c_1 \times(t)^{2} + d_{1}\times(t)^{3} + e_1 \times(t)^{4},
\end{equation}
\noindent where $t$ denotes the time and LC is the light curve. 

We also tested a third-order polynomial but it left significant red noise based on the Allan deviation (a measure of the expected noise versus measured as a function of binsize). 

We model two spot crossing events using Gaussians, which provided better fits (lower red noise) than using triangle spots. There is some evidence for a third spot near the end of the observations, but attempts to add a third spot yielded fits not statistically justified (based on the $\Delta$BIC). We described each spot with three parameters: the spot amplitude ($A$), the spot duration ($\tau$), and the central location in time ($T_{sp}$). The spot parameters are modeled using Equation 2 from \citet{Dai2017}:

\begin{equation}
Spot (t)  = A \times exp \left[ \frac{-(t-T_{sp})^2}{2\tau^{2}}\right]
\end{equation}

The final three parameters were $\sigma$, representing the underestimated errors (added in quadrature with the reported uncertainties), and $x_1$ and $y_1$, which the coefficients governing the shifts in the X and Y-pixel position of the spectra trace, respectively. This is modeled as: 

\begin{equation}
    LC = LC \times (x_{1} \times x_\textrm{shift} + y_{1} \times y_\textrm{shift})
\end{equation}

\noindent where LC is the light curve and $x_{\rm shift}$ and $y_{\rm shift}$ are the change in the x-pixel and y-pixel position of the spectral trace as functions of time. 

We put Gaussian priors on the parameters, $T_0$, $a/R_{*}$, and $i_p$ using the values from the fit to the \tess\, light curves (Table ~\ref{tab:tess_transitfit}). \newedit{Priors on $a/R_{*}$, and $i_p$ from the \tess\ data were important because of the lack of egress in the \jwst\ data}. We also put Gaussian priors on $u_{1}$ and $u_{2}$, which were calculated from the model values derived from BT-SETTL models using the \texttt{pyLDTk} code \citep{Parviainen2015} (Table ~\ref{tab:limb_darkening_priors}).

\begin{figure}[ht]
    \centering
    \includegraphics[width=0.50\textwidth]{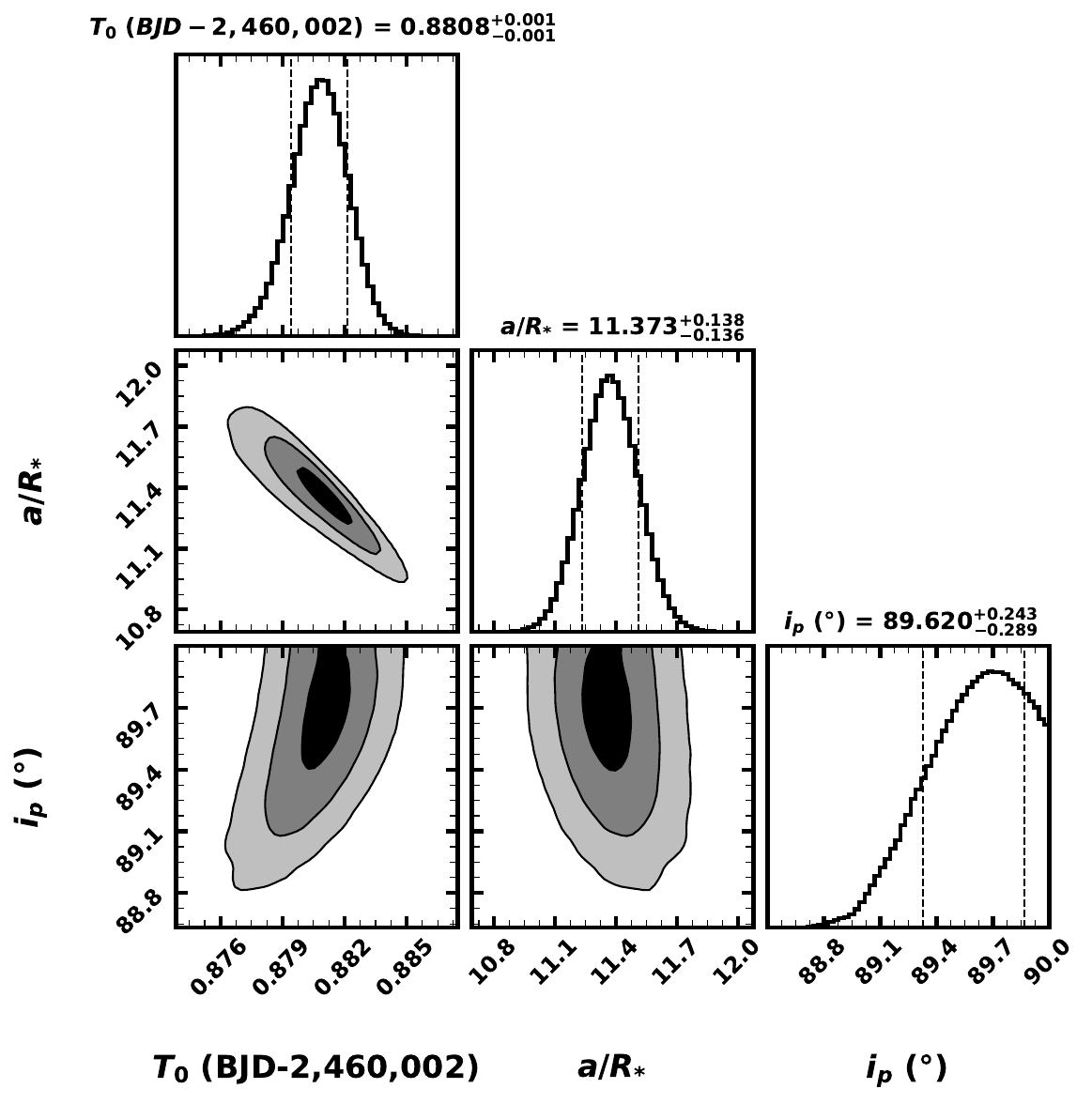}
    \caption{Posteriors from the MCMC fit of the \jwst\, broadband light curve using Pipeline 1 for the parameters, center of transit time ($T_{0}$), ratio of the semi-major axis-to-stellar radius ($a/R{*}$), and inclination ($\textit{inc}$). In the histogram, the dashed lines indicate the 16\% and 84\% percentiles.  Figure made with \texttt{corner.py} \citep{foreman2016corner}. 
    \label{fig:white_lc_corner}}
\end{figure}

\begin{figure}[ht]
    \centering
    \includegraphics[width=0.50\textwidth]{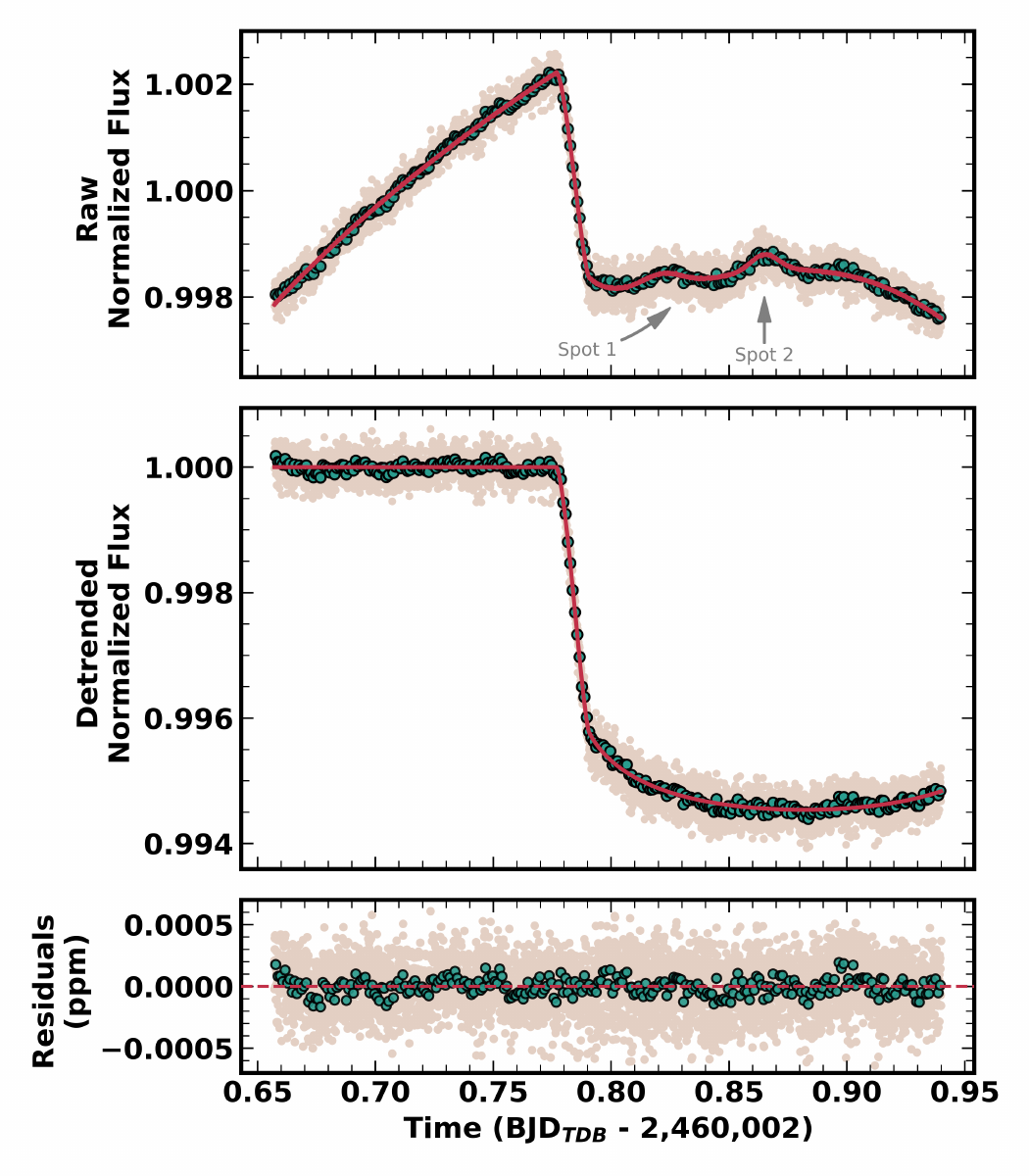}
    \caption{Broadband light curve of \plname\, observed with \jwst\,/NIRSpec, which combines data from both NRS1 and NRS2. The top panel shows the raw normalized data (gray data points) overlaid with the systematic model (red line), as described in Section~\ref{subsec:thao}. During transit, there are two spot crossings, which are highlighted. In the middle panel, the best-fit model from our MCMC fit is overplotted with the systematic-corrected data. The bottom panel shows the residuals. Green data points in each panel represent the binned data at $\sim$1.5 minutes intervals. The egress was missed due to an unknown TTV.
    \label{fig:white_lc}}
\end{figure}

We ran the MCMC using 100 walkers for 100,000 steps including a burn-in of 10,000 steps. All output parameters from the Global Fit analysis are listed in Table ~\ref{tab:white_lc_transitfit}, with a subset of the parameter correlations shown in Figure ~\ref{fig:white_lc_corner}. The white-light curve and best-fit model are shown in Figure~\ref{fig:white_lc}. 

\subsubsection{Spectroscopic Light Curves}

We fixed wavelength-independent parameters, including t$_{0}$, a/$R_{*}$, $inc$, $\tau_{1}$, $T_{sp,1}$, $\tau_{2}$, and $T_{sp,2}$ to the values obtained from the broadband light curve fit (Table~\ref{tab:white_lc_transitfit}). The posteriors for the remaining 11 free parameters for each spectral bin were obtained through MCMC analysis using 30 walkers over 30,000 steps, with a burn-in period of 5,000 steps. The bin size was fixed to a 15-pixel resolution. Nearly all spectroscopic bins exhibited Allan deviation consistent with anticipated noise levels. The exceptions were a single outlier at 4.05\um, and several points at the edges of the detectors. We opted to remove the single outlier but retained the edge points. 

\subsection{Pipeline 2: ExoTiC + Chromatic}\label{subsec:feinstein}

\subsubsection{Data Reduction with ExoTiC}

For our second reduction method, we start with a similar processing of the \jwst\ uncalibrated images as mentioned in Section~\ref{subsec:thao}. That is, we processed both Stage 1 and 2 data products with \texttt{ExoTiC-JEDI}. For our jump detection step, we use the recommended rejection threshold of 15. We used the \texttt{ExoTiC-JEDI} custom background destriping at the group level. For NRS1 and NRS1, we destriped columns [1006, 2042] and [606, 2042], respectively. We masked the same poor data quality flags described in Section~\ref{subsec:thao}.

We removed outliers in the Stage 2 processed images by running \texttt{ExoTiC\_JEDI.CleanOutliersStep}, which uses the optimal extraction method of \cite{horne86} to identify and clean outliers. We used the default data quality flags from the Stage 2 files to identify and interpolate over bad pixels. We developed a custom routine for extracting the spectral orders. We start by finding the location of the spectral trace on the detector. To do this, we identified the peak of the spatial profile in each column. We create a median filter from these peaks as a function of (X, Y) pixel position on the detector; our median filter used a window length of 23 pixels. We identified outliers in the trace location by dividing the trace (X,Y) peaks by the median filter and removing points which deviated from 1 by  $\leq$ or $\geq1.2 \sigma$. We fit a 4\textsuperscript{th}-order polynomial to the spatial profile peaks, masking the columns identified as outliers. This methodology was sufficient for both NRS1 and NRS2.

We identified additional outliers in the images in two ways. First, we identified outliers along the time-axis of our observations. We computed the average for each pixel and removed outliers which were $\geq3.8 \sigma$ from that average. For pixels which were exceptionally noisy ($\sigma_\textrm{std} > 15$), we used a more aggressive outlier threshold of $\geq2.5 \sigma$. Second, we identified outliers along each row of each integration. To do this, we ran a Savitsky-Golay filter with a window length of 101 and a 2\textsuperscript{nd}-order polynomial across each row. We subtracted the Savitsky-Golay filter from the data to identify outliers. We defined outliers as being $\geq 4\sigma$ from the difference between the data and the filter. Once outliers were identified via these two methods, we interpolated over these bad pixels using a linear interpolation with \texttt{scipy.interpolate.griddata}. Once the integrations were sufficiently cleaned, we extracted our spectra with a simple box extraction with a box size of 8 pixels as implemented in \texttt{transitspectroscopy} \citep{espinoza_nestor_2022_6960924}.

\subsubsection{White Light Curve}\label{subsubsec:feinstein}

We fit our white light curve using \texttt{chromatic\_fitting}.\footnote{\url{https://github.com/catrionamurray/chromatic_fitting}} This is an open-source Python package optimized for efficient model fits to spectroscopic light curves to produce transmission spectra. Within our fitting routine, we masked outlier points which were $\geq 2.5\sigma_\textrm{std}$. After the data were masked, we fit the white light curve with 10 free parameters. The first six parameters were associated with the transit model: $T_0$, the stellar radius $R_\star$, the stellar mass $M_\star$, the impact parameter $b$, $R_p/R_\star$, and the quadratic limb-darkening parameters ($u_1$, $u_2$). Additionally, we fit for the baseline flux value, and a second order polynomial to fit for out-of-transit stellar variability expressed similarly to Equation~\ref{eqn:var}, but without the last two terms. We held the period, $P$, constant. We used a No U-Turn Sampler (NUTS) Hamiltonian Monte Carlo method to optimize our parameters. For this fit, we used four chains, 2,000 tuning steps, and 2,000 steps, for a total of 4,000 steps. We fit the light curves from NRS1 and NRS2 separately. We did not find a significant offset between the detectors when fit separately. We define the errors as the 16\textsuperscript{th} and 84\textsuperscript{th} percentile values from our sampling.

\subsubsection{Spectroscopic Light Curves}

For the spectroscopic fits, we fit seven parameters per spectral bin. We fit: $R_p/R_\star$, $u_1$, $u_2$, the baseline flux, and three parameters for the  second-order polynomial to describe the out-of-transit stellar variability. All other parameters which were fit for the white light curve and not fit for the spectroscopic light curves were held constant to the best-fit values from the white light curve fits. These parameters include $T_0$, $R_\star$, $M_\star$, and $b$. The fits were optimized using the same technique described in Section~\ref{subsubsec:feinstein}. We fit 80~spectroscopic bins per detector. The best-fit spectroscopic light curves were optimized separately, e.g. there was no cross-talk between fits in neighboring spectroscopic bins. We define the errors as the 16\textsuperscript{th} and 84\textsuperscript{th} percentile values from our sampling.

For this reduction, we did not fit for any starspot crossing events in the data in either the white-light or spectroscopic light curve fits. The 2\textsuperscript{nd}-order polynomial accounted for only the out-of-transit variability.

\subsection{Pipeline 3: ExoTiC+ custom}\label{subsec:mann}

\subsubsection{White Light Curve}
This analysis started with the reduced \jwst\ spectra from Section~\ref{subsec:feinstein}. We removed bad columns (spectra), identified by $>4\sigma$ higher RMS scatter than the four nearest columns. Two columns from NRS1 and 12 columns from NRS2 were removed this way. 

Measurement uncertainties can be estimated from the spectra, but we opted to adjust these using the time-series, which also allow us to identify any outliers. To start, we flatten the data using a 60-point running median (in time). We then estimate the scatter using the median absolute deviation, and remove any points $>3.5\sigma$ outside the running median. This step typically removed between two and fifteen points (0.1--0.6\%). We then scaled the uncertainties estimates from the combined spectra to match the estimates from the median deviation. This correction was usually $5-10$\%.

\begin{figure*}[ht]
    \centering
    \includegraphics[width=0.95\textwidth]{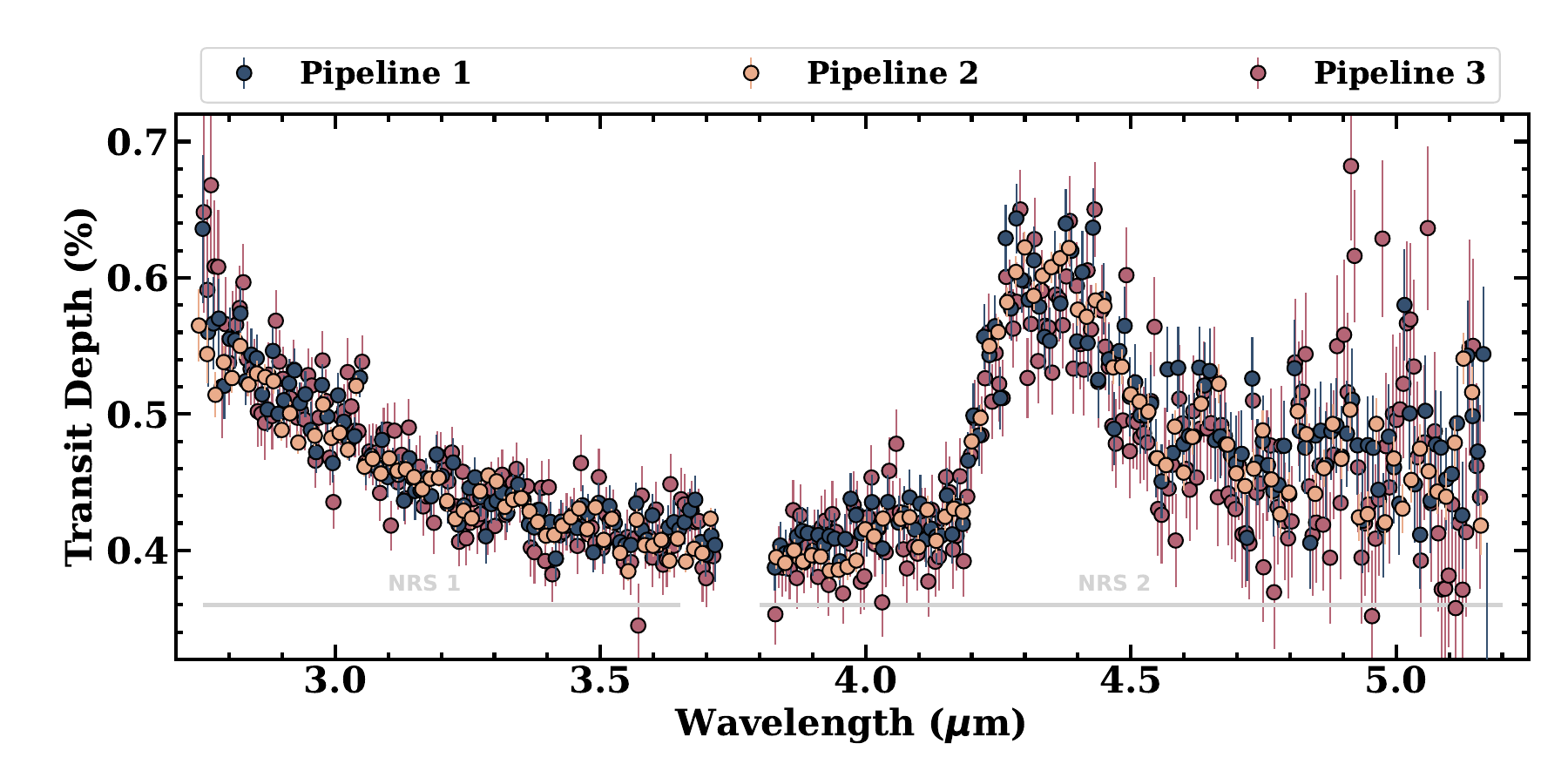}
    \caption{The transmission spectrum of \plname\ from Pipeline 1 (blue data points; Section~\ref{subsec:thao}), Pipeline 2 (orange data points; Section~\ref{subsec:feinstein}) and Pipeline 3 (pink data points; Section~\ref{subsec:mann}). All reduction pipelines are consistent with each other.    
    \label{fig:transspec}}
\end{figure*} 
We fit the broadband light curve with 19 free parameters in total. Six parameters were associated with the \texttt{BATMAN} model: $T_0$, $R_p/R_*$, $a/R_*$, $i_p$, $u_1$, and $u_2$ (as above). The period was locked to the \tess\ value (Table ~\ref{tab:tess_transitfit}), and we fixed eccentricity to zero. We modeled the stellar variability using a 4th-order polynomial with five free parameters. We model two spot crossing events using Gaussians, as described in Section~\ref{subsec:thao}, which adds three free parameters for each spot. 

The light curve shows significant time variability on a timescale of $\simeq$6\,minutes. This is close to the 6.5\,minute periodic signal seen in the trace and FWHM variation from NIRSpec commissioning time-series observations \citep{Espinoza2023}. Similarly, we see a matching time-variability in the y-position of the trace. The effect is clearly visible in NRS1 data, but not NRS2. To correct for this, we fit for two linear coefficients with the x-pixel and y-pixel shifts. Higher-order corrections and corrections in the Y FWHM or x$\times$y offered negligible improvement. 

We fit the broadband data with Gaussian priors on $i_p$ and $a/R_*$ derived from the \tess\, curve, and on the two limb-darkening parameters derived from \texttt{LDTK} \citep{Parviainen2015}. \newedit{As with Pipeline 1, $i_p$ and $a/R_*$ priors were imposed to compensate for the lack of egress, which made the transit duration more difficult to constrain.} All other parameters evolved under uniform priors with only physical limits. 

 \subsubsection{Spectroscopic Light Curves}

For the spectroscopic fits, we fit ten parameters per spectral bin. These were $R_p/R_*$, the five polynomial coefficients, two spot amplitudes ($A_1$ and $A_2)$, and the two coefficients for the x- and y-pixel shifts. All evolved under uniform priors. Spot amplitudes were allowed to go negative to avoid Lucy-Sweeney-type bias. We locked the limb-darkening parameters to the model values derived from BT-SETTL models using the \texttt{LDTK} code. Using a different model grid, non-linear limb-darkening laws, or simply letting the limb-darkening float resulted in a negligible change in the transit depth, primarily because of a trade-off between limb-darkening and the polynomial behavior. All other parameters were locked using the fit to the broadband curve.  

The MCMC fit for each spectral bin was run for 10,000 steps with 100 walkers. The resulting fits were excellent, yielding Allan deviation plots consistent with expectations for almost all bins. The exceptions were a single outlier at 4.05\um, and the four points at the edges of the detectors. We opted to remove the single outlier but retained the edge points. 

\subsection{The transmission spectrum of \plname}

The spectrum from Pipeline 2 was offset from the other two in terms of overall transit depth, most likely due to different treatment of the polynomial fit, $a/R_*$, and $i_p$, and spot handling. We shifted the overall spectrum (by 12\% for NRS1 and 9\% for NRS2), and show the resulting transmission spectra in Figure~\ref{fig:transspec}. The transmission spectra from all three analyses are broadly consistent in terms of spectral shape. 

Some prior analyses have adopted an average of multiple reductions \citep[e.g.,][]{Alderson2023} for a finalized transmission spectrum. However, the output from Pipeline 1 showed the best overall behavior, including the lowest red noise in the residuals and the lowest point-to-point variability (particularly past 4.5\um). The reduction in Pipeline 1 also showed the weakest impact from the spectrum shifting along the y-axis on the resulting light curve. As a result, we use the reduction from Pipeline 1 for all subsequent analysis, but confirm that the model fits and overall conclusions were similar when using the other reductions. 

The \jwst\ transmission spectrum of \plname\ is dominated by features from H$_2$O (2.8-3.5\um) and CO$_2$ (4.2-4.5\um). Both features are exceptionally strong: 30-50\% deeper than the baseline transit depth (0.4\% versus $\simeq0.6$\%). A weak \sotwo\ feature is also present at 4.05\um\ (Figure \ref{fig:transspec}). All three pipelines show a weak double-peak in the \cotwo\ band (near 4.45\um), which is expected as the opacity for \cotwo\ is double-peaked and is apparent in our model comparison below. There is also a small bump at 4.65\um, which is mostly due to CO.

Using the amplitude of the spectral features, we can obtain a preliminary estimate of the mass for a given planet radius and equilibrium temperature \citep{de_wit_constraining_2013}. This is achieve by exploiting the direct relationship between transmission spectral feature amplitude and atmospheric scale height using Equation 1 of \citet{stevenson2016}: 

\begin{equation}
    \Delta D \sim \frac{2HR_p}{R_*^2}
\end{equation}

\noindent where $\Delta D$ is the transit depth difference spanned by a single scale height ($H$). Given that a single large spectral band (e.g. the 4.3 $\mu$m CO$_2$ band) can span $\sim$5$H$ in a clear atmosphere \citep{Seager2000}, we find that $H$ $\sim$ 3000 km. Assuming an atmospheric mean molecular weight of 2.3 g mol$^{-1}$, appropriate for a solar metallicity atmosphere, our estimation yields a mass of only $\sim$15M$_{\oplus}$ -- comparable to the masses of Uranus and Neptune in the Solar System; remarkably, these planets possess radii less than half that of \plname. \newedit{This estimated mass is consistent with the mass derived from detailed atmospheric forward models, which accounts for the impact of the mean molecular weight on scale heights, as described in Section ~\ref{sec:atmosphere}}. This updated mass constraint allows us to shed light on its bulk composition and thermal evolution, which we discuss below. 

\section{Interior Modeling}\label{sec:interior}

\subsection{Model Description}

We construct 1-dimensional planetary thermal evolution models for planet masses of 8, 10, 15, 20, 30, and 50M$_{\oplus}$ that match the observed radius, age, and instellation of \plname. The models are based on the \citet{Thorngren2018} hot Jupiter models, which solve the equations of hydrostatic equilibrium, mass conservation, and an equation of state (EOS) and evolve the interior energy in terms of the specific entropy of the envelope using the \citet{Fortney2007} atmosphere models. For simplicity, we assume that the planet is composed of a gaseous envelope overlying a rocky core, with water vapor acting as a proxy for heavy elements (``metals'') and mixed within the envelope. We use the \citet{Chabrier2019} EOS for the hydrogen/helium mixture, and ANEOS \citep{Thompson1990} for the water and rock. 

\begin{figure*}[ht]
    \centering
    \includegraphics[width=0.45\textwidth]{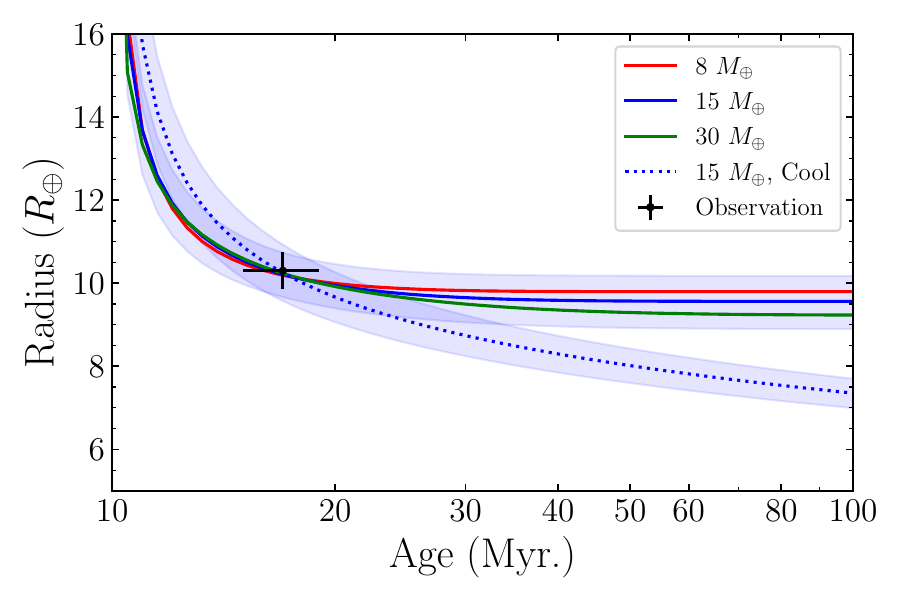}
    \includegraphics[width=0.45\textwidth]{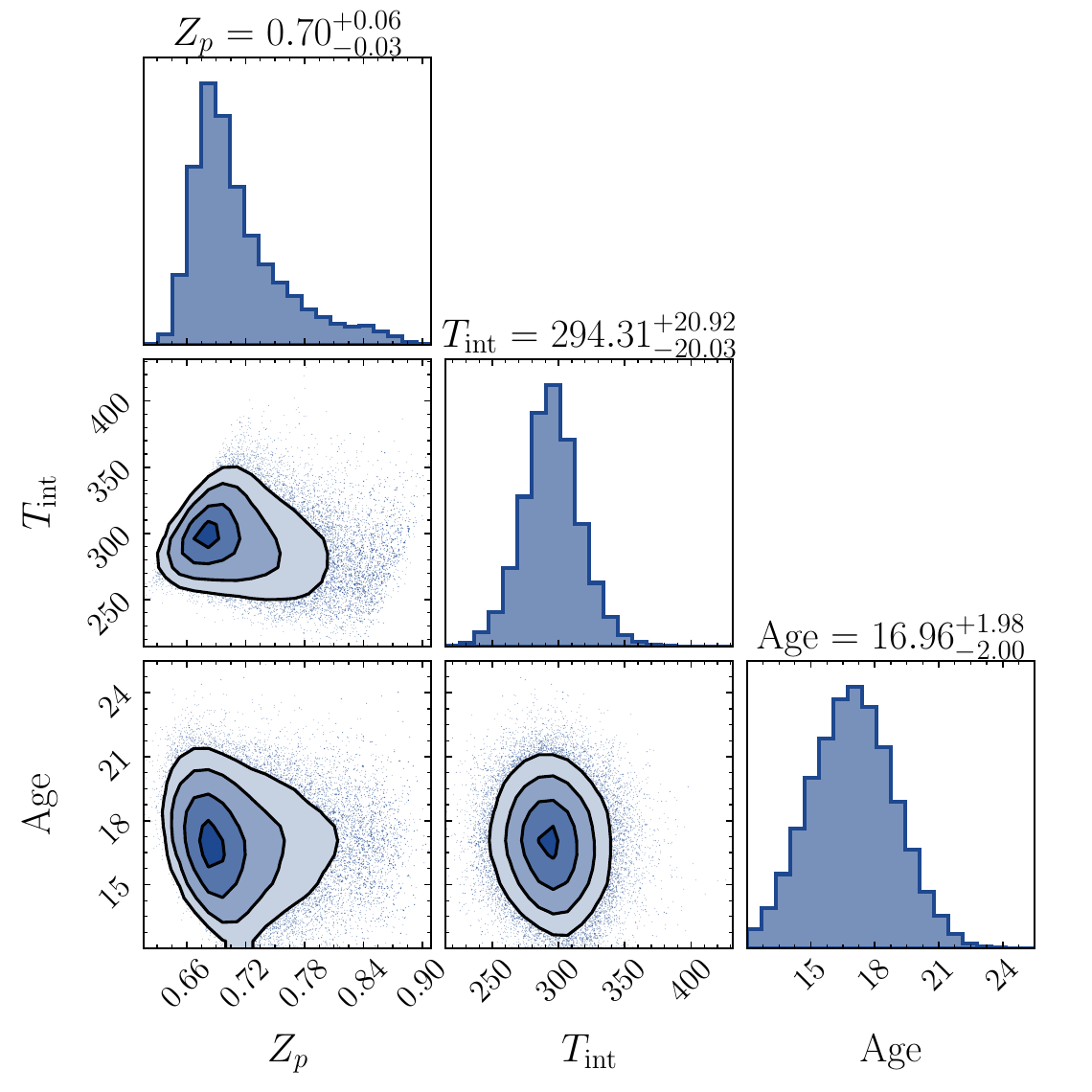}   
    \caption{\textit{Left:} Radius evolution of \plname\ for masses of 8M$_{\oplus}$ (red), 15M$_{\oplus}$ (blue), and 30M$_{\oplus}$ (green) assuming standard hot Jupiter interior heating. The dotted blue line shows the 15M$_{\oplus}$ case without interior heating. Posterior predictive errors on the radius are shown only for the 15M$_{\oplus}$ cases, but those of the other cases are similar in magnitude. \textit{Right: } Corner plot of the heated 15M$_{\oplus}$ posteriors, showing the bulk metallicity as a mass fraction ($Z_p$), the intrinsic temperature of the planet ($T_\mathrm{int}$), and the system age, which is in agreement with the age inferred from observations.
    \label{fig:radius_evolution}}
\end{figure*} 

\plname's updated density is low enough that it necessitated a change to how the models treat the outer layers of the planet. The existing interior model does not extend past the 10 bar pressure level to avoid incorrectly applying an adiabatic temperature gradient above the radiative-convective boundary (RCB); at lower pressures, the atmosphere is assumed to be 
isothermal at the equilibrium temperature using a simple scale-height calculation. We 
account for the change in gravity with altitude in the atmosphere and find that the radius diverges to infinity at small but non-zero pressures. At lower masses, this pressure can reach pressures as high as $\sim$nanobars, indicating the lack of a stable hydrostatic structure. We interpret this as an indication that boil-off \citep[e.g.][]{Owen2016} or core-powered mass loss \citep{Ginzburg2018} could occur in these lower-mass cases for this planet, which can significantly reduce its gas envelope mass and radius on Myr to hundreds of Myr timescales. We will discuss the possible impact of atmospheric loss on \plname\ via the aforementioned processes, as well as XUV photoevaporation, in Section \ref{sec:discuss}. 

To fit the forward models to the observed planet properties, we use a Bayesian retrieval approach similar to the one used in \citet{Thorngren2019}. We used four parameters to describe the state of the planet's interior: the bulk metallicity ($Z_p$), which is the mass ratio of metals in the planet to the total planet mass; the water metal fraction ($f_w$), which is the mass fraction of water mixed into the envelope as a fraction of the total metal mass (which also includes the rocky core)\footnote{This parameter mainly serves to incorporate modeling uncertainty from the metal composition into the inferred $Z_p$ and is not constrained by the data, and so we omit it from our results.};  
the intrinsic temperature, which is a proxy for the specific entropy of the planet; and the age of the planet. Figure \ref{fig:radius_evolution} (right) shows the resulting posteriors for the 15M$_{\oplus}$ case. 

As the equilibrium temperature of \plname\ is greater than 1000 K, the same inflation mechanism responsible for the enlarged radii of hot Jupiters may play a role here, and thus we allow for this extra heating by applying the predictions of \citet{Thorngren2018}. However, it is unknown whether this heating mechanism remains active at ice giant masses or lower, and so we also consider a case in which no interior heating is added.

\subsection{Intrinsic Temperature and Radius Evolution}\label{sec:tint}

The radius evolution implied by the posteriors for the 8, 15, and 30M$_{\oplus}$ cases are shown in Figure \ref{fig:radius_evolution} (left). For all of the masses considered, the thermal inertia of the planet is low enough that cooling from the initial hot state is very rapid. By 50 Myr, even the most massive planet (30M$_\oplus$) has largely reached an equilibrium with its heating term, where the hot Jupiter heating is equal to the intrinsic luminosity and the planet's specific entropy becomes constant in time. This results in a constant intrinsic temperature of 250-300 K, depending on the mass. At the present day, we predict that the intrinsic temperatures are 281, 284, 296, 309, 331, and 375 K for the 8, 10, 15, 20, 30, and 50M$_{\oplus}$ models, respectively -- a few tens of Kelvin hotter than its old-age value, while the planet is around half an Earth radius larger. 

The dotted blue line in Figure \ref{fig:radius_evolution} (left) shows the 15M$_{\oplus}$ case \emph{without} any additional heating. At the planet's current age, the difference in the radius evolution curves is not substantial, but on Gyr timescales, the planet would be significantly smaller. The intrinsic temperature will also perpetually decrease as is seen in the Solar System giants, rather than becoming constant. 

\subsection{Bulk Metallicity}

For masses of 8, 15, and 30M$_\oplus$, we infer bulk metallicities of $0.7908 \pm 0.0483$, $0.713 \pm 0.048$, and $0.624 \pm 0.043$ respectively. For the unheated 15M$_\oplus$ model, we find an only slightly lower value of $0.678 \pm 0.044$. Thus, regardless of the exact mass, we find that \plname\ is mostly metal by mass, with a radius that is inflated by high internal temperatures resulting from youth or a combination of youth and hot Jupiter heating. These inferred bulk metallicities are entirely reasonable under the core-accretion model \citep{Pollack1996}, and imply that the planet's core grew large enough to begin accreting gas in substantial amounts, but not so much that runaway accretion occurred to form a hot Jupiter. If the metals were fully mixed within the planet, then the equivalent atmospheric metallicities would be $260.7^{+90.7}_{-28.9}$, $172.9^{+60.5}_{-20.8}$, $118.9^{+29.6}_{-14.9}$, and $148.9^{+42.5}_{-18.6}$ $\times$ solar for the 8, 15, 30, and unheated 15M$_{\oplus}$ models, respectively. Comparing these values to those inferred from the transmission spectrum (see below) can give us a sense of the stratification of metals in the planet.  

\section{Atmospheric Modeling} \label{sec:atmosphere}

We now interpret the transmission spectrum of \plname\ in detail by comparing it to model spectra grids computed from self-consistent thermal structure models. This effort allows us to more quantitatively constrain the planet's mass, as well as estimate its atmospheric metallicity ([M/H]) and carbon-to-oxygen ratio (C/O). In addition, we consider the impact of simple gray clouds and photochemistry on our mass and atmospheric constraints. We describe the models and detail our modeling results below. 

\subsection{Model Description}\label{sec:modeldes}

We use the radiative-convective-thermochemical equilibrium model \texttt{PICASO}\footnote{\url{https://github.com/natashabatalha/picaso}} \citep{Batalha2019,Mukherjee2023} to generate temperature-pressure (TP) profiles for our self-consistent, clear-sky atmospheric model grid and compute model transmission spectra. We consider the same masses as the interior models, along with atmospheric metallicity [M/H] values of 0, 0.5, 1, 1.5, and 2 (e.g. 1 to 100 $\times$ solar metallicity in half dex increments) and C/O values of 0.25, 0.5, 1, 1.5, and 2 $\times$ the solar value of 0.458 \citep{Lodders2009}, resulting in 150 models. We use the intrinsic temperatures computed by the interior models (Section \ref{sec:tint}) as input to \texttt{PICASO} to set the interior heat flux, assuming the presence of hot Jupiter heating. We also generated a nearly identical model grid that did not include the hot Jupiter heating, but found that they resulted in transmission spectra identical to the heated case; this was expected since the heating mostly affects the TP profile at pressures higher than 1 bar, which the transmission spectra do not probe. Correlated-k coefficients for the thermal structure modeling are taken from \citet{lupu_2023_7542068}, while the opacities used to generate the model spectra are from \citet{natasha_batalha_2020_3759675,natasha_batalha_2022_6928501}. The self-consistent atmospheres span a pressure range from 100 bars to 1 $\mu$bar, but we found that the transmission spectra generated from this setup missed the top of the 4.3 $\mu$m CO$_2$ band, which probes lower pressures. Thus, we extended the model TP profiles upward isothermally to 1 nbar from the temperature at 1 $\mu$bar when computing spectra. 

We fit the model spectra to the observed \jwst\,/NIRSpec/G395H transmission spectrum of \plname\ by varying the reference pressure and minimizing the chi-square ($\chi^2$) using the Nelder-Mead method of \texttt{scipy.optimize.minimize}. We do not consider the simultaneous SOAR point in the fit, as the optical wavelengths could be affected by haze and/or stellar contamination (see Section \ref{sec:spots}), which we do not consider in this current set of models. We do, however, consider the impact of clouds by including a simplified gray cloud with a total optical depth of 100 and asymmetry parameter and single scattering albedo values of 0. \newedit{The base of the cloud is set to 100 bars, while the cloud-top pressure is a free parameter that is allowed to vary from 10 bars to 1 nbar.} We then refit the model grid to the data by varying both the reference pressure and the cloud top pressure to minimize the $\chi^2$.

In addition to CO$_2$ and H$_2$O, the transmission spectrum of \plname\ also exhibits a subtle feature just longward of 4 $\mu$m that is likely due to SO$_2$. In H$_2$/He-dominated atmospheres, SO$_2$ is derived from photochemical destruction of H$_2$S and subsequent oxidation of the freed atomic S \citep{Zahnle2009,Tsai2021}. Photochemical SO$_2$ has been observed in the atmospheres of the hot Jupiter WASP-39b \citep{rustamkulov2023early,Alderson2023,Tsai2023} and the warm inflated Neptune WASP-107b \citep{Dyrek2024}. In particular, \citet{Tsai2023} found that SO$_2$ is a powerful diagnostic of atmospheric metallicity, with higher metallicities leading to larger SO$_2$ features. We thus run a grid of photochemical models for \plname\ using the 1D photochemical code \texttt{VULCAN}\footnote{\url{https://github.com/exoclime/VULCAN}} \citep{Tsai2017,Tsai2021}, which was one of the models used to simulate the photochemistry of WASP-39b in \citet{Tsai2023}. We use the same SCHNO photochemical reaction network as in that study, along with the stellar SED from Section \ref{sec:sed} and the TP profiles from the corresponding \texttt{PICASO} models as input. To fully capture the region of the atmosphere where photochemistry is active, we extend the \texttt{PICASO} TP profiles upwards isothermally to 10 nbar from the temperature at 1 $\mu$bar. We assume a constant-with-pressure eddy diffusion coefficient (K$_{zz}$) of 10$^{10}$ cm$^2$ s$^{-1}$ to parameterize the vertical mixing in the model, which is similar to those predicted by general circulation models \citep[e.g.][]{Parmentier2013,Agundez2014}, though we ignore any pressure dependence for simplicity. Uncertainties in the value of K$_{zz}$ are unlikely to impact our results since the SO$_2$ abundance is mostly insensitive to it \citep{Tsai2023}. 

Since \texttt{VULCAN} takes much longer to run to convergence than \texttt{PICASO}, we chose to only model a single mass (15M$_{\oplus}$) and ignore the 2 $\times$ solar C/O cases due to their low O abundance and bad fit to the data (see below), resulting in 20 total models. We used \texttt{PICASO} to compute the transmission spectra of the atmospheric composition calculated by \texttt{VULCAN}. To capture the full 4.3 $\mu$m CO$_2$ peak, we extend all of the abundance profiles upwards to 1 nbar, assuming constant values above 10 nbar. Although these abundances are not fully consistent with the TP profiles, which were calculated assuming equilibrium chemistry, the impact of the photochemically derived abundances on the actual temperature is minimal, as demonstrated by \citet{Tsai2023}. 

\subsection{Model Results}

\begin{figure*}[ht]
    \centering
    \includegraphics[width=0.93\textwidth]{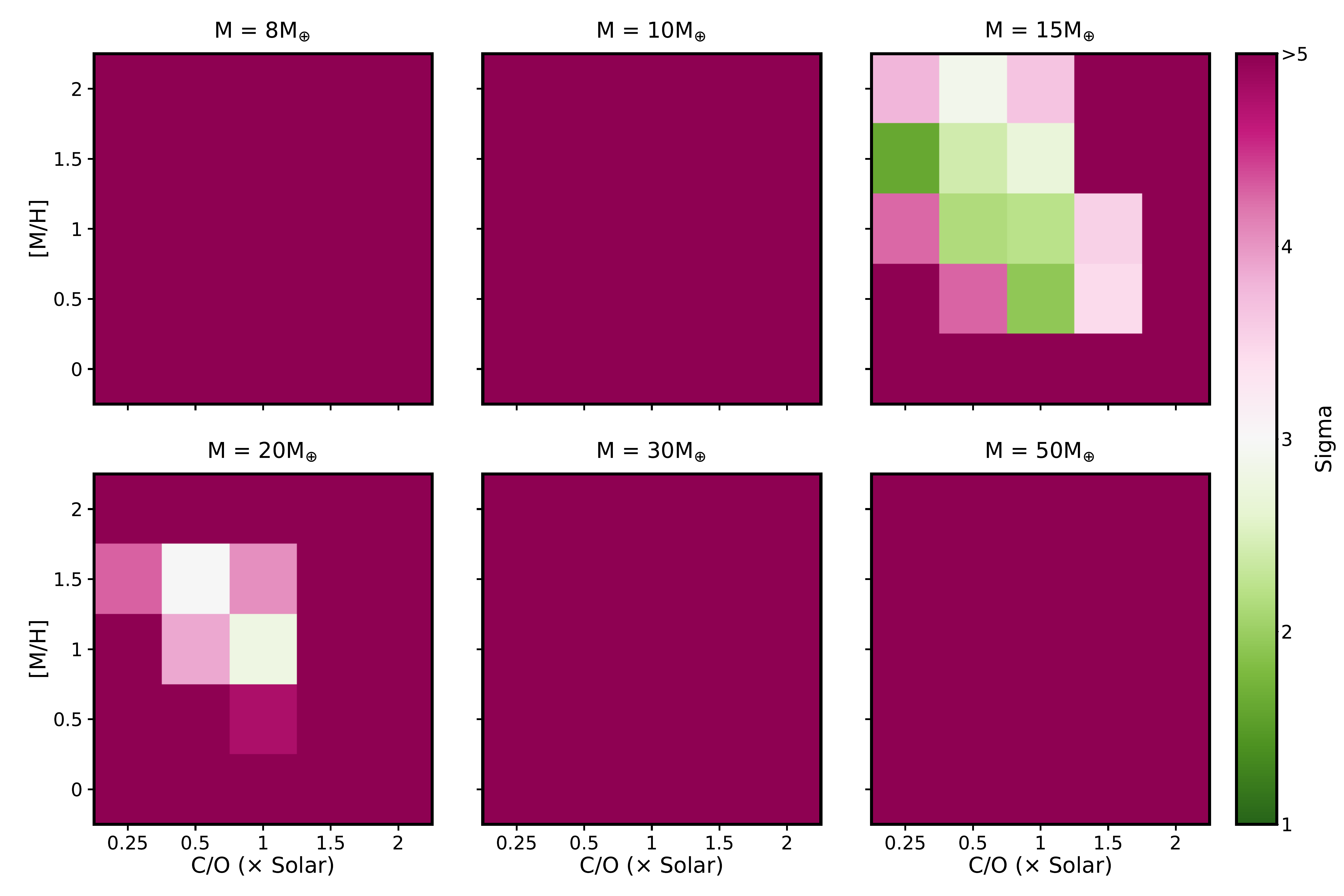}
    \caption{Sigma ($\sigma$) values for the clear atmosphere \texttt{PICASO} models' fits to the \jwst\, transmission spectrum, encompassing masses from 8-50 $M_{\oplus}$,  C/O ratios spanning from 0.25-2 $\times$ Solar, and [M/H] ranging from 0-2. Most of the parameter space is ruled out to a confidence level $>$5$\sigma$ (pink), while the most favorable fit models (green) are 15-20 M$_{\oplus}$, a high metallicity, and a C/O $\le$1.5 $\times$ Solar.}
    \label{fig:clear_chisq}
\end{figure*}

\begin{figure*}[]
    \centering
    \includegraphics[width=0.93\textwidth]{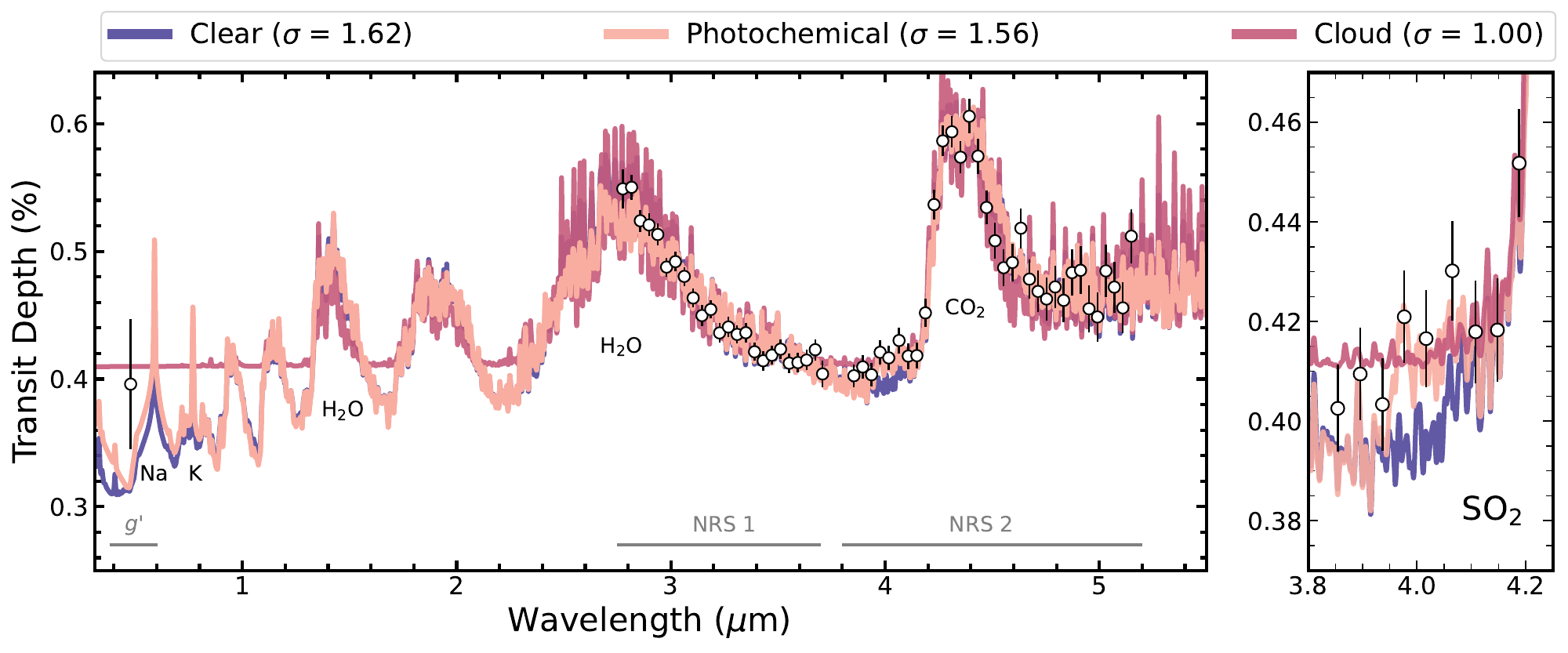}
     \caption{Transmission spectrum of \plname\, with the best (lowest sigma) atmospheric models for each scenario: for the clear atmosphere model, M = 15M$_{\oplus}$, C/O = 0.25, [M/H] = 1.5 (purple); for the cloudy model,  M = 8M$_{\oplus}$, C/O = 0.25, [M/H] = 1.5 (pink); for the photochemical model, M = 15M$_{\oplus}$, C/O = 0.50, [M/H] = 1.0 (red). Models are binned to a spectral resolution of $R$=600 and data points are binned to a bin size of 4. The plot on the right highlights the presence of SO$_{2}$} in the atmosphere of the planet.  \label{fig:best_model}
\end{figure*}

\begin{figure*}[]
    \centering
    \includegraphics[width=0.93\textwidth]{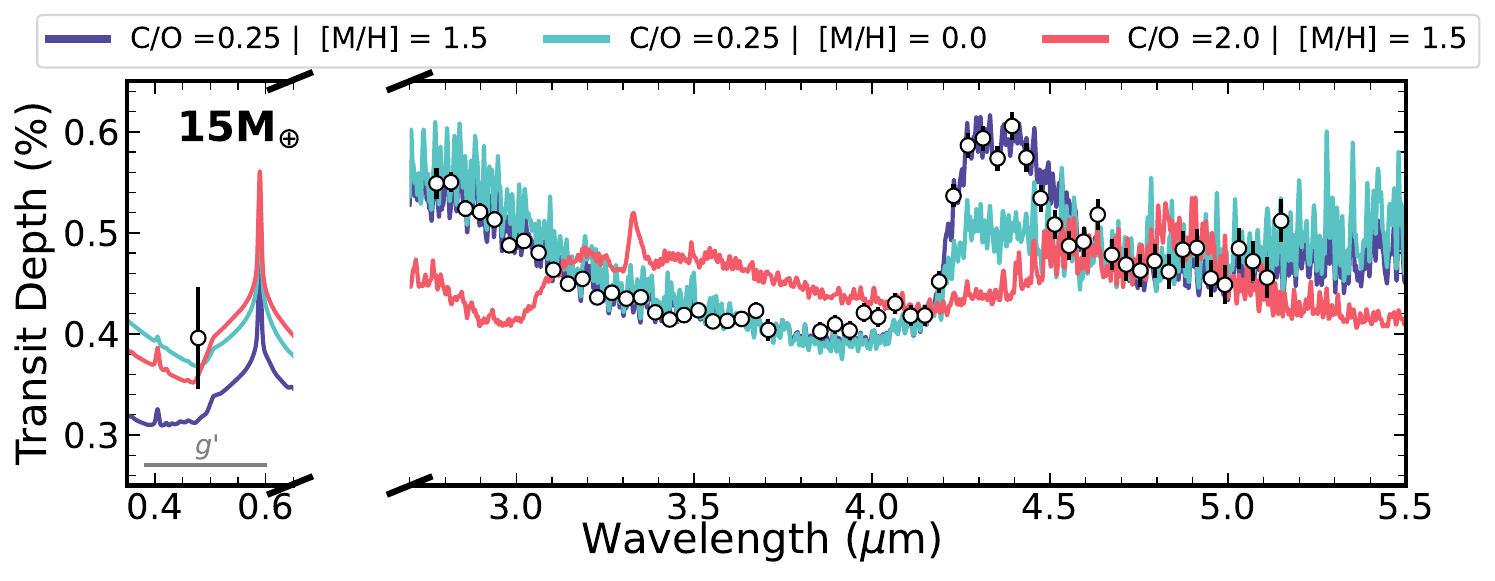}
    \caption{Clear atmospheric models for a mass of $15_{\oplus}$  across various C/O and [M/H] ratios:  C/O=0.25 and [M/H] = 1.5 (purple), C/O=0.25; [M/H] =0.0 (blue);  C/O=2.0 and [M/H] = 1.5 (pink). Models are binned to a spectral resolution of $R$=600. The observational data points are shown in black and have been binned to enhance clarity, with a bin size of 4. The C/O=0.25;[M/H]=1.5 model (purple) demonstrates the closest fit to the observational data. As the C/O ratio increases (as show in pink), a methane feature is introduced (3.4 $\micron$) which is not present in the data; while a decrease in [M/H] (as shown in blue),  the CO$_{2}$ feature decreases. }    \label{fig:best_lowmet_highco}
\end{figure*}

\begin{figure*}[ht]
    \centering
    \includegraphics[width=0.93\textwidth]{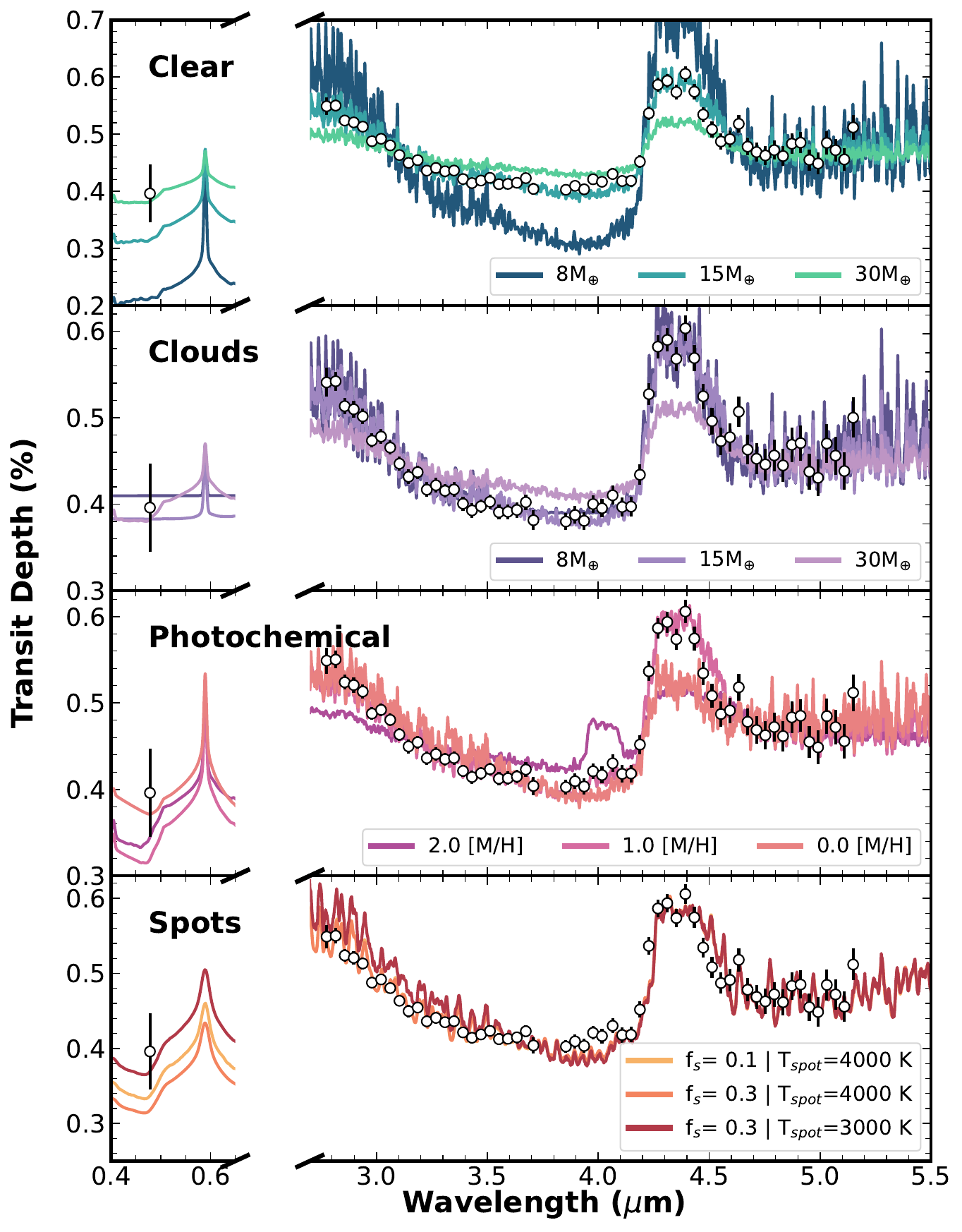}
    \caption{Atmospheric models for each scenario with various masses, metallicity ([M/H]), and C/O.  In clear models, variations in mass influence the intensity of the C/O feature. Cloud models exhibit changes in cloud deck strength with varying masses. Photochemical models show additional features with high [M/H] and a weakened C/O feature with low [M/H]. Spot models highlight significant differences in the optical band} 
    \label{fig:model_ts}
\end{figure*}

\begin{figure*}[ht]
    \centering
    \includegraphics[width=0.93\textwidth]{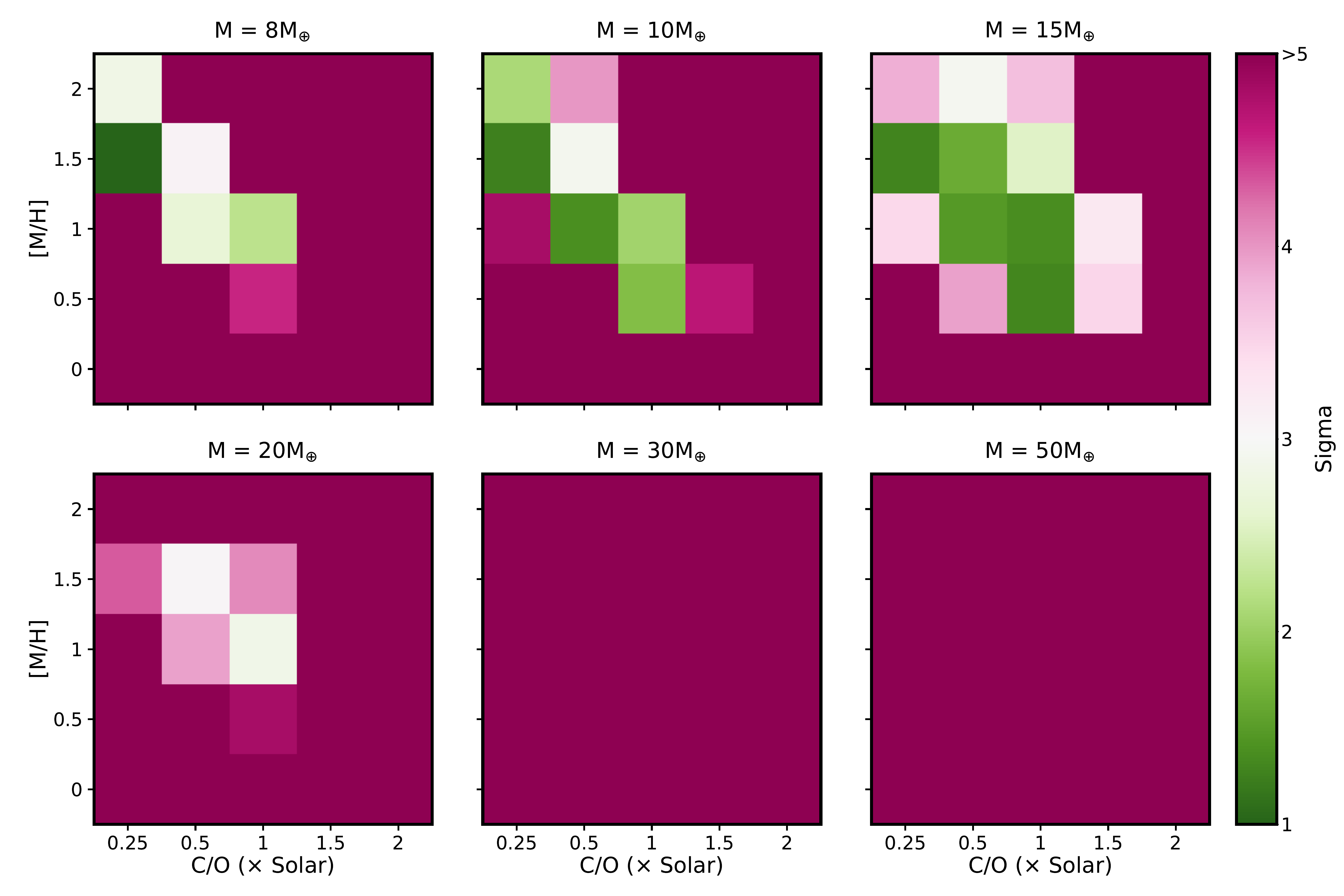}
    \caption{Same as Figure \ref{fig:clear_chisq}, but for the cloudy models.}
    \label{fig:cloud_chisq}
\end{figure*}

For clear atmospheres in radiative-convective-thermochemical equilibrium, we find that the data prefer models with planet masses $\sim$15-20M$_{\oplus}$, supersolar metallicities, and C/O $<$ 2 $\times$ solar, or equivalently C/O $<$ 1; all other models are ruled out to $>$5$\sigma$, while supersolar C/O models in general are ruled out to $>$3$\sigma$ (Figure \ref{fig:clear_chisq} and Table \ref{tab:best_models}). Many of these models are able to reasonably reproduce the overall shape of the spectrum, including the H$_2$O and CO$_2$ features, though as expected the SO$_2$ feature is not captured since photochemistry is not considered in these models. The best fitting model, with $\chi^2/N$ = 1.10, is the 15M$_{\oplus}$ model with [M/H] = 1.5 ($\sim$30 $\times$ solar) and C/O = 0.25 $\times$ solar $\sim$ 0.1 (Figure \ref{fig:best_model}), where $N$ is the number of data points. Both higher and lower mass models result in worse fits due to mismatches in the atmospheric scale height and thus the spectral feature amplitudes (Figure \ref{fig:model_ts}). Solar metallicity models are not preferred due to a lack of the observed prominent 4.3 $\mu$m CO$_2$ feature, while C/O $\sim$ 1 (e.g. 2 $\times$ solar) models badly fit the data since they exhibit a large CH$_4$ feature at 3.3 $\mu$m, which is not observed (Figure \ref{fig:best_lowmet_highco}). 

Including a simplified gray cloud in our model does not change our constraints on the atmospheric metallicity and C/O, but does loosen the constraint on the mass (Figure \ref{fig:cloud_chisq} and Table \ref{tab:best_models}), as the larger spectral feature amplitudes of the lower mass (8 and 10M$_{\oplus}$) models can be reduced by the cloud deck (Figure \ref{fig:model_ts}). As such, our mass constraint of $\sim$15-20M$_{\oplus}$ from this modeling can only be interpreted as an upper limit. The best fitting cloudy model, with $\chi^2/N$ = 1.02, is the same as the clear case, but with a mass of 8 instead of 15M$_{\oplus}$ (Figure \ref{fig:best_model}). In general, the best fit cloud top pressures for the $<$15M$_{\oplus}$ models are in the range of a few $\mu$bar to $\sim$1 mbar, which are somewhat lower than those predicted by cloud physics models for planets with similar equilibrium temperatures as \plname\ \citep[e.g.][]{gao2020}. Such low cloud-top pressures could be achievable due to the low gravity of this planet and/or through rates of vertical mixing higher than those assumed in the cloud physics models. Alternatively, high altitude photochemical hazes could persist in \plname's atmosphere, though the relatively small optical transit depth measured by SOAR may be evidence against the existence of any optical scattering slopes due to hazes.

\begin{figure}[ht]
    \centering
    \includegraphics[width=0.45\textwidth]{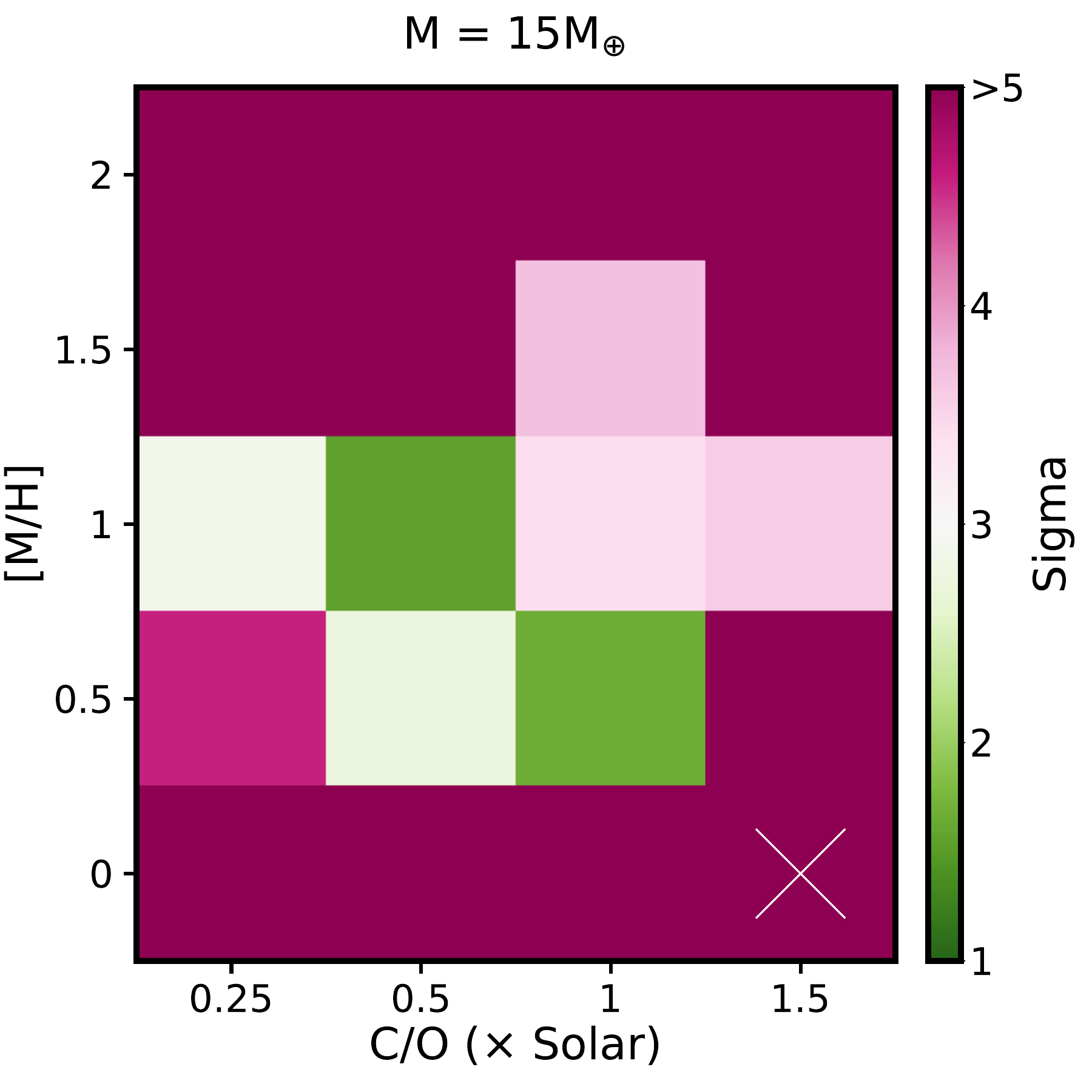}
    \caption{Sigma for the \texttt{VULCAN} photochemical model grid, which considered a subset of the parameter values from the clear and cloudy model grids. The ``X'' in the lower right corner, situated on the solar metallicity, 1.5 $\times$ solar C/O model grid, indicates that that particular \texttt{VULCAN} model did not converge despite our best efforts, and thus its $\sigma$ value may not be correct.}
    \label{fig:photochemistry_chisq}
\end{figure}

\begin{figure}[ht]
    \centering
    \includegraphics[width=0.45\textwidth]{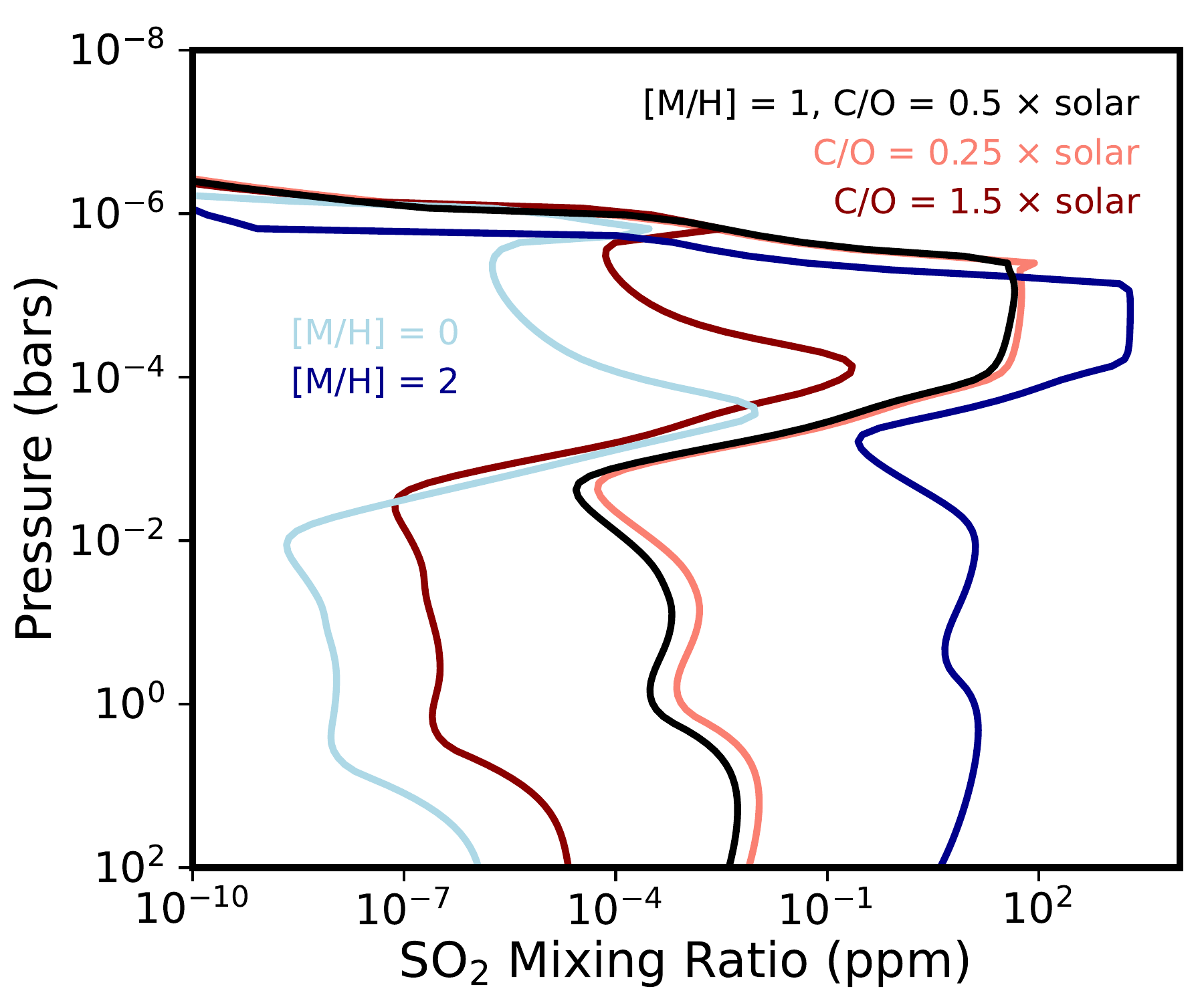}
    \caption{SO$_2$ mixing ratio profiles for selected cases in the \texttt{VULCAN} photochemistry grid, showing the variation in the SO$_2$ abundance with metallicity and C/O. }
    \label{fig:so2abun}
\end{figure}

The consideration of photochemistry tightens our constraints on the atmospheric composition, ruling out to $>$3$\sigma$ models with $>$10 $\times$ solar metallicity (Figure \ref{fig:photochemistry_chisq} and Table \ref{tab:best_models}). The best fitting model, with $\chi^2/N$ = 1.09, possesses an atmospheric metallicity of 10 $\times$ solar and a C/O of 0.5 $\times$ solar (Figure \ref{fig:best_model}). Higher/lower metallicity models result in SO$_2$ features that are larger/smaller than that observed, respectively (Figure \ref{fig:model_ts}), matching the trends seen in the models of \citet{Tsai2023}. A higher metallicity drives greater SO$_2$ production from the increased H$_2$S and H$_2$O abundances, and vice versa for a lower metallicity. Likewise, the lower O abundances in high C/O atmospheres lead to less SO$_2$ production, and vice versa for lower C/O (Figure \ref{fig:so2abun}).

\newedit{\section{Atmospheric Retrievals} \label{sec:retreivals}}

\newedit{We employ a Bayesian inference framework adapted from \texttt{CHIMERA} \citep[e.g.,][]{line_systematic_2013} to further quantify the constraints on \plname's mass, thermal structure, and atmospheric abundances offered by its transmission spectrum. We assume a one-dimensional atmosphere in hydrostatic equilibrium for a plane-parallel transit geometry. The Bayesian inference is done using the MultiNest nested sampling algorithm \citep{feroz2009multinest} via PyMultiNest  \citep{buchner_x-ray_2014}.}

\newedit{We construct an atmospheric model with 99 pressure layers uniformly spaced logarithmically from $10^{-8.7}$ to $10^{1.1}$ bar. We consider opacity contributions due to prominent chemical absorbers in  exoplanet atmospheres \citep[e.g.,][]{madhusudhan_exoplanetary_2019} including H$_2$O \citep{polyansky_exomol_2018}, CO \citep{li_rovibrational_2015}, CO$_2$ \citep{huang_isotopic-independent_2012}, CH$_4$ \citep{hargreaves_accurate_2020}, NH$_3$ \citep{coles_exomol_2019}, H$_2$S \citep{azzam_exomol_2016}, SO$_2$ \citep{underwood_exomol_2016}, as well as Na and K \citep{allard_k-h2_2016,allard_temperature_2023} and H$_2$-H$_2$ and H$_2$-He collision induced absorption \citep{karman_update_2019}. The abundances of these gases are assumed to be constant with height and are free parameters in our model. The vertical temperature structure of the atmosphere is parameterized following \citet{madhusudhan_temperature_2009}. The presence of clouds and hazes follows the two sector prescription of \citet{welbanks_aurora_2021}. In this prescription, the spectroscopic effect of aerosols is parameterized following a deviation from H$_2$-Rayleigh scattering \citep[e.g.,][]{lecavelier_des_etangs_rayleigh_2008} and clouds are included as an additional gray source of opacity \citep[e.g.,][]{mai_exploring_2019, Welbanks2024}. This two sector prescription allows for the presence of inhomogeneities (e.g., `patchy clouds') following the linear combination approach suggested by \citet{line_influence_2016}. We also fit for the planetary mass and place constraints on this parameter by exploiting the mass--scale height dependence, as in Section \ref{sec:atmosphere}. 
The atmospheric models are computed at a resolving power of R=100,000 before being binned down to the resolution of the observations.}

\begin{figure*}[]
    \centering
    \includegraphics[width=0.93\textwidth]{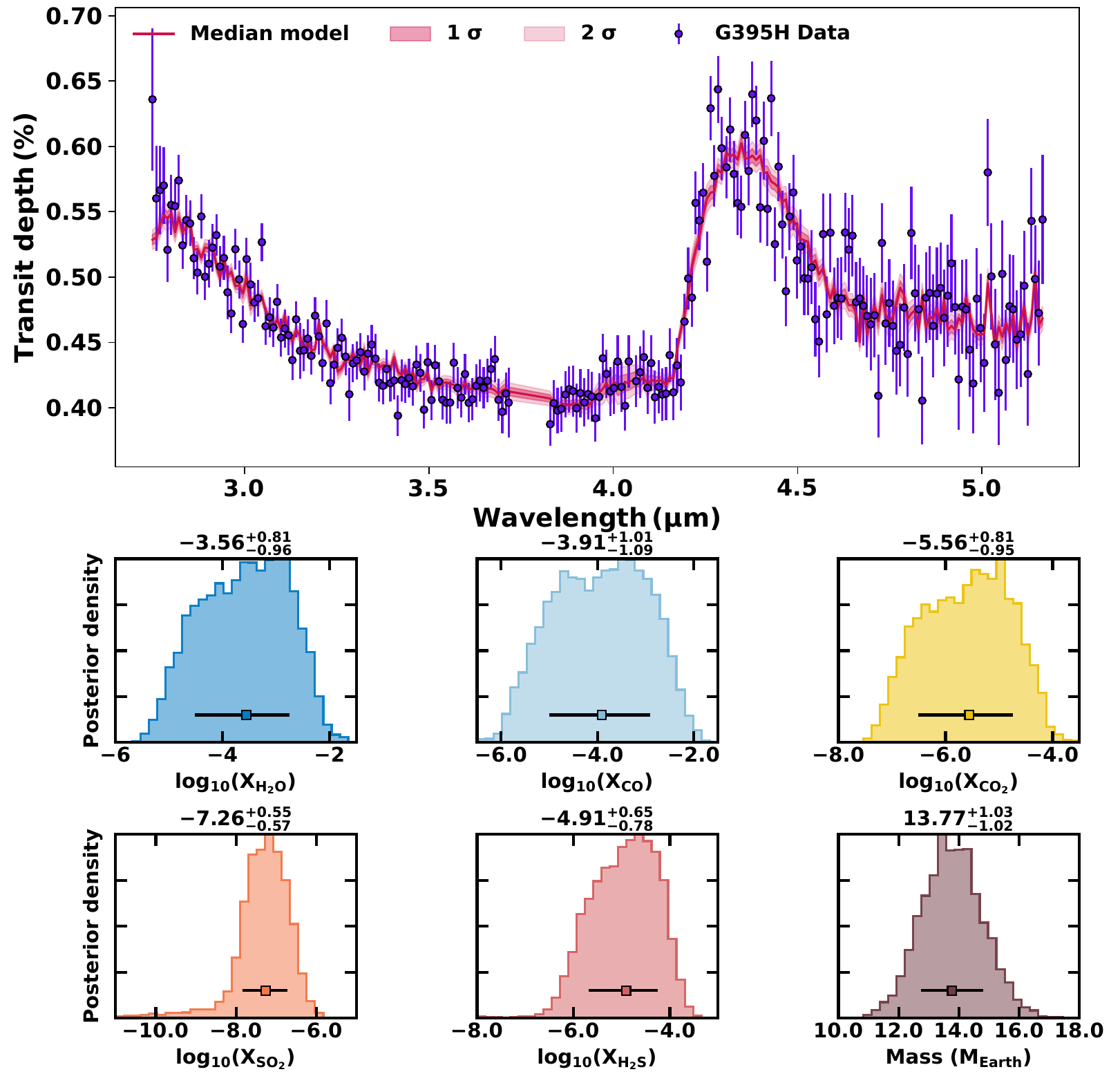}
    \caption{\newedit{Atmospheric retrieval results. The retrieved median model is shown in red while the 1$\sigma$ and 2$\sigma$ envelopes are shown by pink shaded regions. The NIRSpec G395H observations are shown using blue error bars. The retrieved posterior distributions for the chemical species of interest (e.g., H$_2$O, CO, CO$_2$, SO$_2$, and H$_2$S) are shown alongside the retrieved posterior probability density for the planetary mass.}}
    \label{fig:retrieval}
\end{figure*}

\newedit{Our atmospheric retrieval analysis finds strong detections of CO$_2$ (11$\sigma$) and H$_2$O (7$\sigma$), with  a $\chi^{2}/N$ = 1.042 for the best-fit model. The spectrum also suggests a preference for CO at 3.5$\sigma$ with weak suggestions of H$_2$S and SO$_2$ at 2.1 and 1.8$\sigma$, respectively. The retrieved median model alongside its confidence intervals are shown in Figure \ref{fig:retrieval}. Our retrieval finds moderate constraints at the $\sim1$~dex level for the gases in question, namely $\text{log}_{10} \text{X}_{\text{CO}_2} = -5.56 ^{+0.81}_{-0.95}$, $\text{log}_{10} \text{X}_{\text{H}_2\text{O}} = -3.56 ^{+0.81}_{-0.96}$, $\text{log}_{10} \text{X}_{\text{CO}} = -3.91 ^{+1.01}_{-1.09}$, $\text{log}_{10} \text{X}_{\text{H}_2\text{S}} = -4.91 ^{+0.65}_{-0.78}$, and $\text{log}_{10} \text{X}_{\text{SO}_2} = -7.26 ^{+0.55}_{-0.57}$. Assuming these species are representative of the overall oxygen and carbon content in the atmosphere, we derive a constraint on C/O$=0.31^{+0.17}_{-0.11}$. The inferred metallicity from the free retrieval is largely unconstrained with a 99\% upper limit of [M/H]=1.15. These results are consistent with the best-fit \texttt{PICASO} models. }

\newedit{Our analysis does not find a strong preference for an instrumental offset between G395H detectors, retrieving values consistent with an offset of 0~ppm. Similarly, our analysis does not find a strong preference for the presence of clouds or hazes, with the retrieved cloud and haze properties generally unconstrained. The retrieved planetary mass of $\text{M}_\text{p}= 13.8^{+1.0}_{-1.0}~\text{M}_{\oplus}$ is consistent with the findings of the clear atmosphere \texttt{PICASO} models and within the mass upper limit provided by the cloudy models.}

\section{Stellar Contamination Modeling}\label{sec:spots}

An ongoing challenge in studying transmission spectra is contamination by inhomogeneities (spots and faculae) on the stellar surface \citep[e.g.,][]{Rackham2018}. Such surface brightness variations can change the observed signal, whether or not the planet crosses them. In the case of unocculted spots, the transit chord will be brighter than the rest of the star, yielding a transit depth that is deeper than that for a pristine star and wavelength-dependent. In particular, unocculted spots can introduce features and blueward slopes in the transmission spectrum that are not present on the planet \citep[e.g.][]{Barclay2021,Thao2023}. 

Fortunately, nothing on the stellar surface can produce the observed CO$_2$ feature in the \plname's spectrum, as there is no significant CO$_2$ present in a stellar spectrum for reasonable spot temperatures. This can be seen in e.g., Figure~\ref{fig:model_ts}d: adding spots to the model changes the observed transit depth in the optical and near the H$_2$O band (2.8-3.5\um) without any impact on the 4-5\um\ region. Although it would take unrealistically large spot covering fractions ($>80\%$) to reproduce the \water\ band, smaller covering fractions can change the overall strength of the features, which in turn may change which atmospheric model best reproduces the data. We quantify this effect using a model that accounts for stellar surface inhomogeneities.

Our method follows that of \citet{Thao2023}, which was modified from the method outlined in \citet{rackham2017access}, \citet{Thao2020}, and \citet{Libby-Roberts2022}. We fit three free parameters: the transit depth scale factor ($D_{mod}$), the spot temperature ($T_{\rm{spot}}$), and the spot coverage fraction ($f_S$). The scale factor is a modification to the input model transmission spectrum, which only accounts for atmospheric effects. We fix the stellar surface temperature to the spectroscopic value (5650\,K), but the results are unchanged when we adjust this by 50-100\,K in any direction. 

In general, the spot coverage assumes that the transit chord is pristine. This is not true in our case, as we see two clear spot crossings in the data (Section~\ref{sec:jwst}). However, these were fit out in our transit model from Pipeline 1 and hence should not bias our depths. A large number of small spots in the chord could have gone unnoticed. However, adding in another variable for the transit chord spot coverage fraction ($f_C$) had no impact on our results.

We set all parameters to evolve under uniform priors with physical or practical limits. \tspot\ is allowed to go above \teff\ (that is, we include hot faculae) and was only limited by the model grid (3000--6500\,K). $f_S$ was limited to between 0 and 1, and $D_{mod}$ to between 0.85 and 1.15, beyond which the change in inferred radius would invalidate the model. More extreme values of $D_{mod}$ also require large ($>30\%$ if the spot is $<5000$\,K) spot/faculae coverage fractions (which also shift the overall depth), which can be ruled out by other means. For example, $f_S>0.5$ and \tspot$<4000$ would exhibit atomic and molecular features seen in K/M dwarfs in out-of-transit spectrum \citep{Thao2023}.

Rather than treating the model selection as its own free parameter, which would require inaccurate interpolation between models, we performed the fit on each of the 150 base (clear atmosphere) models. We did not repeat this with the full set of models including clouds or photochemistry, as the results are qualitatively similar to that of adding spots to the base models. 

\begin{figure*}[ht]
    \centering
    \includegraphics[width=0.93\textwidth]{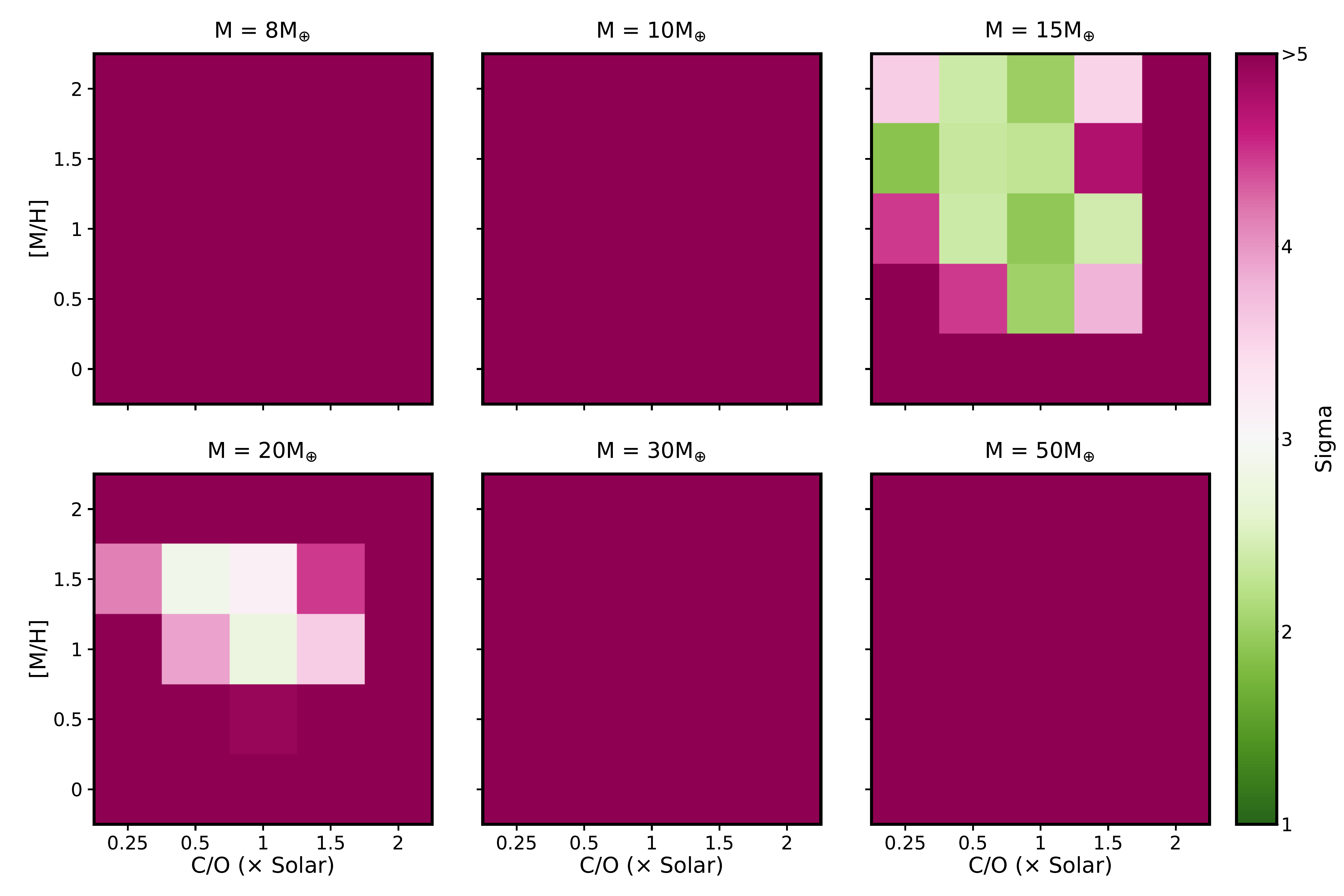}
    \caption{Same as Figure~\ref{fig:clear_chisq} but including the impact of inhomogeneities on the stellar surface for the clear-atmosphere models. The effect is largely negligible, with improvements to the fits isolated to the high metallicity and/or high C/O cases due to better agreement around the \water\ band.}
    \label{fig:spotchisq}
\end{figure*}

We find that the addition of spots had almost no impact on our model fits. The best-fit models were unchanged from the clear fits, and adding spots usually provided an improvement too small to be justified given the additional free parameters (Figure~\ref{fig:spotchisq}). This is particularly true when we consider only good fits ($\sigma<5$). The greatest gains in $\chi^2$ came from the C/O = 1.5 $\times$ solar and 30-100 $\times$ solar metallicity fits. In these cases, cold ($<3500$\,K) spots covering $10-20$\% of the star strengthens the H$_2$O band, providing a better fit to the data. \newedit{This is consistent with the low spot coverage levels from prior Doppler-tomography results \citep{heitzmann2021obliquity}. We also find that no physically realistic spot can reproduce the \cotwo\ and SO$_2$ features in the transmission spectrum, as even the coldest spots do not contain a significant population of either molecule (Figure~\ref{fig:model_ts}).}

\section{Discussion}  \label{sec:discuss}

\subsection{Transit Timing Variations}

\newedit{The transit of \plname\ came later than the expected mid-transit time, resulting in the omission of the egress in the \jwst\, observation. These transit timing variations (TTVs) will be explored in an upcoming paper \citep{LopezMurillo_inprep}. A second planet candidate was reported in \citet{Rizzuto2020} and was initially believed to have gone undetected in subsequent \tess\ data. This second planet was later recovered near a 2:1 resonance and will be further investigated in  \cite{barber2024tess}. Utilizing these TTVs will offer an independent method to constrain the mass of this planet \citep{Thao_inprep}.}  We note that because the TTV signal is small within the \tess\ window, a formal accounting had no impact on the \tess-derived properties in this paper.

\subsection{No significant impact from stellar contamination}

Prior studies of planets transiting low-mass stars have found significant stellar contamination in their final transmission spectrum \citep[e.g.,][]{Barclay2021, Lim2023}. One might expect that the problem is even worse for young stars due to their heightened activity. However, our best-fit models that included the effects of surface inhomogeneities are nearly identical to those that ignored them.  

Our \jwst\, transit observations show at least two starspot crossing events, so clearly there are spots on the surface. However, modest spot levels ($<$20\%) are allowed by our fits, which have no impact on the planet parameters. \newedit{Ultimately, the transmission spectrum features are so large that the required spots would be visible in other ways (e.g., in the stellar spectrum) and in addition would still not be able to explain e.g. the large \cotwo\ feature.}

\newedit{The simultaneous SOAR g' transit also rules out the most extreme spot coverage fractions. For example, $>$50\% $<4000$\,K spots would yield a much deeper transit depth at blue-optical wavelengths. However, these extreme depths were disfavored by the \jwst\ data alone due to the need to reproduce the strong \cotwo\ feature. Nonetheless, we still advise taking such optical counterparts, as it could have proved critical had the planet been higher mass and hence had weaker features in the \jwst~wavelengths.}

\subsection{Uncertainty Estimation}

The $\chi^2$ values from our model fits suggest that our final uncertainties are overestimated. The best-fit equilibrium chemistry models often possess $\chi^2/N$ $\leq$ 1.1, which is unusually low considering that these models did not include SO$_2$. This is most likely due to our conservative approach to fitting the stellar variability. We used a 4\textsuperscript{th}-order polynomial, which was allowed to float freely with wavelength. Stellar variability should vary with wavelength, but in a smooth manner, such that each spectral bin should depend on the points around it. Similarly, the polynomial probably contains both wavelength-independent terms (which can be fixed or evolve under a tighter prior) and wavelength-dependent terms. However, the magnitudes of the transmission spectral features were so large that the more conservative approach was sufficient for the results here.

\subsection{The Formation and Evolution of \plname}

The atmospheric composition of a planet is dependent on where and how the planet formed, where it accreted its atmosphere, its migration history within its natal disk, its atmospheric loss history, and atmosphere-interior interactions. In particular, the C/O value of an atmosphere can reflect the gas composition at the location in the disk where it accreted. For example, \citet{oberg2011effects} showed that the C/O of the gas component of the disk beyond the water snowline is likely supersolar due to freeze out of water ice, and thus planetary gas envelopes accreted there should possess supersolar C/O as well. However, solids should be accreted along with the gas and enhance the envelope metallicity. As the solids are typically oxygen-rich, this would result in a stellar-to-subsolar C/O \citep{mordasini16,espinoza2017, cridland2019}. As such, the supersolar metallicity and solar-to-subsolar C/O we have derived for \plname\, are consistent with formation via simultaneous accretion of gas and oxygen-rich solids, potentially beyond the water snowline. However, significant complexity and nuance exists in connecting disk and atmospheric chemistry \citep[e.g.][]{Madhusudhan2017,Eistrup2018}, and thus other formation scenarios for \plname\, are possible. 
This is one of the limitations of having access to only NIRSpec/G395H data. Future observations of \plname\, with \jwst\, using different instruments across a wider wavelength range (e.g. NIRISS/SOSS and MIRI/LRS) are required to more robustly constrain its formation mechanism. In particular, additional constraints on the \sotwo\, abundance in this planet's atmosphere in the MIRI band \citep[e.g.][]{powell24} can be used to measure the C/S and O/S ratios, which may help break degeneracies in formation and migration pathways \citep{Turrini2021,crossfield23}.

While \plname's atmospheric metallicity is supersolar, it is surprisingly low given its low inferred mass. Mature planets with masses similar to \plname\, tend to host higher ($\sim$100 $\times$ Solar) metallicity atmospheres, such as those of Uranus and Neptune \citep{Sromovsky2011, Karkoschka2011}. Such high metallicities are ruled out by the transmission spectrum to $\geq$3$\sigma$ unless stellar contamination is included (Figure \ref{fig:spotchisq}). If \plname\, is a precursor of mature $\sim$15M$_{\oplus}$ planets, then its atmosphere is likely to become more enriched in metals as it evolves due to some combination of atmosphere-interior mixing and atmospheric loss.

We estimate the atmospheric loss rate of \plname\, using the simulations of \citet{Caldiroli2022A&A...663A.122C}, assuming photoevaporation as the dominant mechanism. Integrating the XUV spectrum (Section \ref{sec:sed}) yields log(F$_{xuv}$) of 6.68 at the stellar surface, consistent with expectations for a star of this mass and age \citep{Johnstone2021A&A...649A..96J}. From these we find mass loss rates of 0.03, 0.023, 0.015, and 0.011 M$_{\oplus}$ Myr$^{-1}$ for the 8, 10, 15, and 20M$_{\oplus}$ models, giving instantaneous planetary lifetimes $M/\dot{M}$ of 0.27, 0.44, 1.0, and 1.8 Gyr, respectively. If only the H/He component were considered, then these values would be reduced by a factor of 3-5 depending on the mass (Section \ref{sec:interior}). Though the mass loss rates should decrease with time due to declining stellar XUV fluxes \citep{Johnstone2021A&A...649A..96J}, it is apparent that \plname\, will undergo significant mass loss that would reduce its gas content and radius, as well as possibly increase its atmospheric metallicity, thereby potentially transforming into a sub-Neptune on Gyr timescales.

\newedit{Future modeling efforts could expand on this to learn about the atmospheric evolution of sub-Neptunes more generally. One route would be to see if evolution/photoevaporation models designed to fit the \kepler\ distribution \citep[e.g.,][]{Rogers2021} can reproduce HIP\,67522\,b and the similar young planet V1298\,Tau\,b \citep[][also see below]{Barat2024_V1298Taub}. This, in turn, may provide some insight on the wider range of planet-formation channels \citep[e.g.][]{Lee2016,Rogers2024}. Unfortunately, the sample of systems is too small (2) and likely biased, as larger radii objects are easier to identify and are treated as higher-priority \jwst\ targets \citep{Kempton2018}. However, approved \jwst\ programs targeting smaller young planets \citep[e.g.][]{Feinstein2024} could provide the broader sample required.}

\subsection{Comparison with V 1298 Tau b}

\newedit{V1298~Tau\,b is the only planetary system similar to \plname{} with a transmission spectrum. It orbits a cooler K-type T-Tauri star in the older population around Taurus-Augira ($\sim$23$\pm$4 Myr) and has a size comparable to \plname{} ($R_{p}=$10.27$^{+0.58}_{-0.53}$ $R_{\oplus}$) \citep{v1298tau_trevor}. Recent \textit{HST} observations of V1298~Tau\,b revealed a significant H$_2$O detection, indicative of a low mass \citep[$M_{p} < 23 \pm 5 M_{\oplus}$;][]{Barat2024_V1298Taub}, as in the case for \plname. In addition, V1298~Tau\,b was found to have a low atmospheric metallicity that is consistent with solar/sub-solar values. This is in contrast with the high atmospheric metallicity observed in mature planets \citep[e.g., GJ 436 b, HD 97658 b; ][]{Madhusudhan2011ApJ_GJ436b, Guo2020AJ_HD97658b}, but again similar to our findings for \plname. }

\subsection{\plname\ in Context}

\begin{figure*}[ht]
    \centering
    \includegraphics[width=0.495\textwidth]{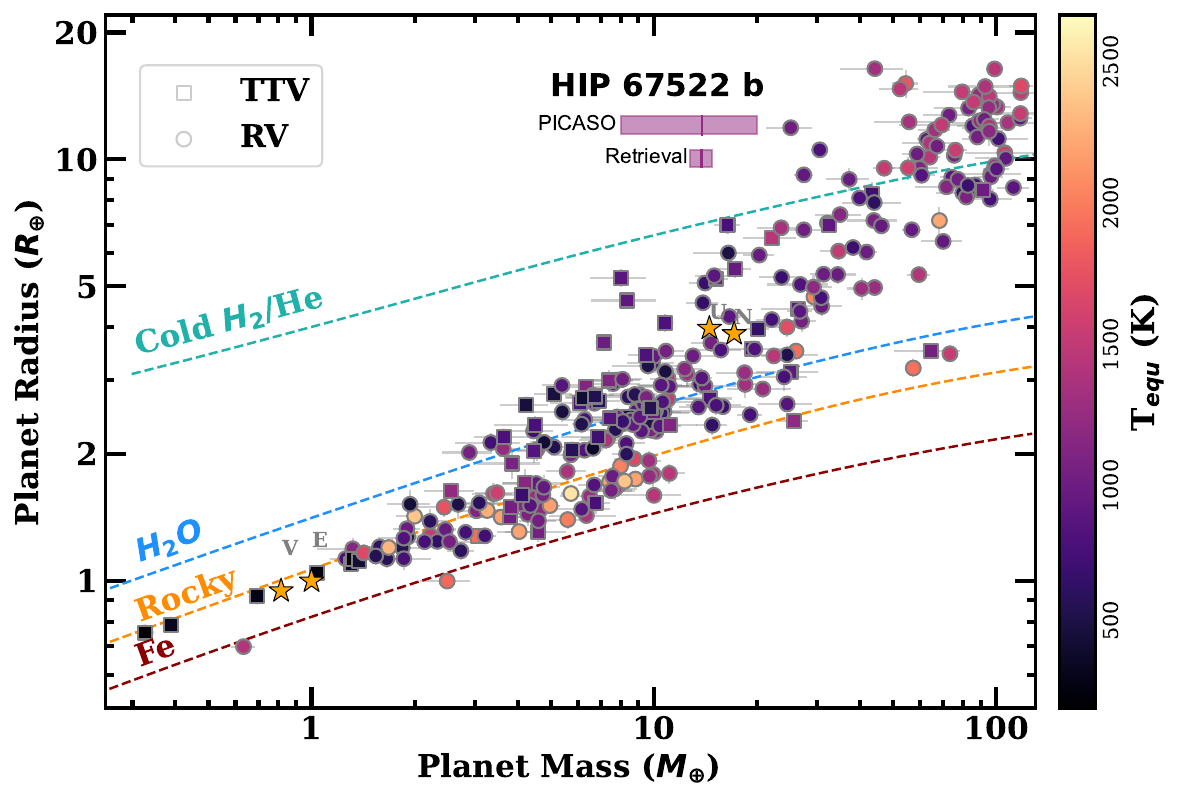}
      \includegraphics[width=0.4950\textwidth]{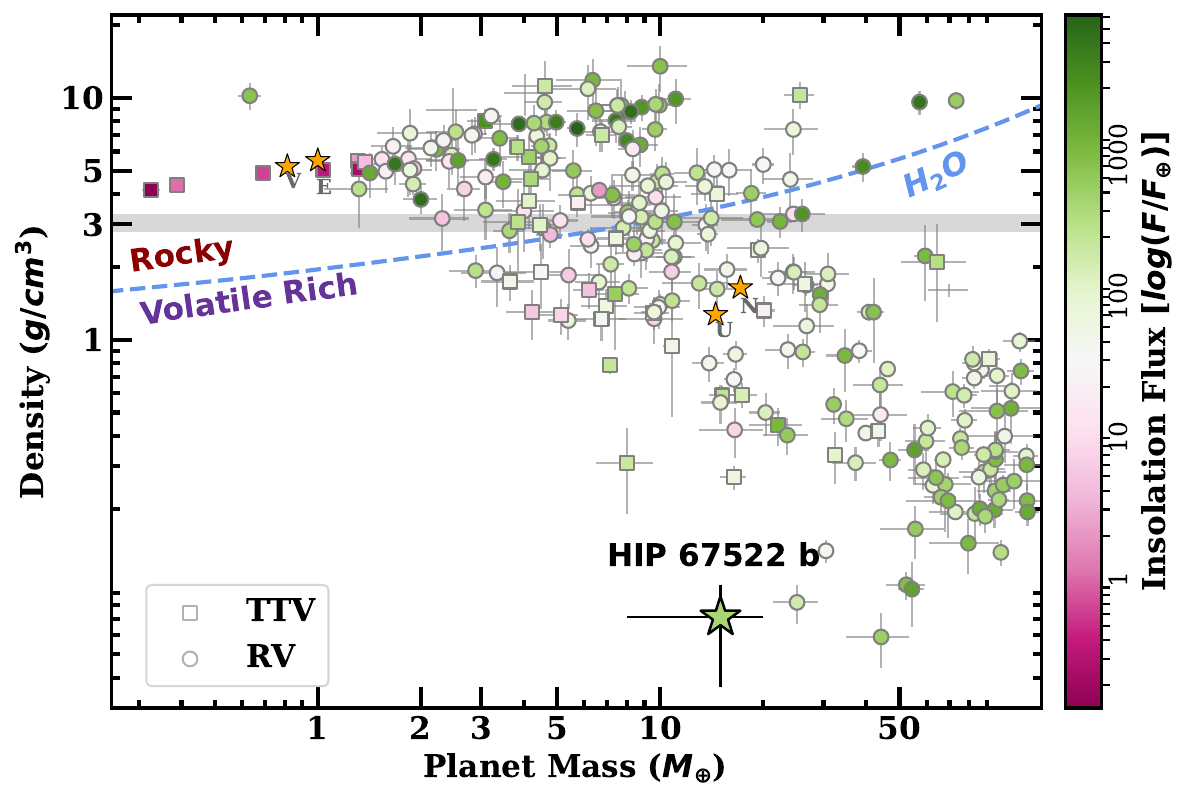}
    \caption{\textit{Left:} Mass-radius plot of planets with reliable mass measurements and relative uncertainties $<$ 25\% for mass, $<$ 8\% for radius, and orbital periods P$<$20 days. Data points are colored by their equilibrium temperature. Square and circle data points represent mass determination from TTVs and RVs, respectively. Composition lines (dashed) indicate iron (red), rocky (orange), water (blue), and cold H2/He (turquoise) \citep{zeng2019growth}.The masses obtained from \texttt{PICASO} and the retrievals are displayed here, with the former offset for clarity. \textit{Right: } Mass-density plot of these same planets, distinguishing between the rocky and volatile-rich planets separated by the composition of water. The data points are colored by the insolation flux. Both of these plots highlight that \plname\, occupies a distinctive position within the M-R diagram, establishing itself as one of the planets with the lowest density ever discovered. Planet properties from \citet{NASAexplanetarchive}. 
    \label{fig:mass_radius_lc}}
\end{figure*}

Our mass constraint for \plname\ makes it one of the lowest-mass planets for its radius and temperature (Figure~\ref{fig:mass_radius_lc}) and debunks the planet's prior classification as a Hot Jupiter. The great majority of mature planets ($>$1 Gyr) with radius $\sim$10R$_{\oplus}$ have masses $>60M_\oplus$ (Figure~\ref{fig:mass_radius_lc}), and thus the mass-radius relations drawn from mature systems is clearly not valid at this age. While more population-level data are needed to see if \plname\, (and V1298 Tau b) is unique, its existence lends significant credence to the expectation that large young planets possess much lower density than their older counterparts. 

The dearth of similarly low density, large, mature planets further suggests that \plname\, will likely undergo significant mass loss, and also that the hot Jupiter anomalous heating may not be (as) active on highly irradiated but lower mass planets as on gas giants. In other words, \plname\, will likely thermally contract significantly with time (Figure \ref{fig:radius_evolution}). A coupled atmospheric loss and thermal evolution model will be needed to better assess \plname's future evolutionary path. 

\plname\, now joins a growing number of exoplanets that have had their atmospheres characterized by \jwst. An emerging theme among the gas-rich planets is the presence of SO$_2$ \citep{rustamkulov2023early,Dyrek2024}, signifying the universality of sulfur photochemistry and its connection to atmospheric metallicity \citep{Tsai2023}. In addition, the solar-to-subsolar C/O of \plname\, also appears to be common \citep{Alderson2023,August2023,Radica2023,Bell2023,Xue2024,Fu2024}, suggesting that close-in orbiting, gas-rich planets may share an origin. Upcoming observations will shed light on whether these findings are a part of a trend or if there is greater diversity in the exoplanet population. 

\section{Summary \& Conclusions} \label{sec:summary}

We observed one simultaneous optical and NIR transit of the 17~Myr old gas giant \plname\ using \jwst/NIRSpec and SOAR. These observations enabled us to construct the planet's transmission spectrum across the wavelength range of 0.5-5.0 $\mu$m. Using three semi-independent reduction and fully independent transit fitting pipelines, we obtained three \jwst\, transmission spectra that were consistent with each other. Our resulting spectrum revealed absorption by \newedit{\water, \cotwo, and CO, with tentative hints of \sotwo\ and H$_2$S.} 

We compared our nominal spectrum to four different atmospheric model grids: (1) equilibrium chemistry models with clear atmospheres; (2) equilibrium chemistry models with simple gray clouds; (3) photochemical models with clear atmospheres; and (4) equilibrium chemistry models with clear atmospheres that include the impact of starspots. For atmospheric model grids (1), (2), and (4), we explored planet masses between 8 and 50M$_{\oplus}$, metallicities between 1 and 100 $\times$ solar, and C/O between 0.25 and 2 $\times$ solar, while for grid (3) we fixed the mass to 15M$_{\oplus}$ and ignored the C/O = 2 $\times$ solar case, ultimately yielding a total of 470 models. The best-fit models from each grid possessed chi-squared per data point $\chi^{2}/N$ values $\leq$1.1. \newedit{We also ran free Bayesian atmospheric retrievals to complement these forward models and further quantify the constraints on the planet mass and atmospheric parameters.} In addition, we generated internal structure models to assess the bulk metallicity and thermal structure of \plname\ and evaluate its future evolution. From these efforts, we can draw the following conclusions: 

\begin{itemize}
  \item Despite a radius of $\sim$10R$_{\oplus}$, the mass of \plname\ is below 20M$_{\oplus}$, \newedit{with a constraint of $14\pm1_{\oplus}$ from the retrieval. This} is significantly lower than the upper limit placed by radial velocity measurements \citep[$<$5$M_J$;][]{Rizzuto2020}. As such, \plname\, is a very young (sub-) Neptune masquerading as a hot Jupiter. 
  \item The atmospheric metallicity of \plname\ is likely in the range of 3-10 $\times$ Solar as suggested by the strength of the SO$_2$ feature. This is considerably lower than its inferred bulk metallicity ($Z_p$ $\sim$ 0.7, or $\sim$200 $\times$ solar if the metals were well-mixed within the planet) and the atmospheric metallicities of Uranus and Neptune \citep[C/H $\sim$ 100 $\times$ Solar;][]{Sromovsky2011,Karkoschka2011}, which likely have comparable masses as \plname. This suggests that the atmosphere could be enriched in metals over time due to atmosphere-interior mixing and/or atmospheric loss. We find that the latter process is likely to significantly impact the planet's mass and radius over Gyr timescales.
  \item The atmospheric C/O of \plname\, is solar-to-subsolar, similar to several other gas-rich planets recently observed by JWST. However, given the complexities in connecting disk and atmospheric composition, more data are needed to make clear statements about the planet's formation process and location within the disk.
  \item Stellar surface inhomogeneities have a negligible impact on our interpretation of the transmission spectrum. This is due to both the strength and type of observed features, which are hard or impossible to reproduce with spots (e.g., \cotwo\ features cannot be explained with reasonable spot temperatures). \starname\ also likely has a low spot fraction compared to late-type dwarfs of almost any age. 
  \item The atmosphere is likely free of significant cloud cover. While we can fit the spectrum with a cloud deck, particularly for low mass models, the required clouds are located at unexpectedly low pressures. Photochemical hazes are more probable in that sense, but are disfavored due to the small transit depth seen in the SDSS $g'$ data ruling out a significant optical scattering slope. \newedit{The retrieval analysis similarly found no evidence for clouds.}
\end{itemize}

The large amplitude spectral features in the \jwst\, transmission spectrum of \plname\
exemplify how young planets are excellent targets for transmission spectroscopy. This is especially important for young planets in the mass range of sub-Neptunes, whose mature ($> 1$~Gyr) counterparts are the most abundant type of planet in the Galaxy, but which often possess featureless transmission spectra due to high mean molecular weight atmospheres and/or absorption by high altitude clouds and hazes \citep{knutson14, kreidberg14, guo20, libbyroberts20, mikalevans21, brande22, kreidberg22, mikalevans23, roy23, brande24}.

Our results also provide strong evidence that masses of young planets can be measured using transmission spectroscopy. While \citet{batalha2017challenges} demonstrated the degeneracies in transmission spectra analysis, highlighting the challenges in accurately determining a planet's mass and consequently its composition, that study focused on the specific case of planets near the rocky-gaseous boundary ($\simeq1.5R_\oplus$) where features are weaker and higher precision masses are required to distinguish between the two scenarios. The mass of \plname\ is robust to changes in the composition, and while changes in cloud-cover do reduce the mass, it still yields far more precise constraints than were previously possible using the radial velocity method. Furthermore, modest mass constraints are still enormously useful for young systems like \plname. Given the challenges in measuring masses of young planets in the presence of stellar noise \citep[e.g.][]{blunt2023overfitting}, this may prove to be the most effective technique for mass determinations of extremely young systems for the foreseeable future. 

Additional transmission spectroscopy of \plname\ would be invaluable for setting tighter limits on its atmospheric composition and mass. Since the strength of the features are large, \plname\ is accessible to ground-based transmission spectroscopy in the optical to look for signs of stellar activity and aerosols, while high-precision data with \jwst/NIRISS and/or MIRI would provide access to a more diverse set of molecular and atomic features. Such future observational efforts would provide much-needed information about abundance variations predicted by different theories of sub-Neptune formation. 

\begin{deluxetable}{lcccr} 
\tabletypesize{\footnotesize} 
\tablecaption{Observation log \label{tab:obslog}}
\tablewidth{0pt}
\tablehead{
\colhead{Telescope} & 
\colhead{Filter} &
\colhead{Exp time (s)} &
\colhead{Start Date (UT)}}
\startdata
    \tess\, Sector 11 \tablenotemark{a}& \tess\, & 120  & 2019 Apr 22 \\
    \tess\, Sector 38 & \tess\, & 120  & 2021 Apr 28  \\
    \tess\, Sector 64 & \tess\, & 20   & 2023 Apr  6 \\
    \hline
    SOAR\tablenotemark{b} & SDSS g' & 2.50 & 2023 Feb 26\\
    \hline
    \jwst\,\tablenotemark{b} & NIRSpec/G395H & 21,943 & 2023 Feb 26\\ 
\enddata
\tablenotetext{a}{Data is from the discovery paper \citet{Rizzuto2020}}
\tablenotetext{b}{Only a partial transit was observed}
\end{deluxetable}

\begin{deluxetable*}{lr|r|r}
\tabletypesize{\footnotesize} 
\tablecaption{Summary of Transit Light Curve Fitting \label{tab:transitfit}}
\tablewidth{0pt}
\tablehead{
\colhead{Reduction Number} &  \colhead{1 (Section~\ref{subsec:thao})} & \colhead{2 (Section~\ref{subsec:feinstein})} & \colhead{3 (Section~\ref{subsec:mann})} }
\startdata
Reduction & JWST Pipeline + \texttt{ExoTIC-JEDI} & \texttt{ExoTIC-JEDI} & \texttt{ExoTIC-JEDI} \\
Spectral Images & custom bias  & custom trace fitting and &  custom trace fitting and \\
 & group-level destriped &  outlier removal & outlier removal \\
 LC Fitting Routine & custom: MCMC, fit for spots & \texttt{chromatic\_fitting} & custom: MCMC, fit for spots \\
& 4\textsuperscript{th} order polynmial OOT,  x and y & 2\textsuperscript{nd} order polynomial OOT  & 4\textsuperscript{th} order polynmial OOT,  x and y\\ 
Limb-darkening & quadratic: fit $u_1$ and $u_2$ & quadratic: fit $u_1$ and $u_2$ & quadratic: fit $u_1$ and $u_2$  \\
\hline 
\enddata
\end{deluxetable*}

\begin{deluxetable*}{lcccr}
\tabletypesize{\footnotesize} 
\tablecaption{Summary of Best Fit Atmospheric Models $<$ 3 $\sigma$ \label{tab:best_models}}
\tablewidth{0pt}
\tablehead{
\colhead{Model} & 
\colhead{Mass} & 
\colhead{C/O} & 
\colhead{[M/H]} &  
\colhead{$\sigma$} \\
\colhead{} & 
\colhead{($M_{\oplus}$)} & 
\colhead{($\times$Solar)} & 
\colhead{} &
\colhead{} }
\startdata
\multirow{9}{8em}{\textbf{Clear}} & 15 & 0.25 & 1.5 & 1.62\\ 
& 15 & 1.00 & 0.5 & 1.94 \\ 
& 15 & 0.50 & 1.0 & 2.15 \\ 
& 15 & 1.00 & 1.0 & 2.22 \\ 
& 15 & 0.50 & 1.5 & 2.41 \\ 
& 15 & 1.00 & 1.5 & 2.72 \\ 
& 20 & 1.00 & 1.0 & 2.81 \\ 
& 15 & 0.50 & 2.0 & 2.88 \\ 
& 20 & 0.50 & 1.5 & 2.99 \\ 
\hline
\multirow{18}{8em}{\textbf{Cloud}} & 8 & 0.25 & 1.5 & 1.00 \\ 
& 10 & 0.25 & 1.5 & 1.25 \\ 
& 15 & 0.25 & 1.5 & 1.29 \\ 
& 15 & 1.00 & 0.5 & 1.31 \\ 
& 15 & 1.00 & 1.0 & 1.37 \\ 
& 10 & 0.50 & 1.0 & 1.38 \\ 
& 15 & 0.50 & 1.0 & 1.48 \\ 
& 15 & 0.50 & 1.5 & 1.65 \\ 
& 10 & 1.00 & 0.5 & 1.83 \\ 
& 10 & 1.00 & 1.0 & 2.06 \\ 
& 10 & 0.25 & 2.0 & 2.12 \\ 
& 8 & 1.00 & 1.0 & 2.24 \\ 
& 15 & 1.00 & 1.5 & 2.56 \\ 
& 8 & 0.50 & 1.0 & 2.68 \\ 
& 8 & 0.25 & 2.0 & 2.84 \\ 
& 20 & 1.00 & 1.0 & 2.85 \\ 
& 10 & 0.50 & 1.5 & 2.91 \\
& 15 & 0.50 & 2.0 & 2.93 \\
\hline 
\multirow{4}{8em}{\textbf{Photochemical}} & 15 & 0.50 & 1.0 & 1.56 \\
& 15 & 1.00 & 0.5 & 1.67 \\ 
& 15 & 0.50 & 0.5 & 2.74 \\ 
& 15 & 0.25 & 1.0 & 2.88 \\ 
\enddata
\end{deluxetable*}

\begin{deluxetable}{lcccccr}
\tabletypesize{\footnotesize} 
\tablecaption{Summary of Best Fit Spot Models $<$ 3 $\sigma$ \label{tab:best_spot_models}}
\tablewidth{0pt}
\tablehead{
\colhead{Mass} & 
\colhead{C/O} & 
\colhead{[M/H]} &
\colhead{Spot Fraction} & 
\colhead{$T_{\rm{Spot}}$} & 
\colhead{Spot Norm} & 
\colhead{$\sigma$} \\
\colhead{$M_{\oplus}$} & 
\colhead{($\times$Solar)} & 
\colhead{} &
\colhead{(\%)} & 
\colhead{[K]} & 
\colhead{Factor} & 
\colhead{}
}
\startdata 
15 & 0.25 & 1.5 & 0.94 & 5700 & 1.01 & 1.89 \\
15 & 1.0 & 1.0 & 1.00 & 6202 & 1.11 & 1.93 \\
15 & 1.0 & 2.0 & 0.23 & 3099 & 0.85 & 2.00 \\
15 & 1.0 & 0.5 & 0.96 & 6001 & 1.07 & 2.04 \\
15 & 1.0 & 1.5 & 0.85 & 6300 & 1.10 & 2.27 \\
15 & 0.5 & 1.5 & 0.86 & 6201 & 1.10 & 2.34 \\
15 & 0.5 & 1.0 & 0.98 & 5702 & 1.01 & 2.38 \\
15 & 0.5 & 2.0 & 0.18 & 2700 & 0.87 & 2.38 \\
15 & 1.5 & 1.0 & 0.99 & 6300 & 1.12 & 2.42 \\
20 & 1.0 & 1.0 & 0.12 & 2701 & 0.91 & 2.76 \\
20 & 0.5 & 1.5 & 0.25 & 3900 & 0.90 & 2.87 
\enddata
\end{deluxetable}

\begin{deluxetable}{lcr}
\tabletypesize{\footnotesize} 
\tablecaption{\jwst\, Broadband Transit Fit from Pipeline 1  \label{tab:white_lc_transitfit}}
\tablewidth{0pt}
\tablehead{
\colhead{Description} &
\colhead{Parameters} & 
\colhead{Value}}
\startdata
Transit mid-point [BJD$_{TDB}$] & $T_{0}$  & $2,460,002.88080_{-0.00138}^{+0.00132}$\\
Planet-to-star radius ratio & $R_p/R_\star$ & $0.06974_{-0.00024}^{+0.00025}$ \\
Semi-major axis-to-stellar radius &  $a/R_{*}$ & $11.37306_{-0.13633}^{+0.13767}$  \\
Inclination (\deg) &  $inc$ & $89.62017_{-0.28870}^{+0.24340}$ \\ 
Limb darkening & $u_{1}$ & $0.237\pm0.014$ \\ 
Limb darkening & $u_{2}$ & $0.170\pm0.017$ \\
Jitter & \newedit{ln($\sigma$)} & $-8.73679_{-0.01323}^{+0.01311}$ \\ 
\hline
Stellar variability parameters \\ 
\hline
& $a_{1}$ & $0.99787\pm0.00001$ \\ 
& $b_{1}$ & $0.04784_{-0.00077}^{+0.00076}$ \\ 
& $c_{1}$ & $-0.15150_{-0.01267}^{+0.01274}$ \\ 
&$d_{1}$ & $0.69870_{-0.08081}^{+0.08169}$ \\ 
& $e_{1}$ & $-1.92855_{-0.16621}^{+0.16335}$\\ 
\hline
Spot Parameters \\ 
\hline
Amplitude of Spot 1 & $A_{1}$ & $0.00026\pm0.00002$ \\ 
Spot duration of Spot 1 & $\tau_{1}$ & $0.00891_{-0.00098}^{+0.00104}$ \\ 
Mid-transit time of Spot 1 & $T_{sp,1}$ & $60002.32103_{-0.00063}^{+0.00069}$ \\ 
Amplitude of Spot 2 & $A_{2}$ & $0.00031_\pm0.00002$ \\ 
Spot duration of Spot 2 & $\tau_{2}$ & $0.00589_{-0.00034}^{+0.00035}$ \\ 
Mid-transit time of Spot 2 & $T_{sp,2}$ & $60002.36527\pm0.00037$ \\ 
\hline
Systematic parameters \\ 
\hline
& $x_1$ & $0.00105_{-0.00036}^{+0.00035}$\\ 
& $y_1$ & $0.00002_{-0.00057}^{+0.00057}$ \\ 
\enddata
\end{deluxetable}

\software{\texttt{misttborn}, \texttt{ExoTiC-JEDI}, \texttt{Eureka!} \citep{bell2022eureka}, \texttt{transitspectroscopy} \citep{espinoza_nestor_2022_6960924}, CIAO \citep{CIAO06}, XSPEC \citep{Arnaud1996}, \texttt{PICASO} \citep{PICASO}, \texttt{VULCAN} \citep{VULCAN}, \texttt{SciPy} \citep{SCIPY}, \texttt{NumPy} \citep{NUMPY}, \texttt{astropy} \citep{ASTROPY, price2018astropy, Astropy2022}, \texttt{emcee} \citep{Foreman-Mackey2013}, \texttt{corner.py} \citep{foreman2016corner}, \texttt{celerite} \citep{celerite}, \texttt{batman}\citep{Kreidberg2015}, \texttt{LDTK} \citep{2015MNRAS.453.3821P}, \texttt{matplotlib} \citep{hunter2007matplotlib}}

\acknowledgements

P.C.T was supported by NSF Graduate Research Fellowship (DGE-1650116), the Zonta International Amelia Earhart Fellowship, the Jack Kent Cooke Foundation Graduate Scholarship, and the NC Space Grant Graduate Research Fellowship. 

A.W.M. and M.L.W were supported by an NSF CAREER grant (AST-2143763), and A.W.M. received additional support from NASA's exoplanet research program (XRP 80NSSC21K0393). 

M.G.B was supported by the NSF Graduate Research Fellowship (DGE-2040435), the NC Space Grant Graduate Research Fellowship, and the TESS guest observer program (21-TESS21-0016). 

ADF acknowledges support by NASA through the NASA Hubble Fellowship grant HST-HF2-51530 awarded by STScI.

These observations are associated with program JWST-GO-02498. Support for program JWST-GO-02498 was provided by NASA through a grant from the Space Telescope Science Institute, which is operated by the Association of Universities for Research in Astronomy, Inc., under NASA contract NAS 5-03127.

This work will be presented in the Extreme Solar Systems V in March 2024 through funding from the UNC Chapel Hill, Department of Physics and Astronomy Silver Award. This work will be presented in Exoplanets V in June 2024 through funding from the Pierazzo International Travel Grant.

Based on observations obtained at the Southern Astrophysical Research (SOAR) telescope, which is a joint project of the Minist\'{e}rio da Ci\^{e}ncia, Tecnologia e Inova\c{c}\~{o}es (MCTI/LNA) do Brasil, the US National Science Foundation’s NOIRLab, the University of North Carolina at Chapel Hill (UNC), and Michigan State University (MSU)

This paper includes data collected with the TESS mission, obtained from the MAST data archive at the Space Telescope Science Institute (STScI). Funding for the TESS mission is provided by the NASA Explorer Program. STScI is operated by the Association of Universities for Research in Astronomy, Inc., under NASA contract NAS 5–26555.

This work is based (in part) on observations made with the NASA/ESA/CSA James Webb Space Telescope. The data were obtained from the Mikulski Archive for Space Telescopes at the Space Telescope Science Institute, which is operated by the Association of Universities for Research in Astronomy, Inc., under NASA contract NAS 5-03127 for JWST. These observations are associated with program \#2498.

Host star SED modeling was funded by grants supporting HST program 16701 and Chandra GO grant GO2-23002X.

This research has used data obtained from the Chandra Data Archive and software provided by the Chandra X-ray Center (CXC) in the CIAO application package.

This research made use of the CHIANTI atomic database, a collaborative project involving George Mason University, the University of Michigan (USA), University of Cambridge (UK) and NASA Goddard Space Flight Center (USA).

\newedit{The TESS and JWST data presented in this article were obtained from the Mikulski Archive for Space Telescopes (MAST) at the Space Telescope Science Institute. The specific observations analyzed can be accessed via \dataset[doi: 10.17909/rs6g-9x65]{https://doi.org/10.17909/rs6g-9x65}.}

\newedit{This paper employs a list of Chandra datasets, obtained by the Chandra X-ray Observatory, contained in~\dataset[DOI: 10.25574/cdc.262]{https://doi.org/10.25574/cdc.262}.}

\facilities{\jwst\,/NIRSpec/BOTS/G395H, \tess\,,  SOAR/Goodman, CXO (ACIS), HST (STIS)}

\clearpage
\bibliography{bib.bib, additionalbib, shortbib, retrieval_citations}

\begin{thebibliography}{}
\expandafter\ifx\csname natexlab\endcsname\relax\def\natexlab#1{#1}\fi

\bibitem[{{Ag{\'u}ndez} {et~al.}(2014){Ag{\'u}ndez}, {Parmentier}, {Venot},
  {Hersant}, \& {Selsis}}]{Agundez2014}
{Ag{\'u}ndez}, M., {Parmentier}, V., {Venot}, O., {Hersant}, F., \& {Selsis},
  F. 2014, \aap, 564, A73

\bibitem[{{Alderson} {et~al.}(2023){Alderson}, {Wakeford}, {Alam}, {Batalha},
  {Lothringer}, {Adams Redai}, {Barat}, {Brande}, {Damiano}, {Daylan},
  {Espinoza}, {Flagg}, {Goyal}, {Grant}, {Hu}, {Inglis}, {Lee}, {Mikal-Evans},
  {Ramos-Rosado}, {Roy}, {Wallack}, {Batalha}, {Bean}, {Benneke},
  {Berta-Thompson}, {Carter}, {Changeat}, {Col{\'o}n}, {Crossfield},
  {D{\'e}sert}, {Foreman-Mackey}, {Gibson}, {Kreidberg}, {Line},
  {L{\'o}pez-Morales}, {Molaverdikhani}, {Moran}, {Morello}, {Moses},
  {Mukherjee}, {Schlawin}, {Sing}, {Stevenson}, {Taylor}, {Aggarwal}, {Ahrer},
  {Allen}, {Barstow}, {Bell}, {Blecic}, {Casewell}, {Chubb}, {Crouzet},
  {Cubillos}, {Decin}, {Feinstein}, {Fortney}, {Harrington}, {Heng}, {Iro},
  {Kempton}, {Kirk}, {Knutson}, {Krick}, {Leconte}, {Lendl}, {MacDonald},
  {Mancini}, {Mansfield}, {May}, {Mayne}, {Miguel}, {Nikolov}, {Ohno}, {Palle},
  {Parmentier}, {Petit dit de la Roche}, {Piaulet}, {Powell}, {Rackham},
  {Redfield}, {Rogers}, {Rustamkulov}, {Tan}, {Tremblin}, {Tsai}, {Turner}, {de
  Val-Borro}, {Venot}, {Welbanks}, {Wheatley}, \& {Zhang}}]{Alderson2023}
{Alderson}, L., {Wakeford}, H.~R., {Alam}, M.~K., {et~al.} 2023, \nat, 614, 664

\bibitem[{Allard {et~al.}(2023)Allard, Myneni, Blakely, \&
  Guillon}]{allard_temperature_2023}
Allard, N.~F., Myneni, K., Blakely, J.~N., \& Guillon, G. 2023, Astronomy and
  Astrophysics, 674, A171, aDS Bibcode: 2023A\&A...674A.171A

\bibitem[{Allard {et~al.}(2016)Allard, Spiegelman, \&
  Kielkopf}]{allard_k-h2_2016}
Allard, N.~F., Spiegelman, F., \& Kielkopf, J.~F. 2016, Astronomy and
  Astrophysics, 589, A21, aDS Bibcode: 2016A\&A...589A..21A

\bibitem[{{Arnaud}(1996)}]{Arnaud1996}
{Arnaud}, K.~A. 1996, in Astronomical Society of the Pacific Conference Series,
  Vol. 101, Astronomical Data Analysis Software and Systems V, ed. G.~H.
  {Jacoby} \& J.~{Barnes}, 17

\bibitem[{{Astropy Collaboration} {et~al.}(2022){Astropy Collaboration},
  {Price-Whelan}, {Lim}, {Earl}, {Starkman}, {Bradley}, {Shupe}, {Patil},
  {Corrales}, {Brasseur}, {N{\"o}the}, {Donath}, {Tollerud}, {Morris},
  {Ginsburg}, {Vaher}, {Weaver}, {Tocknell}, {Jamieson}, {van Kerkwijk},
  {Robitaille}, {Merry}, {Bachetti}, {G{\"u}nther}, {Aldcroft},
  {Alvarado-Montes}, {Archibald}, {B{\'o}di}, {Bapat}, {Barentsen},
  {Baz{\'a}n}, {Biswas}, {Boquien}, {Burke}, {Cara}, {Cara}, {Conroy},
  {Conseil}, {Craig}, {Cross}, {Cruz}, {D'Eugenio}, {Dencheva}, {Devillepoix},
  {Dietrich}, {Eigenbrot}, {Erben}, {Ferreira}, {Foreman-Mackey}, {Fox},
  {Freij}, {Garg}, {Geda}, {Glattly}, {Gondhalekar}, {Gordon}, {Grant},
  {Greenfield}, {Groener}, {Guest}, {Gurovich}, {Handberg}, {Hart},
  {Hatfield-Dodds}, {Homeier}, {Hosseinzadeh}, {Jenness}, {Jones}, {Joseph},
  {Kalmbach}, {Karamehmetoglu}, {Ka{\l}uszy{\'n}ski}, {Kelley}, {Kern},
  {Kerzendorf}, {Koch}, {Kulumani}, {Lee}, {Ly}, {Ma}, {MacBride}, {Maljaars},
  {Muna}, {Murphy}, {Norman}, {O'Steen}, {Oman}, {Pacifici}, {Pascual},
  {Pascual-Granado}, {Patil}, {Perren}, {Pickering}, {Rastogi}, {Roulston},
  {Ryan}, {Rykoff}, {Sabater}, {Sakurikar}, {Salgado}, {Sanghi}, {Saunders},
  {Savchenko}, {Schwardt}, {Seifert-Eckert}, {Shih}, {Jain}, {Shukla}, {Sick},
  {Simpson}, {Singanamalla}, {Singer}, {Singhal}, {Sinha}, {Sip{\H{o}}cz},
  {Spitler}, {Stansby}, {Streicher}, {{\v{S}}umak}, {Swinbank}, {Taranu},
  {Tewary}, {Tremblay}, {de Val-Borro}, {Van Kooten}, {Vasovi{\'c}}, {Verma},
  {de Miranda Cardoso}, {Williams}, {Wilson}, {Winkel}, {Wood-Vasey}, {Xue},
  {Yoachim}, {Zhang}, {Zonca}, \& {Astropy Project Contributors}}]{Astropy2022}
{Astropy Collaboration}, {Price-Whelan}, A.~M., {Lim}, P.~L., {et~al.} 2022,
  \apj, 935, 167

\bibitem[{{August} {et~al.}(2023){August}, {Bean}, {Zhang}, {Lunine}, {Xue},
  {Line}, \& {Smith}}]{August2023}
{August}, P.~C., {Bean}, J.~L., {Zhang}, M., {et~al.} 2023, \apjl, 953, L24

\bibitem[{Azzam {et~al.}(2016)Azzam, Tennyson, Yurchenko, \&
  Naumenko}]{azzam_exomol_2016}
Azzam, A. A.~A., Tennyson, J., Yurchenko, S.~N., \& Naumenko, O.~V. 2016,
  Monthly Notices of the Royal Astronomical Society, 460, 4063

\bibitem[{{Barat} {et~al.}(2024){Barat}, {D{\'e}sert}, {Vazan}, {Baeyens},
  {Line}, {Fortney}, {David}, {Livingston}, {Jacobs}, {Panwar}, {Shivkumar},
  {Todorov}, {Pino}, {Mraz}, \& {Petigura}}]{Barat2024_V1298Taub}
{Barat}, S., {D{\'e}sert}, J.-M., {Vazan}, A., {et~al.} 2024, Nature Astronomy,
  arXiv:2312.16924

\bibitem[{Barber {et~al.}(2024)Barber, Thao, Mann, Vanderburg, Mori,
  Livingston, Fukui, Narita, Kraus, Tofflemire, {et~al.}}]{barber2024tess}
Barber, M.~G., Thao, P.~C., Mann, A.~W., {et~al.} 2024, arXiv preprint
  arXiv:2407.04763

\bibitem[{{Barclay} {et~al.}(2021){Barclay}, {Kostov}, {Col{\'o}n}, {Quintana},
  {Schlieder}, {Louie}, {Gilbert}, \& {Mullally}}]{Barclay2021}
{Barclay}, T., {Kostov}, V.~B., {Col{\'o}n}, K.~D., {et~al.} 2021, \aj, 162,
  300

\bibitem[{Batalha {et~al.}(2022)Batalha, Freedman, Gharib-Nezhad, \&
  Lupu}]{natasha_batalha_2022_6928501}
Batalha, N., Freedman, R., Gharib-Nezhad, E., \& Lupu, R. 2022, Resampled
  Opacity Database for PICASO, doi:10.5281/zenodo.6928501

\bibitem[{Batalha {et~al.}(2020)Batalha, Freedman, Lupu, \&
  Marley}]{natasha_batalha_2020_3759675}
Batalha, N., Freedman, R., Lupu, R., \& Marley, M. 2020, Resampled Opacity
  Database for PICASO v2, doi:10.5281/zenodo.3759675

\bibitem[{Batalha {et~al.}(2017)Batalha, Kempton, \&
  Mbarek}]{batalha2017challenges}
Batalha, N.~E., Kempton, E. M.-R., \& Mbarek, R. 2017, The Astrophysical
  Journal Letters, 836, L5

\bibitem[{{Batalha} {et~al.}(2019){Batalha}, {Marley}, {Lewis}, \&
  {Fortney}}]{Batalha2019}
{Batalha}, N.~E., {Marley}, M.~S., {Lewis}, N.~K., \& {Fortney}, J.~J. 2019,
  \apj, 878, 70

\bibitem[{Batalha {et~al.}(2019)Batalha, Marley, Lewis, \& Fortney}]{PICASO}
Batalha, N.~E., Marley, M.~S., Lewis, N.~K., \& Fortney, J.~J. 2019, The
  Astrophysical Journal, 878, 70

\bibitem[{Bell {et~al.}(2022)Bell, Ahrer, Brande, Carter, Feinstein, Caloca,
  Mansfield, Zieba, Piaulet, Benneke, {et~al.}}]{bell2022eureka}
Bell, T.~J., Ahrer, E.-M., Brande, J., {et~al.} 2022, arXiv preprint
  arXiv:2207.03585

\bibitem[{{Bell} {et~al.}(2023){Bell}, {Welbanks}, {Schlawin}, {Line},
  {Fortney}, {Greene}, {Ohno}, {Parmentier}, {Rauscher}, {Beatty}, {Mukherjee},
  {Wiser}, {Boyer}, {Rieke}, \& {Stansberry}}]{Bell2023}
{Bell}, T.~J., {Welbanks}, L., {Schlawin}, E., {et~al.} 2023, \nat, 623, 709

\bibitem[{{Blunt} {et~al.}(2023){Blunt}, {Carvalho}, {David}, {Beichman},
  {Zink}, {Gaidos}, {Behmard}, {Bouma}, {Cody}, {Dai}, {Foreman-Mackey},
  {Grunblatt}, {Howard}, {Kosiarek}, {Knutson}, {Rubenzahl}, {Beard},
  {Chontos}, {Giacalone}, {Hirano}, {Johnson}, {Lubin}, {Akana Murphy},
  {Petigura}, {Van Zandt}, \& {Weiss}}]{blunt2023overfitting}
{Blunt}, S., {Carvalho}, A., {David}, T.~J., {et~al.} 2023, \aj, 166, 62

\bibitem[{Bouma {et~al.}(2020)Bouma, Hartman, Brahm, Evans, Collins, Zhou,
  Sarkis, Quinn, De~Leon, Livingston, {et~al.}}]{bouma2020cluster}
Bouma, L., Hartman, J., Brahm, R., {et~al.} 2020, The Astronomical Journal,
  160, 239

\bibitem[{{Bouma} {et~al.}(2022){Bouma}, {Kerr}, {Curtis}, {Isaacson},
  {Hillenbrand}, {Howard}, {Kraus}, {Bieryla}, {Latham}, {Petigura}, \&
  {Huber}}]{Bouma2022}
{Bouma}, L.~G., {Kerr}, R., {Curtis}, J.~L., {et~al.} 2022, \aj, 164, 215

\bibitem[{{Brande} {et~al.}(2022){Brande}, {Crossfield}, {Kreidberg},
  {Oklop{\v{c}}i{\'c}}, {Polanski}, {Barman}, {Benneke}, {Christiansen},
  {Dragomir}, {Foreman-Mackey}, {Fortney}, {Greene}, {Howard}, {Knutson},
  {Lothringer}, {Mikal-Evans}, \& {Morley}}]{brande22}
{Brande}, J., {Crossfield}, I. J.~M., {Kreidberg}, L., {et~al.} 2022, \aj, 164,
  197

\bibitem[{{Brande} {et~al.}(2024){Brande}, {Crossfield}, {Kreidberg}, {Morley},
  {Barman}, {Benneke}, {Christiansen}, {Dragomir}, {Fortney}, {Greene},
  {Hardegree-Ullman}, {Howard}, {Knutson}, {Lothringer}, \&
  {Mikal-Evans}}]{brande24}
---. 2024, \apjl, 961, L23

\bibitem[{Buchner {et~al.}(2014)Buchner, Georgakakis, Nandra, Hsu, Rangel,
  Brightman, Merloni, Salvato, Donley, \& Kocevski}]{buchner_x-ray_2014}
Buchner, J., Georgakakis, A., Nandra, K., {et~al.} 2014, Astronomy \&
  Astrophysics, 564, A125, publisher: EDP Sciences

\bibitem[{{Caldiroli} {et~al.}(2022){Caldiroli}, {Haardt}, {Gallo}, {Spinelli},
  {Malsky}, \& {Rauscher}}]{Caldiroli2022A&A...663A.122C}
{Caldiroli}, A., {Haardt}, F., {Gallo}, E., {et~al.} 2022, \aap, 663, A122

\bibitem[{{Chabrier} {et~al.}(2019){Chabrier}, {Mazevet}, \&
  {Soubiran}}]{Chabrier2019}
{Chabrier}, G., {Mazevet}, S., \& {Soubiran}, F. 2019, \apj, 872, 51

\bibitem[{{Clemens} {et~al.}(2004){Clemens}, {Crain}, \&
  {Anderson}}]{Clemens_2004}
{Clemens}, J.~C., {Crain}, J.~A., \& {Anderson}, R. 2004, in Society of
  Photo-Optical Instrumentation Engineers (SPIE) Conference Series, Vol. 5492,
  Ground-based Instrumentation for Astronomy, ed. A.~F.~M. {Moorwood} \&
  M.~{Iye}, 331--340

\bibitem[{Coles {et~al.}(2019)Coles, Yurchenko, \&
  Tennyson}]{coles_exomol_2019}
Coles, P.~A., Yurchenko, S.~N., \& Tennyson, J. 2019, Monthly Notices of the
  Royal Astronomical Society, 490, 4638

\bibitem[{{Cridland} {et~al.}(2019){Cridland}, {van Dishoeck}, {Alessi}, \&
  {Pudritz}}]{cridland2019}
{Cridland}, A.~J., {van Dishoeck}, E.~F., {Alessi}, M., \& {Pudritz}, R.~E.
  2019, \aap, 632, A63

\bibitem[{{Crossfield}(2023)}]{crossfield23}
{Crossfield}, I. J.~M. 2023, arXiv e-prints, arXiv:2303.17622

\bibitem[{{Dai} {et~al.}(2017){Dai}, {Winn}, {Yu}, \& {Albrecht}}]{Dai2017}
{Dai}, F., {Winn}, J.~N., {Yu}, L., \& {Albrecht}, S. 2017, \aj, 153, 40

\bibitem[{Dattilo {et~al.}(2023)Dattilo, Batalha, \& Bryson}]{Dattilo2023}
Dattilo, A., Batalha, N.~M., \& Bryson, S. 2023, The Astronomical Journal, 166,
  122

\bibitem[{Davenport {et~al.}(2019)Davenport, Covey, Clarke, Boeck, Cornet, \&
  Hawley}]{davenport2019evolution}
Davenport, J.~R., Covey, K.~R., Clarke, R.~W., {et~al.} 2019, The Astrophysical
  Journal, 871, 241

\bibitem[{{David} {et~al.}(2019){David}, {Cody}, {Hedges}, {Mamajek},
  {Hillenbrand}, {Ciardi}, {Beichman}, {Petigura}, {Fulton}, {Isaacson},
  {Howard}, {Gagn{\'e}}, {Saunders}, {Rebull}, {Stauffer}, {Vasisht}, \&
  {Hinkley}}]{v1298tau_trevor}
{David}, T.~J., {Cody}, A.~M., {Hedges}, C.~L., {et~al.} 2019, \aj, 158, 79

\bibitem[{{de Wit} \& {Seager}(2013)}]{dewit13}
{de Wit}, J., \& {Seager}, S. 2013, Science, 342, 1473

\bibitem[{de~Wit \& Seager(2013)}]{de_wit_constraining_2013}
de~Wit, J., \& Seager, S. 2013, Science, 342, 1473, aDS Bibcode:
  2013Sci...342.1473D

\bibitem[{{Del Zanna} {et~al.}(2021){Del Zanna}, {Dere}, {Young}, \&
  {Landi}}]{DelZanna:2021}
{Del Zanna}, G., {Dere}, K.~P., {Young}, P.~R., \& {Landi}, E. 2021, \apj, 909,
  38

\bibitem[{{Dere} {et~al.}(1999){Dere}, {Brueckner}, {Howard}, {Michels}, \&
  {Delaboudiniere}}]{Dere1999}
{Dere}, K.~P., {Brueckner}, G.~E., {Howard}, R.~A., {Michels}, D.~J., \&
  {Delaboudiniere}, J.~P. 1999, \apj, 516, 465

\bibitem[{{Duvvuri} {et~al.}(2023){Duvvuri}, {Cauley}, {Aguirre}, {Kilgard},
  {France}, {Berta-Thompson}, \& {Pineda}}]{Duvvuri:2023}
{Duvvuri}, G.~M., {Cauley}, P.~W., {Aguirre}, F.~C., {et~al.} 2023, \aj, 166,
  196

\bibitem[{{Duvvuri} {et~al.}(2021){Duvvuri}, {Pineda}, {Berta-Thompson},
  {Brown}, {France}, {Kowalski}, {Redfield}, {Tilipman}, {Vieytes}, {Wilson},
  {Youngblood}, {Froning}, {Linsky}, {Parke Loyd}, {Mauas}, {Miguel}, {Newton},
  {Rugheimer}, \& {Christian Schneider}}]{Duvvuri:2021}
{Duvvuri}, G.~M., {Pineda}, J.~S., {Berta-Thompson}, Z.~K., {et~al.} 2021,
  \apj, 913, 40

\bibitem[{{Dyrek} {et~al.}(2024){Dyrek}, {Min}, {Decin}, {Bouwman}, {Crouzet},
  {Molli{\`e}re}, {Lagage}, {Konings}, {Tremblin}, {G{\"u}del}, {Pye},
  {Waters}, {Henning}, {Vandenbussche}, {Ardevol Martinez}, {Argyriou},
  {Ducrot}, {Heinke}, {van Looveren}, {Absil}, {Barrado}, {Baudoz},
  {Boccaletti}, {Cossou}, {Coulais}, {Edwards}, {Gastaud}, {Glasse}, {Glauser},
  {Greene}, {Kendrew}, {Krause}, {Lahuis}, {Mueller}, {Olofsson}, {Patapis},
  {Rouan}, {Royer}, {Scheithauer}, {Waldmann}, {Whiteford}, {Colina}, {van
  Dishoeck}, {{\"O}stlin}, {Ray}, \& {Wright}}]{Dyrek2024}
{Dyrek}, A., {Min}, M., {Decin}, L., {et~al.} 2024, \nat, 625, 51

\bibitem[{{Eistrup} {et~al.}(2018){Eistrup}, {Walsh}, \& {van
  Dishoeck}}]{Eistrup2018}
{Eistrup}, C., {Walsh}, C., \& {van Dishoeck}, E.~F. 2018, \aap, 613, A14

\bibitem[{Espinoza(2022)}]{espinoza_nestor_2022_6960924}
Espinoza, N. 2022, TransitSpectroscopy, doi:10.5281/zenodo.6960924

\bibitem[{{Espinoza} {et~al.}(2017){Espinoza}, {Fortney}, {Miguel},
  {Thorngren}, \& {Murray-Clay}}]{espinoza2017}
{Espinoza}, N., {Fortney}, J.~J., {Miguel}, Y., {Thorngren}, D., \&
  {Murray-Clay}, R. 2017, \apjl, 838, L9

\bibitem[{{Espinoza} {et~al.}(2023){Espinoza}, {{\'U}beda}, {Birkmann},
  {Ferruit}, {Valenti}, {Sing}, {Rustamkulov}, {Regan}, {Kendrew}, {Sabbi},
  {Schlawin}, {Beatty}, {Albert}, {Greene}, {Nikolov}, {Karakla}, {Keyes},
  {Alves de Oliveira}, {B{\"o}ker}, {Pena-Guerrero}, {Giardino}, {Kumari},
  {Manjavacas}, {Proffitt}, \& {Rawle}}]{Espinoza2023}
{Espinoza}, N., {{\'U}beda}, L., {Birkmann}, S.~M., {et~al.} 2023, \pasp, 135,
  018002

\bibitem[{{Feinstein} {et~al.}(2024){Feinstein}, {Welbanks}, {Ahrer},
  {Alderson}, {Barat}, {Brande}, {Crossfield}, {Desert}, {Duvvuri}, {Espinoza},
  {France}, {Gao}, {Guzman Caloca}, {Levine}, {Livingston}, {Lunine}, {Luque},
  {Mann}, {Mukherjee}, {Murray}, {Owen}, {Rackham}, {Radica}, {Rockcliffe},
  {Rogers}, {Seager}, {Seligman}, {Shapiro}, {Thao}, \&
  {Vissapragada}}]{Feinstein2024}
{Feinstein}, A., {Welbanks}, L., {Ahrer}, E.-M., {et~al.} 2024, {KRONOS: Keys
  to Revealing the Origin and Nature Of sub-neptune Systems}, JWST Proposal.
  Cycle 3, ID. \#5959

\bibitem[{Feroz {et~al.}(2009)Feroz, Hobson, \& Bridges}]{feroz2009multinest}
Feroz, F., Hobson, M., \& Bridges, M. 2009, Monthly Notices of the Royal
  Astronomical Society, 398, 1601

\bibitem[{Foreman-Mackey(2016)}]{foreman2016corner}
Foreman-Mackey, D. 2016, The Journal of Open Source Software, 1

\bibitem[{{Foreman-Mackey} {et~al.}(2017){Foreman-Mackey}, {Agol},
  {Ambikasaran}, \& {Angus}}]{celerite}
{Foreman-Mackey}, D., {Agol}, E., {Ambikasaran}, S., \& {Angus}, R. 2017, \aj,
  154, 220

\bibitem[{{Foreman-Mackey} {et~al.}(2013){Foreman-Mackey}, {Hogg}, {Lang}, \&
  {Goodman}}]{Foreman-Mackey2013}
{Foreman-Mackey}, D., {Hogg}, D.~W., {Lang}, D., \& {Goodman}, J. 2013, \pasp,
  125, 306

\bibitem[{{Fortney} {et~al.}(2007){Fortney}, {Marley}, \&
  {Barnes}}]{Fortney2007}
{Fortney}, J.~J., {Marley}, M.~S., \& {Barnes}, J.~W. 2007, \apj, 659, 1661

\bibitem[{{Fruscione} {et~al.}(2006){Fruscione}, {McDowell}, {Allen},
  {Brickhouse}, {Burke}, {Davis}, {Durham}, {Elvis}, {Galle}, {Harris},
  {Huenemoerder}, {Houck}, {Ishibashi}, {Karovska}, {Nicastro}, {Noble},
  {Nowak}, {Primini}, {Siemiginowska}, {Smith}, \& {Wise}}]{CIAO06}
{Fruscione}, A., {McDowell}, J.~C., {Allen}, G.~E., {et~al.} 2006, in Society
  of Photo-Optical Instrumentation Engineers (SPIE) Conference Series, Vol.
  6270, Observatory Operations: Strategies, Processes, and Systems, ed. D.~R.
  {Silva} \& R.~E. {Doxsey}, 62701V

\bibitem[{Fu(2024)}]{Fu2024}
Fu, G. 2024

\bibitem[{{Gao} \& {Zhang}(2020)}]{Gao_hazes}
{Gao}, P., \& {Zhang}, X. 2020, \apj, 890, 93

\bibitem[{{Gao} {et~al.}(2020){Gao}, {Thorngren}, {Lee}, {Fortney}, {Morley},
  {Wakeford}, {Powell}, {Stevenson}, \& {Zhang}}]{gao2020}
{Gao}, P., {Thorngren}, D.~P., {Lee}, E. K.~H., {et~al.} 2020, Nature
  Astronomy, 4, 951

\bibitem[{{Ginzburg} {et~al.}(2018){Ginzburg}, {Schlichting}, \&
  {Sari}}]{Ginzburg2018}
{Ginzburg}, S., {Schlichting}, H.~E., \& {Sari}, R. 2018, \mnras, 476, 759

\bibitem[{Grant {et~al.}(2023)Grant, Lothringer, Wakeford, Alam, Alderson,
  Bean, Benneke, D{\'e}sert, Daylan, Flagg, {et~al.}}]{grant2023detection}
Grant, D., Lothringer, J.~D., Wakeford, H.~R., {et~al.} 2023, The Astrophysical
  Journal Letters, 949, L15

\bibitem[{{Guo} {et~al.}(2020{\natexlab{a}}){Guo}, {Crossfield}, {Dragomir},
  {Kosiarek}, {Lothringer}, {Mikal-Evans}, {Rosenthal}, {Benneke}, {Knutson},
  {Dalba}, {Kempton}, {Henry}, {McCullough}, {Barman}, {Blunt}, {Chontos},
  {Fortney}, {Fulton}, {Hirsch}, {Howard}, {Isaacson}, {Matthews}, {Mocnik},
  {Morley}, {Petigura}, \& {Weiss}}]{Guo2020AJ_HD97658b}
{Guo}, X., {Crossfield}, I. J.~M., {Dragomir}, D., {et~al.} 2020{\natexlab{a}},
  \aj, 159, 239

\bibitem[{{Guo} {et~al.}(2020{\natexlab{b}}){Guo}, {Crossfield}, {Dragomir},
  {Kosiarek}, {Lothringer}, {Mikal-Evans}, {Rosenthal}, {Benneke}, {Knutson},
  {Dalba}, {Kempton}, {Henry}, {McCullough}, {Barman}, {Blunt}, {Chontos},
  {Fortney}, {Fulton}, {Hirsch}, {Howard}, {Isaacson}, {Matthews}, {Mocnik},
  {Morley}, {Petigura}, \& {Weiss}}]{guo20}
---. 2020{\natexlab{b}}, \aj, 159, 239

\bibitem[{Hargreaves {et~al.}(2020)Hargreaves, Gordon, Rey, Nikitin, Tyuterev,
  Kochanov, \& Rothman}]{hargreaves_accurate_2020}
Hargreaves, R.~J., Gordon, I.~E., Rey, M., {et~al.} 2020, The Astrophysical
  Journal Supplement Series, 247, 55, publisher: The American Astronomical
  Society

\bibitem[{Harris {et~al.}(2020)Harris, Millman, Van Der~Walt, Gommers,
  Virtanen, Cournapeau, Wieser, Taylor, Berg, Smith, {et~al.}}]{NUMPY}
Harris, C.~R., Millman, K.~J., Van Der~Walt, S.~J., {et~al.} 2020, Nature, 585,
  357

\bibitem[{Heitzmann {et~al.}(2021)Heitzmann, Zhou, Quinn, Marsden, Wright,
  Petit, Vanderburg, Bouma, Mann, \& Rizzuto}]{heitzmann2021obliquity}
Heitzmann, A., Zhou, G., Quinn, S.~N., {et~al.} 2021, The Astrophysical Journal
  Letters, 922, L1

\bibitem[{{Horne}(1986)}]{horne86}
{Horne}, K. 1986, \pasp, 98, 609

\bibitem[{Huang {et~al.}(2012)Huang, Schwenke, Tashkun, \&
  Lee}]{huang_isotopic-independent_2012}
Huang, X., Schwenke, D.~W., Tashkun, S.~A., \& Lee, T.~J. 2012, The Journal of
  Chemical Physics, 136, 124311

\bibitem[{Hunter(2007)}]{hunter2007matplotlib}
Hunter, J.~D. 2007, Computing in science \& engineering, 9, 90

\bibitem[{Jenkins {et~al.}(2016)Jenkins, Twicken, McCauliff, Campbell,
  Sanderfer, Lung, Mansouri-Samani, Girouard, Tenenbaum, Klaus,
  {et~al.}}]{jenkins2016tess}
Jenkins, J.~M., Twicken, J.~D., McCauliff, S., {et~al.} 2016, in Software and
  Cyberinfrastructure for Astronomy IV, Vol. 9913, SPIE, 1232--1251

\bibitem[{{Johnson} {et~al.}(2018){Johnson}, {Dai}, {Justesen}, {Gandolfi},
  {Hatzes}, {Nowak}, {Endl}, {Cochran}, {Hidalgo}, {Watanabe}, {Parviainen},
  {Hirano}, {Villanueva}, {Prieto-Arranz}, {Narita}, {Palle}, {Guenther},
  {Barrag{\'a}n}, {Trifonov}, {Niraula}, {MacQueen}, {Cabrera}, {Csizmadia},
  {Eigm{\"u}ller}, {Grziwa}, {Korth}, {P{\"a}tzold}, {Smith}, {Albrecht},
  {Alonso}, {Deeg}, {Erikson}, {Esposito}, {Fridlund}, {Fukui}, {Kusakabe},
  {Kuzuhara}, {Livingston}, {Monta{\~n}es Rodriguez}, {Nespral}, {Persson},
  {Purismo}, {Raimundo}, {Rauer}, {Ribas}, {Tamura}, {Van Eylen}, \&
  {Winn}}]{MISTTBORN}
{Johnson}, M.~C., {Dai}, F., {Justesen}, A.~B., {et~al.} 2018, \mnras, 481, 596

\bibitem[{{Johnstone} {et~al.}(2021){Johnstone}, {Bartel}, \&
  {G{\"u}del}}]{Johnstone2021A&A...649A..96J}
{Johnstone}, C.~P., {Bartel}, M., \& {G{\"u}del}, M. 2021, \aap, 649, A96

\bibitem[{{Karkoschka} \& {Tomasko}(2011)}]{Karkoschka2011}
{Karkoschka}, E., \& {Tomasko}, M.~G. 2011, \icarus, 211, 780

\bibitem[{Karman {et~al.}(2019)Karman, Gordon, van~der Avoird, Baranov, Boulet,
  Drouin, Groenenboom, Gustafsson, Hartmann, Kurucz, Rothman, Sun, Sung,
  Thalman, Tran, Wishnow, Wordsworth, Vigasin, Volkamer, \& van~der
  Zande}]{karman_update_2019}
Karman, T., Gordon, I.~E., van~der Avoird, A., {et~al.} 2019, Icarus, 328, 160

\bibitem[{{Kempton} {et~al.}(2018){Kempton}, {Bean}, {Louie}, {Deming}, {Koll},
  {Mansfield}, {Christiansen}, {L{\'o}pez-Morales}, {Swain}, {Zellem},
  {Ballard}, {Barclay}, {Barstow}, {Batalha}, {Beatty}, {Berta-Thompson},
  {Birkby}, {Buchhave}, {Charbonneau}, {Cowan}, {Crossfield}, {de Val-Borro},
  {Doyon}, {Dragomir}, {Gaidos}, {Heng}, {Hu}, {Kane}, {Kreidberg}, {Mallonn},
  {Morley}, {Narita}, {Nascimbeni}, {Pall{\'e}}, {Quintana}, {Rauscher},
  {Seager}, {Shkolnik}, {Sing}, {Sozzetti}, {Stassun}, {Valenti}, \& {von
  Essen}}]{Kempton2018}
{Kempton}, E. M.~R., {Bean}, J.~L., {Louie}, D.~R., {et~al.} 2018, \pasp, 130,
  114401

\bibitem[{{Knutson} {et~al.}(2014){Knutson}, {Benneke}, {Deming}, \&
  {Homeier}}]{knutson14}
{Knutson}, H.~A., {Benneke}, B., {Deming}, D., \& {Homeier}, D. 2014, \nat,
  505, 66

\bibitem[{{Kreidberg}(2015)}]{Kreidberg2015}
{Kreidberg}, L. 2015, \pasp, 127, 1161

\bibitem[{{Kreidberg} {et~al.}(2014){Kreidberg}, {Bean}, {D{\'e}sert},
  {Benneke}, {Deming}, {Stevenson}, {Seager}, {Berta-Thompson}, {Seifahrt}, \&
  {Homeier}}]{kreidberg14}
{Kreidberg}, L., {Bean}, J.~L., {D{\'e}sert}, J.-M., {et~al.} 2014, \nat, 505,
  69

\bibitem[{{Kreidberg} {et~al.}(2022){Kreidberg}, {Molli{\`e}re}, {Crossfield},
  {Thorngren}, {Kawashima}, {Morley}, {Benneke}, {Mikal-Evans}, {Berardo},
  {Kosiarek}, {Gorjian}, {Ciardi}, {Christiansen}, {Dragomir}, {Dressing},
  {Fortney}, {Fulton}, {Greene}, {Hardegree-Ullman}, {Howard}, {Howell},
  {Isaacson}, {Krick}, {Livingston}, {Lothringer}, {Morales}, {Petigura},
  {Rodriguez}, {Schlieder}, \& {Weiss}}]{kreidberg22}
{Kreidberg}, L., {Molli{\`e}re}, P., {Crossfield}, I. J.~M., {et~al.} 2022,
  \aj, 164, 124

\bibitem[{Lecavelier~des Etangs {et~al.}(2008)Lecavelier~des Etangs, Pont,
  Vidal-Madjar, \& Sing}]{lecavelier_des_etangs_rayleigh_2008}
Lecavelier~des Etangs, A., Pont, F., Vidal-Madjar, A., \& Sing, D. 2008,
  Astronomy \& Astrophysics, 481, L83, number: 2 Publisher: EDP Sciences

\bibitem[{{Lee} \& {Chiang}(2016)}]{Lee2016}
{Lee}, E.~J., \& {Chiang}, E. 2016, \apj, 817, 90

\bibitem[{Li {et~al.}(2015)Li, Gordon, Rothman, Tan, Hu, Kassi, Campargue, \&
  Medvedev}]{li_rovibrational_2015}
Li, G., Gordon, I.~E., Rothman, L.~S., {et~al.} 2015, The Astrophysical Journal
  Supplement Series, 216, 15, publisher: The American Astronomical Society

\bibitem[{{Libby-Roberts} {et~al.}(2020){Libby-Roberts}, {Berta-Thompson},
  {D{\'e}sert}, {Masuda}, {Morley}, {Lopez}, {Deck}, {Fabrycky}, {Fortney},
  {Line}, {Sanchis-Ojeda}, \& {Winn}}]{libbyroberts20}
{Libby-Roberts}, J.~E., {Berta-Thompson}, Z.~K., {D{\'e}sert}, J.-M., {et~al.}
  2020, \aj, 159, 57

\bibitem[{{Libby-Roberts} {et~al.}(2022){Libby-Roberts}, {Berta-Thompson},
  {Diamond-Lowe}, {Gully-Santiago}, {Irwin}, {Kempton}, {Rackham},
  {Charbonneau}, {D{\'e}sert}, {Dittmann}, {Hofmann}, {Morley}, \&
  {Newton}}]{Libby-Roberts2022}
{Libby-Roberts}, J.~E., {Berta-Thompson}, Z.~K., {Diamond-Lowe}, H., {et~al.}
  2022, \aj, 164, 59

\bibitem[{{Lim} {et~al.}(2023){Lim}, {Benneke}, {Doyon}, {MacDonald},
  {Piaulet}, {Artigau}, {Coulombe}, {Radica}, {L'Heureux}, {Albert}, {Rackham},
  {de Wit}, {Salhi}, {Roy}, {Flagg}, {Fournier-Tondreau}, {Taylor}, {Cook},
  {Lafreni{\`e}re}, {Cowan}, {Kaltenegger}, {Rowe}, {Espinoza}, {Dang}, \&
  {Darveau-Bernier}}]{Lim2023}
{Lim}, O., {Benneke}, B., {Doyon}, R., {et~al.} 2023, \apjl, 955, L22

\bibitem[{Line \& Parmentier(2016)}]{line_influence_2016}
Line, M.~R., \& Parmentier, V. 2016, The Astrophysical Journal, 820, 78,
  publisher: The American Astronomical Society

\bibitem[{Line {et~al.}(2013)Line, Wolf, Zhang, Knutson, Kammer, Ellison,
  Deroo, Crisp, \& Yung}]{line_systematic_2013}
Line, M.~R., Wolf, A.~S., Zhang, X., {et~al.} 2013, The Astrophysical Journal,
  775, 137, publisher: The American Astronomical Society

\bibitem[{{Lodders} {et~al.}(2009){Lodders}, {Palme}, \& {Gail}}]{Lodders2009}
{Lodders}, K., {Palme}, H., \& {Gail}, H.~P. 2009, Landolt B\&ouml;rnstein, 4B,
  712

\bibitem[{{Lopez} \& {Fortney}(2014)}]{Lopez2014}
{Lopez}, E.~D., \& {Fortney}, J.~J. 2014, \apj, 792, 1

\bibitem[{Lopez~Murillo(in prep)}]{LopezMurillo_inprep}
Lopez~Murillo, A.~I. in prep

\bibitem[{Lupu {et~al.}(2023)Lupu, Freedman, Gharib-Nezhad, Visscher, \&
  Molliere}]{lupu_2023_7542068}
Lupu, R., Freedman, R., Gharib-Nezhad, E., Visscher, C., \& Molliere, P. 2023,
  {Correlated k coefficients for H2-He atmospheres; 196 spectral windows and
  1460 pressure-temperature points}, doi:10.5281/zenodo.7542068

\bibitem[{Madhusudhan(2019)}]{madhusudhan_exoplanetary_2019}
Madhusudhan, N. 2019, Annual Review of Astronomy and Astrophysics, 57, 617, aDS
  Bibcode: 2019ARA\&A..57..617M

\bibitem[{{Madhusudhan} {et~al.}(2017){Madhusudhan}, {Bitsch}, {Johansen}, \&
  {Eriksson}}]{Madhusudhan2017}
{Madhusudhan}, N., {Bitsch}, B., {Johansen}, A., \& {Eriksson}, L. 2017,
  \mnras, 469, 4102

\bibitem[{Madhusudhan \& Seager(2009)}]{madhusudhan_temperature_2009}
Madhusudhan, N., \& Seager, S. 2009, The Astrophysical Journal, 707, 24, aDS
  Bibcode: 2009ApJ...707...24M

\bibitem[{{Madhusudhan} \& {Seager}(2011)}]{Madhusudhan2011ApJ_GJ436b}
{Madhusudhan}, N., \& {Seager}, S. 2011, \apj, 729, 41

\bibitem[{Mai \& Line(2019)}]{mai_exploring_2019}
Mai, C., \& Line, M.~R. 2019, The Astrophysical Journal, 883, 144, publisher:
  The American Astronomical Society

\bibitem[{{Mann} {et~al.}(2016){Mann}, {Gaidos}, {Mace}, {Johnson}, {Bowler},
  {LaCourse}, {Jacobs}, {Vanderburg}, {Kraus}, {Kaplan}, \&
  {Jaffe}}]{Mann2016a}
{Mann}, A.~W., {Gaidos}, E., {Mace}, G.~N., {et~al.} 2016, \apj, 818, 46

\bibitem[{{Mikal-Evans} {et~al.}(2021){Mikal-Evans}, {Crossfield}, {Benneke},
  {Kreidberg}, {Moses}, {Morley}, {Thorngren}, {Molli{\`e}re},
  {Hardegree-Ullman}, {Brewer}, {Christiansen}, {Ciardi}, {Dragomir},
  {Dressing}, {Fortney}, {Gorjian}, {Greene}, {Hirsch}, {Howard}, {Howell},
  {Isaacson}, {Kosiarek}, {Krick}, {Livingston}, {Lothringer}, {Morales},
  {Petigura}, {Schlieder}, \& {Werner}}]{mikalevans21}
{Mikal-Evans}, T., {Crossfield}, I. J.~M., {Benneke}, B., {et~al.} 2021, \aj,
  161, 18

\bibitem[{{Mikal-Evans} {et~al.}(2023){Mikal-Evans}, {Madhusudhan}, {Dittmann},
  {G{\"u}nther}, {Welbanks}, {Van Eylen}, {Crossfield}, {Daylan}, \&
  {Kreidberg}}]{mikalevans23}
{Mikal-Evans}, T., {Madhusudhan}, N., {Dittmann}, J., {et~al.} 2023, \aj, 165,
  84

\bibitem[{{Mordasini} {et~al.}(2016){Mordasini}, {van Boekel}, {Molli{\`e}re},
  {Henning}, \& {Benneke}}]{mordasini16}
{Mordasini}, C., {van Boekel}, R., {Molli{\`e}re}, P., {Henning}, T., \&
  {Benneke}, B. 2016, \apj, 832, 41

\bibitem[{{Mukherjee} {et~al.}(2023){Mukherjee}, {Batalha}, {Fortney}, \&
  {Marley}}]{Mukherjee2023}
{Mukherjee}, S., {Batalha}, N.~E., {Fortney}, J.~J., \& {Marley}, M.~S. 2023,
  \apj, 942, 71

\bibitem[{{Nardiello}(2020)}]{2020MNRAS.498.5972N}
{Nardiello}, D. 2020, \mnras, 498, 5972

\bibitem[{{NASA Exoplanet Science Institute}(2020)}]{NASAexplanetarchive}
{NASA Exoplanet Science Institute}. 2020, Planetary Systems Table,
  doi:10.26133/NEA12

\bibitem[{{\"O}berg {et~al.}(2011){\"O}berg, Murray-Clay, \&
  Bergin}]{oberg2011effects}
{\"O}berg, K.~I., Murray-Clay, R., \& Bergin, E.~A. 2011, The Astrophysical
  Journal Letters, 743, L16

\bibitem[{Ohno \& Tanaka(2021)}]{Ohno2021}
Ohno, K., \& Tanaka, Y.~A. 2021, The Astrophysical Journal, 920, 124

\bibitem[{Ohno {et~al.}(2022)Ohno, Thao, Mann, \&
  Fortney}]{ohno2022circumplanetary}
Ohno, K., Thao, P.~C., Mann, A.~W., \& Fortney, J.~J. 2022, The Astrophysical
  Journal Letters, 940, L30

\bibitem[{{Owen} \& {Wu}(2016)}]{Owen2016}
{Owen}, J.~E., \& {Wu}, Y. 2016, \apj, 817, 107

\bibitem[{Owen \& Wu(2017)}]{owen2017evaporation}
Owen, J.~E., \& Wu, Y. 2017, The Astrophysical Journal, 847, 29

\bibitem[{{Parmentier} {et~al.}(2013){Parmentier}, {Showman}, \&
  {Lian}}]{Parmentier2013}
{Parmentier}, V., {Showman}, A.~P., \& {Lian}, Y. 2013, \aap, 558, A91

\bibitem[{{Parviainen} \& {Aigrain}(2015{\natexlab{a}})}]{2015MNRAS.453.3821P}
{Parviainen}, H., \& {Aigrain}, S. 2015{\natexlab{a}}, \mnras, 453, 3821

\bibitem[{{Parviainen} \& {Aigrain}(2015{\natexlab{b}})}]{Parviainen2015}
---. 2015{\natexlab{b}}, \mnras, 453, 3821

\bibitem[{{Pizzolato} {et~al.}(2003){Pizzolato}, {Maggio}, {Micela},
  {Sciortino}, \& {Ventura}}]{Pizzolato2003}
{Pizzolato}, N., {Maggio}, A., {Micela}, G., {Sciortino}, S., \& {Ventura}, P.
  2003, \aap, 397, 147

\bibitem[{{Pollack} {et~al.}(1996){Pollack}, {Hubickyj}, {Bodenheimer},
  {Lissauer}, {Podolak}, \& {Greenzweig}}]{Pollack1996}
{Pollack}, J.~B., {Hubickyj}, O., {Bodenheimer}, P., {et~al.} 1996, \icarus,
  124, 62

\bibitem[{Polyansky {et~al.}(2018)Polyansky, Kyuberis, Zobov, Tennyson,
  Yurchenko, \& Lodi}]{polyansky_exomol_2018}
Polyansky, O.~L., Kyuberis, A.~A., Zobov, N.~F., {et~al.} 2018, Monthly Notices
  of the Royal Astronomical Society, 480, 2597

\bibitem[{{Powell} {et~al.}(2024){Powell}, {Feinstein}, {Lee}, {Zhang}, {Tsai},
  {Taylor}, {Kirk}, {Bell}, {Barstow}, \& {Gao}}]{powell24}
{Powell}, D., {Feinstein}, A.~D., {Lee}, E. K.~H., {et~al.} 2024, Nature

\bibitem[{{Predehl} \& {Schmitt}(1995)}]{Predehl1995}
{Predehl}, P., \& {Schmitt}, J.~H.~M.~M. 1995, \aap, 293, 889

\bibitem[{Price-Whelan {et~al.}(2018)Price-Whelan, Sip{\H{o}}cz, G{\"u}nther,
  Lim, Crawford, Conseil, Shupe, Craig, Dencheva, Ginsburg,
  {et~al.}}]{price2018astropy}
Price-Whelan, A.~M., Sip{\H{o}}cz, B., G{\"u}nther, H., {et~al.} 2018, The
  Astronomical Journal, 156, 123

\bibitem[{Rackham {et~al.}(2017)Rackham, Espinoza, Apai, L{\'o}pez-Morales,
  Jord{\'a}n, Osip, Lewis, Rodler, Fraine, Morley,
  {et~al.}}]{rackham2017access}
Rackham, B., Espinoza, N., Apai, D., {et~al.} 2017, The Astrophysical Journal,
  834, 151

\bibitem[{{Rackham} {et~al.}(2018){Rackham}, {Apai}, \&
  {Giampapa}}]{Rackham2018}
{Rackham}, B.~V., {Apai}, D., \& {Giampapa}, M.~S. 2018, \apj, 853, 122

\bibitem[{{Radica} {et~al.}(2023){Radica}, {Welbanks}, {Espinoza}, {Taylor},
  {Coulombe}, {Feinstein}, {Goyal}, {Scarsdale}, {Albert}, {Baghel}, {Bean},
  {Blecic}, {Lafreni{\`e}re}, {MacDonald}, {Zamyatina}, {Allart1}, {Artigau},
  {Batalha}, {Cook}, {Cowan}, {Dang}, {Doyon}, {Fournier-Tondreau},
  {Johnstone}, {Line}, {Moran}, {Mukherjee}, {Pelletier}, {Roy}, {Talens},
  {Filippazzo}, {Pontoppidan}, \& {Volk}}]{Radica2023}
{Radica}, M., {Welbanks}, L., {Espinoza}, N., {et~al.} 2023, \mnras, 524, 835

\bibitem[{Ribas {et~al.}(2005)Ribas, Guinan, G{\"u}del, \& Audard}]{Ribas2005}
Ribas, I., Guinan, E.~F., G{\"u}del, M., \& Audard, M. 2005, The Astrophysical
  Journal, 622, 680

\bibitem[{{Ricker}(2014)}]{Ricker2014}
{Ricker}, G.~R. 2014, Journal of the American Association of Variable Star
  Observers (JAAVSO), 42, 234

\bibitem[{{Rizzuto} {et~al.}(2020){Rizzuto}, {Newton}, {Mann}, {Tofflemire},
  {Vanderburg}, {Kraus}, {Wood}, {Quinn}, {Zhou}, {Thao}, {Law}, {Ziegler}, \&
  {Briceno}}]{Rizzuto2020}
{Rizzuto}, A.~C., {Newton}, E.~R., {Mann}, A.~W., {et~al.} 2020, arXiv
  e-prints, arXiv:2005.00013

\bibitem[{Robitaille {et~al.}(2013)Robitaille, Tollerud, Greenfield,
  Droettboom, Bray, Aldcroft, Davis, Ginsburg, Price-Whelan, Kerzendorf,
  {et~al.}}]{ASTROPY}
Robitaille, T.~P., Tollerud, E.~J., Greenfield, P., {et~al.} 2013, Astronomy \&
  Astrophysics, 558, A33

\bibitem[{{Rogers} \& {Owen}(2021)}]{Rogers2021}
{Rogers}, J.~G., \& {Owen}, J.~E. 2021, \mnras, 503, 1526

\bibitem[{{Rogers} {et~al.}(2024){Rogers}, {Owen}, \&
  {Schlichting}}]{Rogers2024}
{Rogers}, J.~G., {Owen}, J.~E., \& {Schlichting}, H.~E. 2024, \mnras, 529, 2716

\bibitem[{{Roy} {et~al.}(2023){Roy}, {Benneke}, {Piaulet}, {Gully-Santiago},
  {Crossfield}, {Morley}, {Kreidberg}, {Mikal-Evans}, {Brande}, {Delisle},
  {Greene}, {Hardegree-Ullman}, {Barman}, {Christiansen}, {Dragomir},
  {Fortney}, {Howard}, {Kosiarek}, \& {Lothringer}}]{roy23}
{Roy}, P.-A., {Benneke}, B., {Piaulet}, C., {et~al.} 2023, \apjl, 954, L52

\bibitem[{Rustamkulov {et~al.}(2023)Rustamkulov, Sing, Mukherjee, May, Kirk,
  Schlawin, Line, Piaulet, Carter, Batalha, {et~al.}}]{rustamkulov2023early}
Rustamkulov, Z., Sing, D., Mukherjee, S., {et~al.} 2023, Nature, 614, 659

\bibitem[{Schwarz(1978)}]{schwarz1978_bic}
Schwarz, G. 1978, The annals of statistics, 461

\bibitem[{{Seager} \& {Sasselov}(2000)}]{Seager2000}
{Seager}, S., \& {Sasselov}, D.~D. 2000, \apj, 537, 916

\bibitem[{{Sromovsky} {et~al.}(2011){Sromovsky}, {Fry}, \&
  {Kim}}]{Sromovsky2011}
{Sromovsky}, L.~A., {Fry}, P.~M., \& {Kim}, J.~H. 2011, \icarus, 215, 292

\bibitem[{{Stevenson}(2016)}]{stevenson2016}
{Stevenson}, K.~B. 2016, \apjl, 817, L16

\bibitem[{Thao(in prep)}]{Thao_inprep}
Thao, P.~C. in prep

\bibitem[{{Thao} {et~al.}(2020){Thao}, {Mann}, {Johnson}, {Newton}, {Guo},
  {Kain}, {Rizzuto}, {Charbonneau}, {Dalba}, {Gaidos}, {Irwin}, \&
  {Kraus}}]{Thao2020}
{Thao}, P.~C., {Mann}, A.~W., {Johnson}, M.~C., {et~al.} 2020, \aj, 159, 32

\bibitem[{{Thao} {et~al.}(2023){Thao}, {Mann}, {Gao}, {Owens}, {Vanderburg},
  {Newton}, {Tang}, {Fields}, {David}, {Irwin}, {Husser}, {Charbonneau}, \&
  {Ballard}}]{Thao2023}
{Thao}, P.~C., {Mann}, A.~W., {Gao}, P., {et~al.} 2023, \aj, 165, 23

\bibitem[{Thompson(1990)}]{Thompson1990}
Thompson, S.~L. 1990, {ANEOS} analytic equations of state for shock physics
  codes input manual, doi:10.2172/6939284

\bibitem[{{Thorngren} \& {Fortney}(2019)}]{Thorngren2019}
{Thorngren}, D., \& {Fortney}, J.~J. 2019, \apjl, 874, L31

\bibitem[{{Thorngren} \& {Fortney}(2018)}]{Thorngren2018}
{Thorngren}, D.~P., \& {Fortney}, J.~J. 2018, \aj, 155, 214

\bibitem[{Tran {et~al.}(2021)Tran, Bowler, Cochran, Endl, Stef{\'a}nsson,
  Mahadevan, Ninan, Bender, Halverson, Roy, {et~al.}}]{tran2021epoch}
Tran, Q.~H., Bowler, B.~P., Cochran, W.~D., {et~al.} 2021, The Astronomical
  Journal, 161, 173

\bibitem[{{Tsai} {et~al.}(2017){Tsai}, {Lyons}, {Grosheintz}, {Rimmer},
  {Kitzmann}, \& {Heng}}]{Tsai2017}
{Tsai}, S.-M., {Lyons}, J.~R., {Grosheintz}, L., {et~al.} 2017, \apjs, 228, 20

\bibitem[{Tsai {et~al.}(2017)Tsai, Lyons, Grosheintz, Rimmer, Kitzmann, \&
  Heng}]{VULCAN}
Tsai, S.-M., Lyons, J.~R., Grosheintz, L., {et~al.} 2017, The Astrophysical
  Journal Supplement Series, 228, 20

\bibitem[{{Tsai} {et~al.}(2021){Tsai}, {Malik}, {Kitzmann}, {Lyons}, {Fateev},
  {Lee}, \& {Heng}}]{Tsai2021}
{Tsai}, S.-M., {Malik}, M., {Kitzmann}, D., {et~al.} 2021, \apj, 923, 264

\bibitem[{{Tsai} {et~al.}(2023){Tsai}, {Lee}, {Powell}, {Gao}, {Zhang},
  {Moses}, {H{\'e}brard}, {Venot}, {Parmentier}, {Jordan}, {Hu}, {Alam},
  {Alderson}, {Batalha}, {Bean}, {Benneke}, {Bierson}, {Brady}, {Carone},
  {Carter}, {Chubb}, {Inglis}, {Leconte}, {Line}, {L{\'o}pez-Morales},
  {Miguel}, {Molaverdikhani}, {Rustamkulov}, {Sing}, {Stevenson}, {Wakeford},
  {Yang}, {Aggarwal}, {Baeyens}, {Barat}, {de Val-Borro}, {Daylan}, {Fortney},
  {France}, {Goyal}, {Grant}, {Kirk}, {Kreidberg}, {Louca}, {Moran},
  {Mukherjee}, {Nasedkin}, {Ohno}, {Rackham}, {Redfield}, {Taylor}, {Tremblin},
  {Visscher}, {Wallack}, {Welbanks}, {Youngblood}, {Ahrer}, {Batalha}, {Behr},
  {Berta-Thompson}, {Blecic}, {Casewell}, {Crossfield}, {Crouzet}, {Cubillos},
  {Decin}, {D{\'e}sert}, {Feinstein}, {Gibson}, {Harrington}, {Heng},
  {Henning}, {Kempton}, {Krick}, {Lagage}, {Lendl}, {Lothringer}, {Mansfield},
  {Mayne}, {Mikal-Evans}, {Palle}, {Schlawin}, {Shorttle}, {Wheatley}, \&
  {Yurchenko}}]{Tsai2023}
{Tsai}, S.-M., {Lee}, E. K.~H., {Powell}, D., {et~al.} 2023, \nat, 617, 483

\bibitem[{{Turrini} {et~al.}(2021){Turrini}, {Schisano}, {Fonte}, {Molinari},
  {Politi}, {Fedele}, {Pani{\'c}}, {Kama}, {Changeat}, \&
  {Tinetti}}]{Turrini2021}
{Turrini}, D., {Schisano}, E., {Fonte}, S., {et~al.} 2021, \apj, 909, 40

\bibitem[{{Twicken} {et~al.}(2010){Twicken}, {Clarke}, {Bryson}, {Tenenbaum},
  {Wu}, {Jenkins}, {Girouard}, \& {Klaus}}]{Twicken2010}
{Twicken}, J.~D., {Clarke}, B.~D., {Bryson}, S.~T., {et~al.} 2010, in Society
  of Photo-Optical Instrumentation Engineers (SPIE) Conference Series, Vol.
  7740, Software and Cyberinfrastructure for Astronomy, ed. N.~M. {Radziwill}
  \& A.~{Bridger}, 774023

\bibitem[{Underwood {et~al.}(2016)Underwood, Tennyson, Yurchenko, Huang,
  Schwenke, Lee, Clausen, \& Fateev}]{underwood_exomol_2016}
Underwood, D.~S., Tennyson, J., Yurchenko, S.~N., {et~al.} 2016, Monthly
  Notices of the Royal Astronomical Society, 459, 3890

\bibitem[{{Vach} {et~al.}(2024){Vach}, {Zhou}, {Huang}, {Rogers}, {Bouma},
  {Douglas}, {Kunimoto}, {Mann}, {Barber}, {Quinn}, {Latham}, {Bieryla}, \&
  {Collins}}]{Vach2024}
{Vach}, S., {Zhou}, G., {Huang}, C.~X., {et~al.} 2024, arXiv e-prints,
  arXiv:2403.03261

\bibitem[{Vanderburg {et~al.}(2019)Vanderburg, Huang, Rodriguez, Becker,
  Ricker, Vanderspek, Latham, Seager, Winn, Jenkins,
  {et~al.}}]{vanderburg2019tess}
Vanderburg, A., Huang, C.~X., Rodriguez, J.~E., {et~al.} 2019, The
  Astrophysical Journal Letters, 881, L19

\bibitem[{{Vilhu} \& {Walter}(1987)}]{VilhuWalter1987}
{Vilhu}, O., \& {Walter}, F.~M. 1987, \apj, 321, 958

\bibitem[{Virtanen {et~al.}(2020)Virtanen, Gommers, Oliphant, Haberland, Reddy,
  Cournapeau, Burovski, Peterson, Weckesser, Bright, {et~al.}}]{SCIPY}
Virtanen, P., Gommers, R., Oliphant, T.~E., {et~al.} 2020, Nature methods, 17,
  261

\bibitem[{{Wang} \& {Dai}(2019)}]{Wang2019}
{Wang}, L., \& {Dai}, F. 2019, \apj, 873, L1

\bibitem[{Welbanks \& Madhusudhan(2021)}]{welbanks_aurora_2021}
Welbanks, L., \& Madhusudhan, N. 2021, The Astrophysical Journal, 913, 114,
  publisher: The American Astronomical Society

\bibitem[{{Welbanks} {et~al.}(2024){Welbanks}, {Bell}, {Beatty}, {Line},
  {Ohno}, {Fortney}, {Schlawin}, {Greene}, {Rauscher}, {McGill}, {Murphy},
  {Parmentier}, {Tang}, {Edelman}, {Mukherjee}, {Wiser}, {Lagage}, {Dyrek}, \&
  {Arnold}}]{Welbanks2024}
{Welbanks}, L., {Bell}, T.~J., {Beatty}, T.~G., {et~al.} 2024, \nat, 630, 836

\bibitem[{Wood {et~al.}(2005)Wood, Redfield, Linsky, M{\"u}ller, \&
  Zank}]{Wood2005}
Wood, B.~E., Redfield, S., Linsky, J.~L., M{\"u}ller, H.-R., \& Zank, G.~P.
  2005, The Astrophysical Journal Supplement Series, 159, 118

\bibitem[{{Wood} {et~al.}(2023){Wood}, {Mann}, {Barber}, {Bush}, {Kraus},
  {Tofflemire}, {Vanderburg}, {Newton}, {Feiden}, {Zhou}, {Bouma}, {Quinn},
  {Armstrong}, {Osborn}, {Adibekyan}, {Mena}, {Sousa}, {Gagn{\'e}}, {Fields},
  {Milburn}, {Thao}, {Schmidt}, {Gnilka}, {Howell}, {Law}, {Ziegler},
  {Brice{\~n}o}, {Ricker}, {Vanderspek}, {Latham}, {Seager}, {Winn}, {Jenkins},
  {Schlieder}, {Osborn}, {Twicken}, {Ciardi}, \& {Huang}}]{wood23}
{Wood}, M.~L., {Mann}, A.~W., {Barber}, M.~G., {et~al.} 2023, \aj, 165, 85

\bibitem[{{Wright} {et~al.}(2018){Wright}, {Newton}, {Williams}, {Drake}, \&
  {Yadav}}]{Wright2018}
{Wright}, N.~J., {Newton}, E.~R., {Williams}, P. K.~G., {Drake}, J.~J., \&
  {Yadav}, R.~K. 2018, \mnras, 479, 2351

\bibitem[{{Xue} {et~al.}(2024){Xue}, {Bean}, {Zhang}, {Welbanks}, {Lunine}, \&
  {August}}]{Xue2024}
{Xue}, Q., {Bean}, J.~L., {Zhang}, M., {et~al.} 2024, \apjl, 963, L5

\bibitem[{Youngblood {et~al.}(2016)Youngblood, France, Loyd, Linsky, Redfield,
  Schneider, Wood, Brown, Froning, Miguel, {et~al.}}]{Youngblood2016}
Youngblood, A., France, K., Loyd, R.~P., {et~al.} 2016, The Astrophysical
  Journal, 824, 101

\bibitem[{{Zahnle} {et~al.}(2009){Zahnle}, {Marley}, {Freedman}, {Lodders}, \&
  {Fortney}}]{Zahnle2009}
{Zahnle}, K., {Marley}, M.~S., {Freedman}, R.~S., {Lodders}, K., \& {Fortney},
  J.~J. 2009, \apjl, 701, L20

\bibitem[{Zeng {et~al.}(2019)Zeng, Jacobsen, Sasselov, Petaev, Vanderburg,
  Lopez-Morales, Perez-Mercader, Mattsson, Li, Heising,
  {et~al.}}]{zeng2019growth}
Zeng, L., Jacobsen, S.~B., Sasselov, D.~D., {et~al.} 2019, Proceedings of the
  National Academy of Sciences, 116, 9723

\end{thebibliography}

\end{document}